\documentclass[12pt]{article}
\usepackage[nodisplayskipstretch]{setspace}
\usepackage[stretch=10,shrink=10]{microtype}
\usepackage{multirow}
\usepackage{amsmath,amsthm,amssymb,comment,paralist,graphicx,mathbbol,nicefrac,bm,mathtools,theoremref,xcolor,tikz,xurl,pgfplots,enumitem}
\usetikzlibrary{patterns.meta}
\usepackage[margin=1in]{geometry}

\usepackage[T1]{fontenc}  
\usepackage{hyperref}
\hypersetup{
  colorlinks=true,
  linkcolor={red!80!black},
  citecolor={black!50!blue},
  filecolor=magenta,      
  urlcolor=blue,
}
\pgfplotsset{compat=1.18}
\usepgfplotslibrary{groupplots}

\setlist[enumerate]{itemsep=0pt, topsep=3pt}
\usepackage{titlesec}

\titlespacing*{\section}
  {0pt}    
  {2.0ex plus .2ex minus .2ex} 
  {1.2ex plus .2ex minus .2ex}            
\titlespacing*{\subsection}
  {0pt}     
  {2.0ex plus .2ex minus .2ex}   
  {1.2ex plus .2ex minus .2ex}

\definecolor{hazyblue}{RGB}{135, 150, 230}

\usepackage[nameinlink]{cleveref}
\usepackage{subcaption}
\makeatletter
\renewcommand{\p@subfigure}{\thefigure}
\makeatother
\usetikzlibrary{calc,positioning,patterns,arrows.meta}
\theoremstyle{definition}
\newtheorem{definition}{Definition}
\newtheorem*{definition*}{Definition}
\newtheorem{example}{Example}
\newtheorem*{example*}{Example}

\theoremstyle{plain}
\newtheorem{lemma}{Lemma}
\newtheorem{proposition}{Proposition}

\newtheorem{corollary}{Corollary}
\newtheorem{theorem}{Theorem}
\newtheorem{conjecture}{Conjecture}
\newtheorem*{theorem*}{Theorem}
\theoremstyle{definition}
\newtheorem*{lemma*}{Lemma}

\usepackage{newtxtext}

\usepackage[authoryear]{natbib}  

\begin{document}
\title{Herding with Multiple States\thanks{This paper is based on the author's master's thesis at the University of Tokyo. I am indebted to Yuichi Yamamoto for his guidance and support. I thank Navin Kartik for helpful comments. This paper has benefited from discussions with Yu Awaya, Tianxiang Cong, Benjamin Golub, Michihiro Kandori, Fuhito Kojima, Hongcheng Li, Xuanguang Li, Stephen Morris, Shinpei Noguchi, Daisuke Oyama, Hiroto Sato, Bruno Strulovici, Kenji Utagawa, Zhang Xu, Wei Zhao and Renjie Zhong. I thank the participants in the 2026 Asian School in Economic Theory. All errors are my own. }}

\author{Xiaoyu Chen\thanks{Graduate School of Economics, The University of Tokyo. Email: xychen@g.ecc.u-tokyo.ac.jp.}}

\date{December 1, 2026}

\maketitle

\begin{abstract}
In sequential observational learning, does society eventually take the correct action? With two states, the answer does not depend on the full-support initial prior. We show that with more than two states it can. The same signals and preferences can lead to correct learning for some full-support priors and to incorrect herding for others. We ask when herding occurs with multiple states. The key condition, \textit{escapability}, jointly restricts signals and preferences. Escapability asks whether signals can provide arbitrarily strong evidence against any given action. If it fails, herding occurs for some prior; if its pairwise variant fails, herding occurs for every full-support prior. When states are ordered, each action is optimal on an interval of states, and signals satisfy the monotone likelihood ratio property, escapability completely characterizes learning and herding, regardless of the full-support prior. Within this class, with location-shift signals, escapability reduces to a simple pairwise distinguishability condition.

\end{abstract}

\section{Introduction}
Rational reliance on others' decisions can prevent society from aggregating the information needed to correct those decisions. This is the key message from the economic literature on observational learning, initiated by \citet{banerjee_simple_1992} and \citet{BHW1992}. In the canonical model, there is an unknown payoff-relevant state. Agents act sequentially, each observing a private signal and the actions of her predecessors, and choosing an action to maximize expected utility. The central question is: \textit{do agents asymptotically take the correct action (learning) or not (herding)?}

With binary states and finitely many actions, the answer depends on the information structure—specifically, on whether signals induce unbounded or bounded beliefs. The former leads to learning, while the latter leads to (incorrect) herding for nontrivial preferences \citep{smith_pathological_2000}.

This paper studies the canonical model with more than two states. Unbounded beliefs still imply learning \citep{arieli_general_2021}, but this condition is too demanding. \emph{Preferences also matter for learning with multiple states}. In their published article, \citet*[hereafter KLLR]{kartik_beyond_2024} introduce \emph{excludability}, a condition that jointly restricts preferences and information. This condition is weaker than unbounded beliefs and implies learning for every initial prior, but its failure need not imply herding.

The initial prior has received less attention than information and preferences, partly because it has been viewed as irrelevant to whether society learns. In the classic model with binary states and finitely many actions, if herding occurs for some initial prior, it occurs for every full-support prior. In their working paper, KLLR \citeyearpar{kartik_beyond_arxiv_v5_2024} conjecture a property of infinite action histories that would imply the same prior independence with multiple states and two actions.

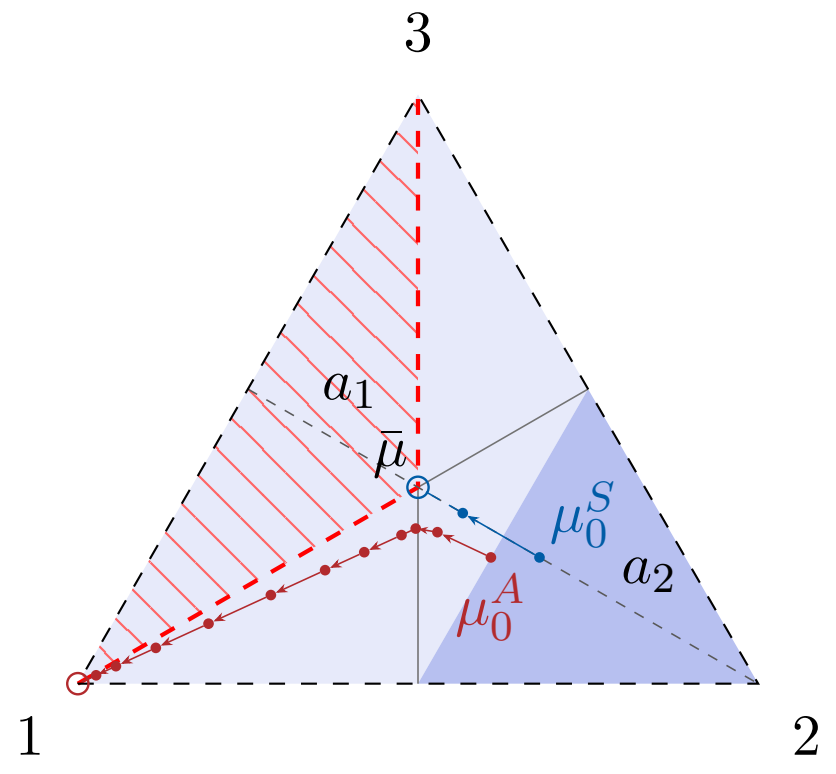
\begin{figure}[htbp]
\centering
\begin{tikzpicture}[x=4.5cm,y=4.5cm,font=\small,
  path arrow/.style={-{Stealth[length=2.2pt,width=1.7pt]},line width=0.3pt,
    shorten <=1pt,shorten >=1pt},
  region boundary/.style={black!55,line width=0.3pt},
  belief label/.style={inner sep=1pt,font=\scriptsize}]
  \definecolor{learningred}{RGB}{178,42,45}
  \definecolor{herdingblue}{RGB}{0,92,166}
  \coordinate (S1) at (0,0);
  \coordinate (S2) at (1,0);
  \coordinate (S3) at (0.5,{sqrt(3)/2});
  \coordinate (B) at ($(S1)!0.5!(S2)$);
  \coordinate (T) at ($(S2)!0.5!(S3)$);
  \coordinate (M) at ($(S1)!0.5!(S3)$);
  \coordinate (C) at (0.5,{sqrt(3)/6});

  \fill[hazyblue!20] (S1)--(S3)--(T)--(B)--cycle;
  \fill[hazyblue!60] (B)--(T)--(S2)--cycle;
  \fill[pattern={Lines[angle=135,distance=6pt]},pattern color=red!60]
    (S1)--(C)--(S3)--cycle;
  \draw[region boundary] (B)--(C);
  \draw[region boundary] (C)--(T);
  \draw[thick,dashed,red] (S1)--(C)--(S3);
  \draw[black!65,dash pattern=on 2pt off 2pt,line width=0.3pt] (M)--(S2);


  \coordinate (A0) at (0.607142857143,0.185576872240); 
  \coordinate (A1) at (0.528571428571,0.222692246687); 
  \coordinate (A2) at (0.496640465602,0.227982591499); 
  \coordinate (A3) at (0.475727387029,0.218382451802); 
  \coordinate (A4) at (0.420987238791,0.193254010366); 
  \coordinate (A5) at (0.363445825351,0.166839649348); 
  \coordinate (A6) at (0.283740470286,0.130250940488); 
  \coordinate (A7) at (0.192087756518,0.088177801769); 
  \coordinate (A8) at (0.114620211181,0.052616358499); 
  \coordinate (A9) at (0.055984818889,0.025699806961); 
  \coordinate (A10) at (0.026847107356,0.012324153051); 
  \coordinate (Alimit) at (0.000000000000,0.000000000000); 
  \draw[path arrow,learningred] (A0)--(A1);
  \draw[path arrow,learningred,shorten <=0.5pt,shorten >=0.5pt,-{Stealth[length=1.8pt,width=1.4pt]}] (A1)--(A2);
  \draw[learningred,line width=0.3pt] (A2)--(A3);
  \draw[path arrow,learningred] (A3)--(A4);
  \draw[path arrow,learningred] (A4)--(A5);
  \draw[path arrow,learningred] (A5)--(A6);
  \draw[path arrow,learningred] (A6)--(A7);
  \draw[path arrow,learningred] (A7)--(A8);
  \draw[path arrow,learningred] (A8)--(A9);
  \draw[path arrow,learningred,shorten <=0.5pt,shorten >=0.5pt,-{Stealth[length=1.8pt,width=1.4pt]}] (A9)--(A10);
  \draw[learningred,line width=0.3pt,densely dashed] (A10)--(Alimit);
  \foreach \i in {0,...,10} \fill[learningred] (A\i) circle[radius=1pt];
  \node[below=2pt,text=learningred,belief label,font=\small] at (A0) {$\mu_0^{A}$};
  \coordinate (H0) at (0.678571428571,0.185576872240); 
  \coordinate (H1) at (0.565789473684,0.250691564253); 
  \draw[path arrow,herdingblue] (H0)--(H1);
  \draw[herdingblue,line width=0.3pt,densely dashed] (H1)--(C);
  \foreach \i in {0,1} \fill[herdingblue] (H\i) circle[radius=1pt];
  \node[above right=1pt,text=herdingblue,belief label,font=\small] at (H0) {$\mu_0^{S}$};

  \draw[dashed] (S1)--(S2)--(S3)--cycle;
  \draw[learningred,fill=none,line width=0.45pt] (S1) circle[radius=2pt];
  \draw[herdingblue,fill=none,line width=0.45pt] (C) circle[radius=2pt];
  \node[above left=1pt,belief label, font= \small] at (C)
    {$\bar\mu$};
  \node[font=\small] at (0.4,0.43) {$a_1$};
  \node[font=\small] at (0.84,0.16) {$a_2$};
  \node[below left=3pt] at (S1) {$1$};
  \node[below right=3pt] at (S2) {$2$};
  \node[above=4pt] at (S3) {$3$};

\end{tikzpicture}
\caption{An example where learning or herding depends on the initial prior (\Cref{ex:counter}). }
\label{fig:counterexample_path}
\end{figure}

We disprove this conjecture. \emph{With multiple states, the initial prior matters}. Consider an example with three states $\{1,2,3\}$ and two actions $\{a_1,a_2\}$. Action $a_1$ is optimal in states $1$ and $3$, while $a_2$ is optimal in state $2$. \Cref{fig:counterexample_path} depicts the belief simplex and each action's optimality region. The hatched region consists of ``cascade beliefs,'' at which $a_1$ is optimal regardless of the received signal and society herds on $a_1$. If KLLR's conjecture were true, the public belief\footnote{The public belief is an agent's belief about the state after observing her predecessors' actions but before receiving her private signal.} would converge to this cascade region with positive probability from every full-support prior.

What happens is almost the opposite: learning occurs for generic full-support priors outside the cascade region. \Cref{fig:counterexample_path} depicts two belief paths under repeated $a_1$ observations. Starting from the symmetric prior $\mu_0^S$, where $\mu_0^S(1)=\mu_0^S(3)$, the public belief converges to the cascade belief $\bar\mu$. Starting from the asymmetric prior $\mu_0^A$, it converges instead to certainty that the state is $1$, without entering the cascade region. More generally, herding for priors outside the cascade region can occur only if some finite action history leads to a symmetric public belief satisfying $\mu(1)=\mu(3)$. The priors for which this can happen form a meager set of Lebesgue measure zero.

The counterexample highlights two features of social learning with multiple states. First, cascade beliefs can occupy a region with
nonempty interior even though learning occurs for generic full-support priors outside it. With multiple states, identifying cascade beliefs and identifying herding can therefore be different problems. Second, learning the correct action can require society to distinguish between states that are equivalent for the decision problem. In this example, even though states $1$ and $3$ share the same optimal action (in fact, the payoffs are the same), the true state is learned almost surely from generic full-support priors outside the cascade region. 

Given that learning can depend on the initial prior, we ask how preferences and information jointly determine \emph{herding}. We provide sufficient conditions for herding (equivalently, necessary conditions for learning)  that distinguish between herding for some initial prior and for every full-support prior. The conditions ask whether private signals can provide arbitrarily strong evidence that an action is ex post suboptimal. For a class of environments, they yield a complete characterization of learning and herding that is independent of the full-support initial prior. 

\begin{figure}[htbp]
  \centering
  \begin{subfigure}[t]{0.3\textwidth}
    \centering
    \resizebox{\linewidth}{!}{%
      \begin{tikzpicture}[scale=4.5]
        \coordinate (S1) at (0,0);
        \coordinate (S2) at (1,0); 
        \coordinate (S3) at (0.5, 0.866);
        \coordinate (B_05) at ($(S1)!0.5!(S2)$);
        \coordinate (T_05) at ($(S3)!0.5!(S2)$);
        
        \fill[hazyblue!20] (S1) -- (S3) -- (T_05) -- (B_05) -- cycle;
        \fill[hazyblue!60] (B_05) -- (T_05) -- (S2) -- cycle;
        
        \draw[dashed] (S1) -- (S2) -- (S3) -- cycle;
        \node[align=center, font=\large, text=black] at (0.45, 0.3) {\textbf{$a_1$}};
        \node[align=center, font=\large, text=black] at (0.8, 0.15) {\textbf{$a_2$}};
        \node[below left] at (S1) {$1$};
        \node[below right] at (S2) {$2$}; 
        \node[above] at (S3) {$3$};      
        \draw[] plot[smooth] coordinates {
          (0.001,0.000) (0.002,0.000) (0.003,0.000) (0.005,0.000) 
          (0.009,0.001) (0.017,0.002) (0.029,0.005) (0.048,0.012) 
          (0.081,0.030) (0.136,0.071) (0.220,0.157) (0.326,0.301) 
          (0.403,0.433) (0.460,0.559) (0.494,0.661) (0.510,0.733) 
          (0.515,0.781) (0.514,0.818) (0.510,0.839) (0.506,0.852) 
          (0.503,0.860) (0.502,0.863) (0.501,0.864) (0.501,0.865)
        };
        \filldraw[] (0.325, 0.39) circle (0.3pt) node[above] {$\mu_0$};
      \end{tikzpicture}%
    }
    \caption{$A = \{a_1,a_2\}$}
    \label{fig:scenario_2}
  \end{subfigure}
  \hspace{1em}
    \begin{subfigure}[t]{0.3\textwidth}
    \centering
\resizebox{\linewidth}{!}{%
      \begin{tikzpicture}[scale=4.5]
        \coordinate (S1) at (0,0);
        \coordinate (S2) at (1,0); 
        \coordinate (S3) at (0.5, 0.866025);
        
        \coordinate (R23) at (0.75, 0.433013);
        
        \coordinate (B12) at (0.5, 0);
        
        \coordinate (L23) at (0.375, 0.649519);
        
        \fill[hazyblue!20] (S1) -- (B12) -- (R23) -- (L23) -- cycle;
        \fill[hazyblue!60] (B12) -- (S2) -- (R23) -- cycle;
        \fill[hazyblue] (L23) -- (R23) -- (S3) -- cycle;
        
        
        \draw[dashed] (S1) -- (S2) -- (S3) -- cycle;
        
        \node[align=center, font=\large] at (0.45, 0.3) {\textbf{$a_1$}};
        \node[align=center, font=\large] at (0.8, 0.15) {\textbf{$a_2$}};
        \node[align=center, font=\large] at (0.52, 0.64) {\textbf{$a_3$}};
        
        \node[below left] at (S1) {$1$};
        \node[below right] at (S2) {$2$}; 
        \node[above] at (S3) {$3$};    
          \draw[] plot[smooth] coordinates {
          (0.001,0.000) (0.002,0.000) (0.003,0.000) (0.005,0.000) 
          (0.009,0.001) (0.017,0.002) (0.029,0.005) (0.048,0.012) 
          (0.081,0.030) (0.136,0.071) (0.220,0.157) (0.326,0.301) 
          (0.403,0.433) (0.460,0.559) (0.494,0.661) (0.510,0.733) 
          (0.515,0.781) (0.514,0.818) (0.510,0.839) (0.506,0.852) 
          (0.503,0.860) (0.502,0.863) (0.501,0.864) (0.501,0.865)
        };
        \filldraw[] (0.325, 0.39) circle (0.3pt) node[above] {$\mu_0$};
      \end{tikzpicture}%
    }
    \caption{$\varepsilon \sim N(0,1)$}
    \label{fig:scenario_triangle_a3}
  \end{subfigure}
    \hspace{1em}
  \begin{subfigure}[t]{0.3\textwidth}
    \centering
    \resizebox{\linewidth}{!}{%
      \begin{tikzpicture}[scale=4.5]
        \coordinate (S1) at (0,0);
        \coordinate (S2) at (1,0);
        \coordinate (S3) at (0.5, 0.866025);
        \coordinate (R23) at (0.75, 0.433013);
        \coordinate (B12) at (0.5, 0);
        \coordinate (L23) at (0.375, 0.649519);
        \fill[hazyblue!20] (S1) -- (B12) -- (R23) -- (L23) -- cycle;
        \fill[hazyblue!60] (B12) -- (S2) -- (R23) -- cycle;
        \fill[hazyblue] (L23) -- (R23) -- (S3) -- cycle;
        \draw[dashed] (S1) -- (S2) -- (S3) -- cycle;
        \node[align=center, font=\large] at (0.45, 0.3) {\textbf{$a_1$}};
        \node[align=center, font=\large] at (0.8, 0.15) {\textbf{$a_2$}};
        \node[align=center, font=\large] at (0.52, 0.64) {\textbf{$a_3$}};
        \node[below left] at (S1) {$1$};
        \node[below right] at (S2) {$2$};
        \node[above] at (S3) {$3$};

        \filldraw[] (0.325, 0.39) circle (0.3pt) node[above] {$\mu_0$};
        \coordinate (CB) at (0.26894,0.00000);

\coordinate (CB) at (0.26894,0.00000); 
\coordinate (CJ) at (0.34006,0.02776); 
\coordinate (CC) at (0.51060,0.24136); 
\coordinate (CL) at (0.14438,0.25008); 

\begin{scope}
    \clip (S1) -- (B12) -- (R23) -- (L23) -- cycle;

    \fill[pattern={Lines[angle=135, distance=6pt]}, pattern color=red!60]
        (S1)
        -- (CB)
        -- (CJ)
        plot[smooth, tension=0.15] coordinates {
            (0.34006,0.02776)
            (0.34550,0.02996)
            (0.35099,0.03232)
            (0.35652,0.03485)
            (0.36208,0.03757)
            (0.36767,0.04049)
            (0.37329,0.04361)
            (0.37894,0.04696)
            (0.38461,0.05054)
            (0.39030,0.05438)
            (0.39600,0.05847)
            (0.40171,0.06284)
            (0.40742,0.06750)
            (0.41313,0.07247)
            (0.41883,0.07776)
            (0.42451,0.08338)
            (0.43018,0.08935)
            (0.43581,0.09569)
            (0.44141,0.10241)
            (0.44696,0.10952)
            (0.45245,0.11703)
            (0.45789,0.12496)
            (0.46325,0.13331)
            (0.46853,0.14210)
            (0.47371,0.15134)
            (0.47879,0.16102)
            (0.48376,0.17115)
            (0.48860,0.18174)
            (0.49331,0.19278)
            (0.49788,0.20427)
            (0.50229,0.21620)
            (0.50653,0.22857)
            (0.51060,0.24136)
        }
        -- (CL)
        -- (S1)
        -- cycle;


    \draw[red!80!black, thick]
        (S1) -- (CB) -- (CJ);


    \draw[red!80!black, thick]
        plot[smooth, tension=0.15] coordinates {
            (0.34006,0.02776)
            (0.34550,0.02996)
            (0.35099,0.03232)
            (0.35652,0.03485)
            (0.36208,0.03757)
            (0.36767,0.04049)
            (0.37329,0.04361)
            (0.37894,0.04696)
            (0.38461,0.05054)
            (0.39030,0.05438)
            (0.39600,0.05847)
            (0.40171,0.06284)
            (0.40742,0.06750)
            (0.41313,0.07247)
            (0.41883,0.07776)
            (0.42451,0.08338)
            (0.43018,0.08935)
            (0.43581,0.09569)
            (0.44141,0.10241)
            (0.44696,0.10952)
            (0.45245,0.11703)
            (0.45789,0.12496)
            (0.46325,0.13331)
            (0.46853,0.14210)
            (0.47371,0.15134)
            (0.47879,0.16102)
            (0.48376,0.17115)
            (0.48860,0.18174)
            (0.49331,0.19278)
            (0.49788,0.20427)
            (0.50229,0.21620)
            (0.50653,0.22857)
            (0.51060,0.24136)
        };

    \draw[red!80!black, thick]
        (CC) -- (CL) -- (S1);
\end{scope}

    \draw[black, thin]
    plot[smooth, tension=0.25] coordinates {
        (0.00000,0.00000)
        (0.00150,0.00000)
        (0.00303,0.00000)
        (0.00563,0.00000)
        (0.00970,0.00000)
        (0.01565,0.00001)
        (0.02382,0.00004)
        (0.03441,0.00016)
        (0.04753,0.00056)
        (0.06319,0.00164)
        (0.08140,0.00421)
        (0.10222,0.00959)
        (0.12576,0.01965)
        (0.15210,0.03652)
        (0.18106,0.06207)
        (0.21203,0.09718)
        (0.24395,0.14122)
        (0.27549,0.19202)
        (0.30531,0.24644)
        (0.33239,0.30118)
        (0.35617,0.35355)
        (0.37651,0.40173)
        (0.39358,0.44479)
        (0.40772,0.48251)
        (0.41937,0.51510)
        (0.42893,0.54302)
        (0.43678,0.56683)
        (0.44323,0.58708)
        (0.44855,0.60431)
        (0.45296,0.61897)
        (0.45663,0.63148)
        (0.45970,0.64217)
        (0.46228,0.65132)
        (0.46445,0.65917)
        (0.46629,0.66594)
        (0.46786,0.67177)
        (0.46920,0.67681)
        (0.47035,0.68118)
        (0.47133,0.68497)
        (0.47218,0.68827)
        (0.47292,0.69114)
        (0.47355,0.69364)
        (0.47411,0.69583)
        (0.47459,0.69774)
        (0.47501,0.69941)
        (0.47537,0.70088)
        (0.47569,0.70216)
        (0.47621,0.70428)
        (0.47661,0.70591)
        (0.47692,0.70717)
        (0.47726,0.70855)
        (0.47755,0.70974)
        (0.47784,0.71097)
        (0.47799,0.71156)
    };
      \node[text=black, font=\footnotesize, align=right, overlay] at (-0.07, 0.35) {Cascade Beliefs};
        \draw[->, red!80!black, shorten >=2pt, overlay] (0, 0.3) -- (0.1, 0.05);
      \end{tikzpicture}%
    }
    \caption{$\varepsilon \sim \mathrm{Gumbel}(0,1)$}
    \label{fig:scenario_exact_cascade}
  \end{subfigure}

  \caption{The investor example. Payoff vectors are $u(a_1,\cdot)=(1,-1,1)$, $u(a_2,\cdot)=(0,0,0)$, and $u(a_3,\cdot)=(-2,-2,2)$. Black curves show posteriors attainable from $\mu_0$.}
  \label{fig:opening example}
\end{figure}
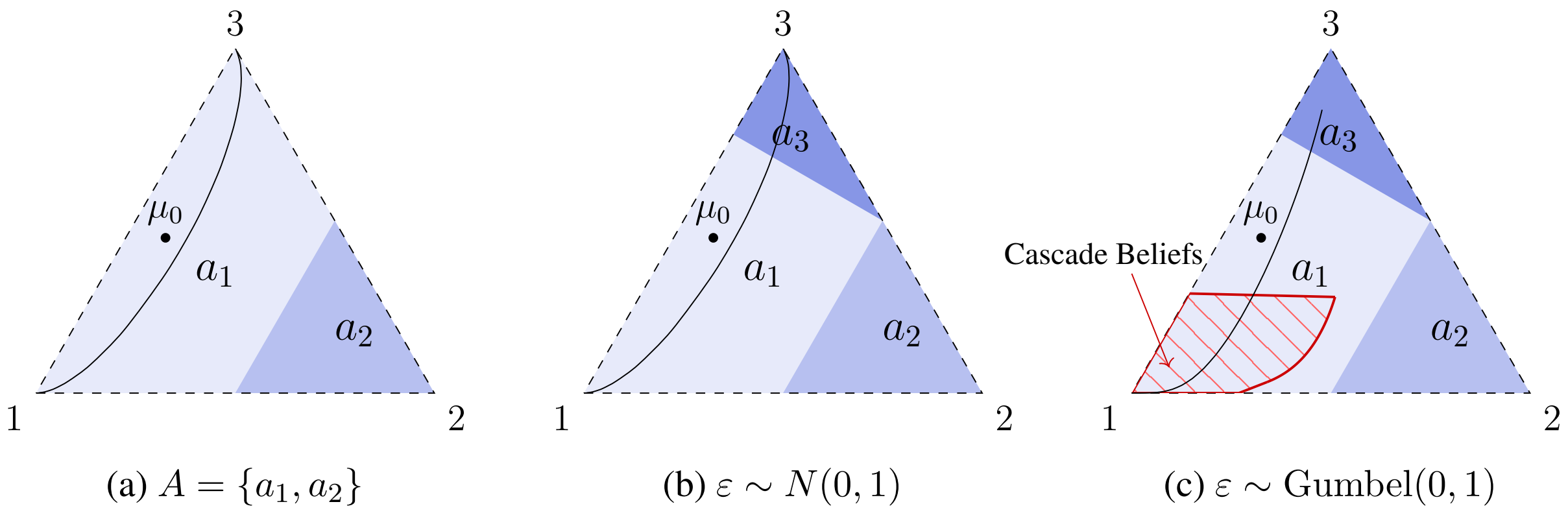

To fix ideas, consider an example adapted from \Cref{fig:counterexample_path}. Speculative investors sequentially choose whether to buy a large amount ($a_1$), not buy ($a_2$), or sell a small amount ($a_3$). The state space is $\Omega=\{1,2,3\}$, representing undervaluation, fair valuation, and overvaluation, respectively. Each investor observes a private signal $s=\omega+\varepsilon$, where $\varepsilon\sim N(0,1)$. Investors prefer $a_1$ to $a_2$ in states $1$ and $3$, anticipating a rebound or riding a bubble. Among the three actions, $a_1$ is optimal in state $1$, $a_2$ in state $2$, and $a_3$ in state $3$.\footnote{For empirical evidence, see \citet{lakonishok_contrarian_1994} and \citet{k_brunnermeier_hedge_2004}.} \Cref{fig:scenario_2,fig:scenario_triangle_a3} show the optimal action regions in the belief simplex for $A = \{a_1, a_2\}$ and $A = \{a_1, a_2, a_3\}$.

This information structure is neither bounded nor unbounded.\footnote{With finitely many states, unbounded beliefs require each state to be distinguishable from its complement; bounded beliefs require that no state be distinguishable from any other.} As \Cref{fig:scenario_2} shows, from any full-support prior such as $\mu_0$, posteriors can approach certainty on state $1$ or $3$, but not on state $2$. Nevertheless, panel (a) illustrates the learning failure predicted by KLLR. At the prior $\mu_0$, $a_1$ remains optimal after every private signal. All subsequent investors therefore choose $a_1$, and society herds on action $a_1$, regardless of the realized state.

The ability of signals to make one set of states arbitrarily likely relative to another is called \emph{distinguishability} (\Cref{def:distinguishability}). Let $\Omega_{a,a'}=\{\omega\in\Omega:u(a,\omega)>u(a',\omega)\}$. KLLR's \emph{excludability} condition requires $\Omega_{a,a'}$ to be distinguishable from $\Omega_{a',a}$ for every ordered pair of actions (\Cref{def: excludability}). In \Cref{fig:scenario_2}, $\Omega_{a_2,a_1}=\{2\}$ is not distinguishable from $\Omega_{a_1,a_2}=\{1,3\}$, so excludability fails. With binary actions, Theorem 2 of KLLR then implies herding for some initial prior.

This result leaves two gaps. First, as the counterexample illustrates, herding for some initial prior does not imply herding for every full-support prior. Second, with more than two actions, failure of excludability need not imply herding. Indeed, \emph{adding a single action can restore learning}. Adding $a_3$ to \Cref{fig:scenario_2} gives \Cref{fig:scenario_triangle_a3}, where learning occurs for every initial prior, as we show below. Excludability still fails, because the comparison between $a_1$ and $a_2$ is unaffected.\footnote{KLLR's Example 1 establishes that learning can obtain without excludability using a continuum of actions under quadratic loss. Our example provides a finite-action analogue.}

Now replace the Gaussian noise with Gumbel noise (\Cref{fig:scenario_exact_cascade}).\footnote{The standard Gumbel density is $g(x)=\exp(-x-e^{-x})$ for $x\in\mathbb{R}$.} Excludability again fails, but it does not tell us what happens. In the Gumbel example, incorrect herds on $a_1$ occur with positive probability for every full-support prior. To understand why herding is possible, note that $a_1$ is suboptimal in states $\{2,3\}$ and strictly optimal in state $1$. For each full-support prior, such as $\mu_0$, the posterior probability of $\{2,3\}$ is bounded away from one after any private signal. That is, $\{2,3\}$ is not distinguishable from $\{1\}$: private signals cannot provide arbitrarily strong evidence that $a_1$ is ex post suboptimal. Thus, when the public belief assigns sufficiently high probability to state $1$, no private signal can overturn the choice of $a_1$. Such beliefs form the hatched cascade region.

We formalize this intuition through two necessary conditions for learning. With finite state and action spaces, \textbf{\emph{escapability}} requires, for every action $a$, that the states where it is suboptimal, $\cup_{a'}\Omega_{a',a}$, be distinguishable from the states where it is strictly optimal, $\Omega_a$. \textbf{\emph{Pairwise escapability}} requires each state in $\cup_{a'} \Omega_{a',a}$ to be distinguishable from each state in $\Omega_a$ (\Cref{def: escapability}). Failure of escapability implies herding for some initial prior, while failure of pairwise escapability implies herding for every full-support initial prior (\Cref{thm1}). The Gumbel example violates pairwise escapability because $\{2\}$ is not distinguishable from $\{1\}$, establishing the every-prior result above. In the opening counterexample in \Cref{fig:counterexample_path}, escapability fails but pairwise escapability holds: herding occurs for some priors, but not necessarily for every full-support prior.

In a natural class of environments, learning and herding admit a complete characterization independent of the full-support initial prior, as in the binary-state model. States and signals are one-dimensional and ordered, the action set is finite, and each state has a unique optimal action. Preferences satisfy \emph{interval optimality}: each action's optimality region $\Omega_a$ is an interval. The information structure satisfies the monotone likelihood ratio property (MLRP). Under these assumptions, (pairwise) escapability implies learning for every initial prior, while its failure implies herding for every full-support initial prior (\Cref{thm: escapability_necc_suff}). The environment in \Cref{fig:scenario_triangle_a3} belongs to this class. Since each state has a distinct optimal action and is distinguishable from every other state, pairwise escapability holds, and learning occurs for every initial prior.

With location-shift information (i.e., additive noise signals $s = \omega + \varepsilon$) and nontrivial preferences, the characterization reduces to \textit{pairwise distinguishability} (\Cref{def: pairwise_dist} and \Cref{prop: pairwise}). Pairwise distinguishability is an extension of unbounded beliefs to the multi-state environment: it requires each state to be distinguishable from every other state. This result identifies when the familiar binary-state conditions extend to a multi-state model. It implies learning for every prior in \Cref{fig:scenario_triangle_a3} and herding for every full-support prior in \Cref{fig:scenario_exact_cascade}.

If preferences further satisfy KLLR's single-crossing differences (SCD) condition (\Cref{def:SCD}), then excludability, (pairwise) escapability, and pairwise distinguishability are all \emph{equivalent} (\Cref{prop:equiv_excludability_necess_MLRP}). Moreover, interval optimality itself is necessary for learning at every prior: if it fails under MLRP information, herding occurs for some initial prior (\Cref{prop: non-interval}), as in \Cref{fig:scenario_2}.

Methodologically, this paper analyzes the public belief process and exploits its martingale property (\Cref{lem: martingale}). The main challenge is to determine which limiting beliefs can arise from a given prior. To establish herding for every full-support prior, we show that when pairwise escapability fails, the limiting public belief lies in a particular set with probability zero (\Cref{lem: ball around AK}). This argument yields herding even when all cascade beliefs lack full support and cannot be reached in finite time. In settings where this argument does not apply, such as the opening counterexample, a further difficulty is to analyze the public belief process quantitatively to determine whether the cascade beliefs can be reached from a given prior.

\paragraph{Related Literature.}
This paper contributes to the literature on sequential observational learning, or social learning. In binary-state settings, this literature studies how learning and herding depend on information structures \citep{smith_pathological_2000}, observational networks \citep{CELEN200472,acemoglu_bayesian_2011,Lobel_Sadler_2015}, the size of the action space \citep{LEE1993395,ALI201879}, and correlation across signals \citep{awaya_krishna_2026_social_learning_correlated_information}. Related work examines the efficiency \citep{rosenberg_efficiency_2019} and speed \citep{tamuz_speed2018383} of learning. \citet{bikhchandani_information_2024} provide a comprehensive survey.

This paper is most closely related to work on social learning with general state spaces. \citet{arieli_general_2021} show that ``totally unbounded'' beliefs suffice for asymptotic learning under all preferences. \citet{kartik_beyond_2024} introduce excludability, a joint restriction on preferences and information that guarantees learning for every initial prior for every choice set in a model with general observational networks. When the number of states is finite, the failure of excludability implies herding for some binary choice set and some initial prior, but need not imply herding when there are more than two actions. In their working paper, \citet[Proposition~SA.3]{kartik_beyond_arxiv_v5_2024} show that if their Conjecture~SA.1 holds and excludability fails, there is a binary choice set for which herding occurs for every full-support prior.

Our first contribution concerns the initial prior. With preferences and information fixed, different full-support priors can lead to learning or herding when there are more than two states. To our knowledge, this is new in the standard observational learning model, where, with binary states and finitely many actions, the outcome is the same for every full-support prior. \citet{danenberg2026sequential} obtain prior dependence with two states when agents observe others' outcomes rather than actions. Prior dependence offers new perspectives on cascade beliefs and on how social learning happens. In particular, identifying cascade beliefs and identifying herding can be different problems, and learning the correct action may require society to learn the state. 

Our second contribution consists of two necessary conditions for learning, escapability and pairwise escapability. Unlike KLLR's herding result, they apply to any fixed action set and distinguish herding for some prior from herding for every full-support prior. In a natural class of environments with interval optimality and MLRP, they completely characterize learning independently of the full-support initial prior. The only previous complete characterizations with multiple states that we know of are Theorem 3 of \citet{arieli_general_2021} and the binary-action characterization implied by Claim SA.1 of \citet{kartik_beyond_arxiv_v5_2024}. Our characterization allows preferences beyond matching utility and action sets beyond the binary case; \Cref{sec: application} provides a detailed comparison.

The paper proceeds as follows. In \Cref{sec: model}, we present the model. In \Cref{sec: characterization}, we introduce escapability and state the main herding result, \Cref{thm1}, followed by a sketch of the proof.  In \Cref{sec:prior_dependence}, we discuss how learning depends on the initial prior. In \Cref{sec: application}, we give applications to one-dimensional environments. Finally, \Cref{sec: conclusion} concludes with further discussion. Proofs omitted from the main text are in \Cref{sec: appendix}.

\section{Preliminaries}\label{sec: model}

\subsection{The Model}

There is a countable (finite or infinite) state space $\Omega$ endowed with the discrete topology, along with standard Borel spaces of actions $(A, \mathcal{A})$ and signals $(S, \mathcal{S})$. An information structure $F := \{F(\cdot \mid \omega)\}_{\omega \in \Omega}$ is a family of probability measures over $S$, one for each state. We assume that for any $\omega$ and $\omega'$, $F(\cdot \mid \omega)$ and $F(\cdot \mid \omega')$ are mutually absolutely continuous. Each $F(\cdot \mid \omega)$ hence has a density $f(\cdot \mid \omega)$, which is defined as the Radon--Nikodym derivative of $F(\cdot \mid \omega)$ with respect to a reference measure on $S$ that is mutually absolutely continuous with respect to all $F(\cdot \mid \omega)$. We include extensions to Polish state spaces in \Cref{app:general_model}.

We consider the canonical social learning game. At $t = 0$, a state $\omega$ is drawn according to an initial prior $\mu_0 \in \Delta(\Omega)$. Then at each time $t = 1,2,\ldots$, agent $t$ chooses an action $a_t \in A$ after observing the actions of her predecessors $(a_1, \ldots, a_{t-1})$ and receiving a private signal $s_t$ drawn from $F(\cdot \mid \omega)$. Conditional on the state, private signals are independent across time. Each agent $t$ aims to maximize her expected utility $\mathbb{E}[u(a_t, \omega) \mid a_1, \ldots, a_{t-1}, s_t]$, where $u: A \times \Omega \to \mathbb{R}$ is a bounded and measurable utility function common to all agents. We assume nontrivial preferences: no action is optimal in every state.

A strategy for agent $t$ is a measurable function $\sigma_t: A^{t-1} \times S \to \Delta(A)$. We study the Bayes Nash equilibria—hereafter simply equilibria—of this game. We assume that an optimal action exists for every belief, so that an equilibrium exists. This induces an overarching probability space $(X_\infty, \mathcal{X}_\infty, \mathbb{P})$. Specifically, $X_\infty = \Omega \times (S \times A)^\infty$ is the set of infinite histories of states, signals, and actions; $\mathcal{X}_\infty$ is the corresponding Borel sigma-algebra; and $\mathbb{P}$ is the probability measure induced by $\mu_0$, the strategy profile $\sigma = (\sigma_1, \sigma_2, \ldots)$, and $F$. 

We introduce two concepts regarding the value of information. As in KLLR, let $c(\mu) := \arg\max_{a \in A} \mathbb{E}_{\mu}[u(a, \omega)]$ be the set of optimal actions under belief $\mu$. Write $\mu_s$ for the posterior after observing signal $s$ under prior $\mu$, and $c(\omega)$ for the set of optimal actions at state $\omega$. A belief $\mu$ is \textit{stationary} (or a cascade belief) if there is $a \in c(\mu)$ such that $a \in c(\mu_s)$ for $\mu$-a.e.\ $s$. It has \textit{adequate knowledge} if there is $a \in c(\mu)$ such that $a \in c(\omega)$ for all $\omega$ in its support. At a stationary belief, observing signals has no value; with adequate knowledge, the first-best utility is already attained. Write $\Delta:=\Delta(\Omega)$, and let $\Delta^S$ and $\Delta^{AK}$ denote the sets of stationary beliefs and beliefs with adequate knowledge, respectively.

\subsection{Adequate Learning and Herding}\label{subsec: public belief}

Fix the prior $\mu_0$, an equilibrium $\sigma$, and the information structure $F$. Each history $h_n \in A^{n-1}$ induces, by Bayes' rule, a public belief $\mu_n(\cdot \mid h_n) \in \Delta(\Omega)$. Write $\tilde{\mu}_n$ for this belief evaluated at the random history, and let $\mathcal{F}_n$ be the sigma-algebra generated by $(a_1, \ldots, a_{n-1})$. Then $\tilde{\mu}_n(E) = \mathbb{P}(\omega \in E \mid \mathcal{F}_n)$ for every Borel set $E$. By iterated expectations, $\langle\tilde{\mu}_n(E)\rangle$ is a bounded martingale with respect to $(\mathbb{P}, \mathcal{F}_n)$. We denote $\sigma(\cup_n \mathcal{F}_n)$ by $\mathcal{F}_\infty$. The martingale convergence theorem therefore gives $\mathbb{P}$-almost-sure convergence to $\tilde{\mu}_\infty(E):= \mathbb{P}(\omega \in E \mid \mathcal{F}_\infty)$. The following lemma characterizes the public belief process.

\begin{lemma}\label{lem: martingale}
$\tilde{\mu}_n$ converges to $\tilde{\mu}_\infty$ weakly $\mathbb{P}$-almost surely. Moreover, $\tilde{\mu}_\infty \in \Delta^S$ $\mathbb{P}$-almost surely.
\end{lemma}

\Cref{lem: martingale} shows that the public belief process converges to a stationary belief. The second statement is similar to Lemma 2 of \citet{arieli_general_2021}, but adapted to our setting. 

As in KLLR, our notion of learning follows \citet{aghion1991}. Fix any belief $\mu$. We denote the first-best utility under this belief by $u^*(\mu) := \sum_{\omega \in \Omega} \mu(\omega) \max_{a \in A} u(a, \omega)$. Let $u(\mu):=\sup_{a\in A}\mathbb{E}_{\mu}[u(a,\omega)]$ denote the optimal expected utility under belief $\mu$. 

\begin{definition}
Fix an initial prior $\mu_0$.
\begin{enumerate}[label=(\roman*)]
\item There is \textit{adequate learning for $\mu_0$} if, in every equilibrium $\sigma$, the ex ante expected utility $\mathbb{E}_{\sigma, \mu_0}[u(a_t, \omega)]$ converges to the first-best utility $u^*(\mu_0)$ as $t$ goes to infinity.
\item There is \textit{(incorrect) herding for $\mu_0$} if, in some equilibrium $\sigma$,\footnote{We distinguish between ``every equilibrium'' and ``some equilibrium'' so that herding for $\mu_0$ is the negation of adequate learning for $\mu_0$. Our main results continue to hold under the stronger definition requiring herding in every equilibrium. In any case, there is no strategic interaction among agents.} the limiting public belief satisfies $u(\tilde\mu_\infty)<u^*(\tilde\mu_\infty)$ with \textit{positive probability}.
\end{enumerate}
\end{definition}

We say there is \textit{adequate learning} if there is adequate learning for every prior belief $\mu_0$.

The definition of incorrect herding generalizes the familiar notion in the literature. Consider a simple model with $\Omega = \{0,1\}$, $A = \{0,1\}$, and $u(a, \omega) = \mathbb{1}_{\{\omega = a\}}$. Let $(1-p,p)$ denote a limiting public belief, where $p$ denotes the probability of state $1$. Suppose that $p > 1/2$. Then action $1$ is optimal, and the resulting utility is $u(1-p,p) = p$, whereas the first-best utility $u^*(1-p,p)$ is $1$. The payoff gap $1-p$ is therefore the conditional probability that action $1$ is wrong. When positive payoff gaps exist with positive probability, agents herd on the wrong action with positive probability ex ante. In the general model, multiple actions may tie, so our definition captures learning failure without requiring eventual agreement on a single action. For conciseness, we also incorporate the positive-probability requirement into the definition.

Though learning and herding are defined through utilities, \Cref{lem: martingale} enables us to analyze learning through the public belief process. The following lemma relates learning and herding to the limiting public belief.

\begin{lemma}\label{lem:learning_limit}
Fix an initial prior $\mu_0$.
\begin{enumerate}[label=(\roman*)]
\item There is adequate learning for $\mu_0$ if and only if, in every equilibrium, $\tilde\mu_\infty\in\Delta^{AK}$ $\mathbb{P}$-a.s.
\item There is herding for $\mu_0$ if and only if, in some equilibrium, $\tilde\mu_\infty\in\Delta^S\setminus\Delta^{AK}$ with positive probability.
\end{enumerate}
\end{lemma}

Thus, inadequate learning for a given prior is equivalent to herding for that prior, and there is adequate learning if and only if every stationary belief has adequate knowledge.\footnote{Theorem 1 of \citet{kartik_beyond_2024} generalizes this observation to general state spaces and network structures, in which public beliefs may not converge.}

\section{Characterization}\label{sec: characterization}

\subsection{Escapability}
A key concept for our analysis is \textit{distinguishability}, defined as in KLLR. Let $\mathrm{Pr}_\mu(S')$ be the probability of observing signals in $S'$ under belief $\mu$.\footnote{For a set of signals $S' \subseteq S$ and a belief $\mu$, $\mathrm{Pr}_{\mu}(S') := \int_{\Omega}\int_{S'} F(\mathrm{d}s \mid \omega)\mu(\mathrm{d}\omega)$.}

\begin{definition}[Distinguishability]\label{def:distinguishability}
Let $\Omega',\Omega'' \subseteq \Omega$. We say that $\Omega'$ is \emph{distinguishable} from $\Omega''$ if for any belief $\mu \in \Delta(\Omega' \cup \Omega'')$ with $\mu(\Omega')>0$ and every $\varepsilon>0$, $\mathrm{Pr}_{\mu}\left(s:\mu_s(\Omega') > 1-\varepsilon\right) > 0$.\footnote{This definition, as in KLLR, does not require $\Omega'$ and $\Omega''$ to be disjoint or nonempty. This generality is convenient for some proofs. In our definitions, the sets are typically disjoint by construction and are empty only in trivial cases. We therefore retain the original definition for consistency.}

\end{definition}
Distinguishability has two useful properties. First, $\Omega'$ is distinguishable from $\Omega''$ if every $\omega \in \Omega'$ is distinguishable from $\Omega''$. Second, for any disjoint $\Omega', \Omega'' \subseteq \Omega$, if $\Omega'$ is distinguishable from $\Omega''$, then every subset of $\Omega'$ is distinguishable from every subset of $\Omega''$.

The key condition of this paper is \textbf{\textit{escapability}}, which is a joint restriction on preferences and the information structure.  We introduce the following notation. For actions $a_1, a_2 \in A$, let $\Omega_{a_1, a_2} := \{\omega \in \Omega: u(a_1, \omega) > u(a_2, \omega)\}$ be the set of states where action $a_1$ is preferred to action $a_2$. Let $\Omega_a := \cap_{a' \neq a} \Omega_{a, a'}$ be the set of states where $a$ is strictly optimal, and let $\Omega_a^n := \{\omega \in \Omega: u(a, \omega) > u(a', \omega) + 1/n, \forall a' \neq a\}$ be the set of states where $a$ is strictly optimal by a uniform margin of $1/n$. The margin rules out cases where $a$ is optimal by an arbitrarily small amount.

\begin{definition}
[Escapability] A utility function and an information structure satisfy

  \begin{enumerate}[label=(\roman*)]
    \item \textbf{\textit{escapability}} if for every $a \in A$ and $n \in \mathbb{N}$, $\cup_{a'} \Omega_{a',a}$ is distinguishable from $\Omega_a^n$;
    \item \textbf{\textit{pairwise escapability}} if for every $a \in A$ and $n \in \mathbb{N}$, for every $\omega \in \Omega_a^n$ and every $\omega' \in \cup_{a'} \Omega_{a',a}$, $\{\omega'\}$ is distinguishable from $\{\omega\}$.\footnote{By the subset property of distinguishability, requiring distinguishability for all $n \in \mathbb{N}$ is weaker than requiring distinguishability for $\Omega_a$ and is equivalent to requiring it for all sufficiently large $n$.}
  \end{enumerate}
  \label{def: escapability}
\end{definition}

Escapability asks whether private signals can provide arbitrarily strong evidence that any given action is ex post suboptimal. Clearly, by the subset property of distinguishability, escapability implies pairwise escapability. The investor  example in \Cref{fig:scenario_2} illustrates why the converse can fail: for $\Omega = \{1,2,3\}$, every pair of states is distinguishable, but $\{2\}$ is not distinguishable from $\{1,3\}$. For finite $\Omega$ and $A$, we can omit the superscript $n$.

Although the definition quantifies over every action $a\in A$, both conditions impose a nontrivial restriction only when $\cup_{a'}\Omega_{a',a}$ and $\Omega_a^n$ are both nonempty for some $n$. In particular, an action that is never strictly optimal at any state satisfies its own escapability restriction vacuously. Such an action may nevertheless be chosen in a herd, as in the binary-state environment with bounded beliefs \citep{smith_pathological_2000}. The failure of escapability instead identifies an action that is strictly optimal at some state and can support an incorrect herd for a suitable initial prior, as our opening investor example illustrates. More generally, however, the action for which escapability fails need not coincide with the action on which agents eventually herd.

\begin{theorem}\label{thm1}
  If escapability fails, then there is herding for some initial prior. If pairwise escapability fails, then there is herding for every full-support initial prior. 
\end{theorem}

\Cref{thm1} generalizes existing herding results. In the binary-state setting with finitely many actions where each state has a unique optimal action \citep{smith_pathological_2000}, failure of (pairwise) escapability reduces to the negation of unbounded beliefs. This is weaker than bounded beliefs,\footnote{Bounded beliefs require that the two states are not distinguishable from each other, whereas \Cref{thm1} in a binary-state model only requires that one state is not distinguishable from the other.} but still implies herding for every full-support prior. When action sets are binary, suppose $A=\{a_1,a_2\}$ and $\Omega$ is finite. For nontrivial preferences, $\Omega_{a_1,a_2}=\Omega_{a_1}$ and $\Omega_{a_2,a_1}=\Omega_{a_2}$ are nonempty. Thus the failure of escapability is equivalent to failure of excludability (\Cref{def: excludability}). Therefore, if $\Omega_{a_1,a_2}$ is not distinguishable from $\Omega_{a_2,a_1}$, then there is herding for some initial prior. This is the sense in which excludability of KLLR is necessary for adequate learning. For $\Omega = A$ finite and utility $u(a,\omega)=\mathbb{1}_{\{\omega = a\}}$, pairwise escapability is equivalent to pairwise distinguishability, which requires that any two states are distinguishable from each other. Hence its failure implies herding for every full-support prior, as in \citet[Theorem 3]{arieli_general_2021}.

With finitely many actions, pairwise escapability fails as soon as one state where an action is suboptimal is not distinguishable from a state where it is strictly optimal, regardless of the size of $\Omega$. We compare it with bounded beliefs, generalized to the multi-state setting as follows.

\begin{definition}\label{def: bounded_belief}
  Suppose that $\Omega$ is finite. The information structure has bounded beliefs if for any $\omega, \omega' \in \Omega$ with $\omega \neq \omega'$, $\{\omega\}$ is not distinguishable from $\{\omega'\}$.\footnote{When $\Omega$ is finite, this definition is equivalent to the following ``totally bounded beliefs'' condition: for any $E \in \mathcal{B}(\Omega)$ and prior $\mu$ with $0 < \mu(E) < 1$, there exists some $\varepsilon > 0$ such that for $\mu$-a.e.\ $s$, $\mu_s(E) \le 1-\varepsilon$. }
\end{definition}

Clearly, if $\Omega_a^n$ is nonempty for some $a \in A$ and $n \in \mathbb{N}$, bounded beliefs imply herding for every full-support prior by \Cref{thm1}. However, for a state space of size $N < \infty$, the bounded beliefs condition requires failure of distinguishability for all $N(N-1)$ ordered pairs of distinct states. By contrast, failure of pairwise escapability requires only one pair, given different optimal actions.

Moreover, failure of pairwise escapability can imply herding for every full-support prior even in cases without full-support stationary beliefs. For example, applying \Cref{thm1} to Figure~SA.1 of KLLR's working paper yields herding for every full-support prior because $\{2\}$ is not distinguishable from $\{1\}$. This argument does not require the additional restrictions on preferences and information used by KLLR.

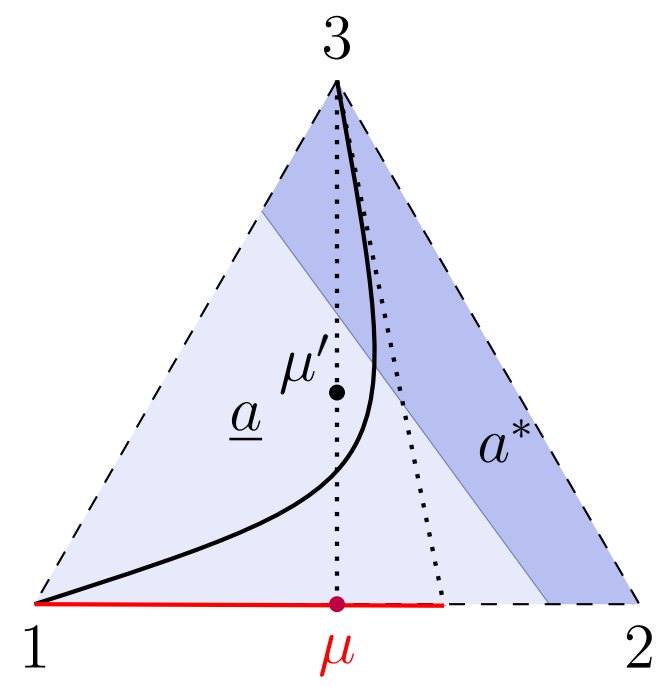
\begin{figure}[h] 
\centering 
\begin{tikzpicture}[scale=4]
  \coordinate (A) at (0,0);
  \coordinate (B) at (1,0);
  \coordinate (C) at (0.5,0.866);

  \coordinate (prior)   at (0.5,0.35);
  \coordinate (midpriorB) at (0.65,0.13);
  \coordinate (muend)   at ($(C)!1.18!(midpriorB)$);
  \coordinate (a32)     at (0.85,0);
  \coordinate (a31)     at ($(A)!0.75!(C)$);
  \coordinate (aunder)  at (0.45,0.30);
  \coordinate (astar)   at (0.78,0.38);
  \coordinate (prior12) at ($(A)!0.5!(B)$);

  \fill[hazyblue!60] (a31) -- (a32) -- (B) -- (C) -- cycle;
  \draw[color=hazyblue!80!black] (a31) -- (a32);
  \node[label={[black] below:$a^*$}] at (astar) {};

  \fill[hazyblue!20] (B) -- (A) -- (a31) -- (a32) -- cycle;
  \node[label={[black] left:$\underline{a}$}] at (aunder) {};

  \draw[dashed] (A) node[below] {$1$} -- (B) node[below] {$2$} -- (C) node[above] {$3$} -- (A);

  \draw[thick,dotted] (C) -- (muend);	
  \draw[thick,dotted] (C) -- (prior12);
  \draw[thick] (A) .. controls (0.62,0.20) .. (C);

  \node[circle,fill=black,inner sep=0pt,minimum size=3pt,label={[xshift=-6, yshift=-6]:{$\mu'$}}] at (prior) {};
  \draw[red,thick] (muend) -- (A);
  \node[circle,fill=purple,inner sep=0pt,minimum size=3pt,label={[red] below:{$\mu$}}] at (prior12) {};
\end{tikzpicture}
\caption{Figure SA.1 of \citet{kartik_beyond_arxiv_v5_2024}. Blue regions mark preferences between $\underline a$ and $a^*$. From belief $\mu'$, posteriors follow the black curve; from belief $\mu$, they lie on the red line.}
\label{fig: figure_sa1_kllr} 
\end{figure}

Because of the margin $1/n$ in $\Omega_a^n$, \Cref{thm1} does not require $A$ to be finite, as illustrated in the following example. Indeed, social learning can fail even with infinitely many actions if the decision problem is ``unresponsive'' \citep{LEE1993395, ALI201879}.

\begin{example}\label{ex: responsive} Let $\Omega = \{0,1\}$, $A = \{0\} \cup [1/2,1]$, and $u(a, \omega) = -(\omega - a)^2$. The set $A$ is infinite. For any information structure that is bounded in the sense of \citet{smith_pathological_2000}, $\cup_{a' \in A} \Omega_{a',0} = \{1\}$ is not distinguishable from $\Omega_0^n = \{0\}$ for $n > 4$, so learning fails for every full-support prior by \Cref{thm1}. 
\end{example}

We compare escapability with KLLR's excludability condition, defined as follows.

\begin{definition}
  [\citealp{kartik_beyond_2024}] A utility function and an information structure jointly satisfy \textit{excludability} if for any $a_1, a_2 \in A$, $\Omega_{a_1, a_2}$ is distinguishable from $\Omega_{a_2, a_1}$.
  \label{def: excludability}
\end{definition}

Excludability is sufficient for adequate learning because it allows every wrong action to be displaced. If $a_1$ is optimal under a public belief without adequate knowledge, excludability ensures that the posterior can concentrate on $\Omega_{a_2,a_1}$ relative to $\Omega_{a_1,a_2}$ enough to make some $a_2$ strictly preferable to $a_1$. Here $a_2$ need not be optimal at the true state; its role is to overturn the wrong herd.

Escapability and pairwise escapability are necessary for adequate learning. Thus excludability implies both, but the converse generally fails. Intuitively, escapability only requires signals to concentrate the posterior on $\cup_{a'}\Omega_{a',a_1}$ relative to $\Omega_{a_1}^n$. Different states in this union may favor different alternatives to $a_1$. Such evidence need not make any single alternative preferable to $a_1$ under the posterior, so it need not overturn a wrong herd.

Both conditions have a partial converse. Theorem 2 of KLLR shows that, with two actions and finitely many states, failure of excludability implies herding for some prior, as illustrated in \Cref{fig:scenario_2}. Notably, failure of excludability need not imply herding with more than two actions, as shown in \Cref{fig:scenario_triangle_a3}. Although (pairwise) escapability is generally not sufficient for adequate learning, it is sufficient on every binary subset of states under the following assumptions.

\begin{proposition}
  Suppose that $A$ is finite and each state has a unique optimal action. If (pairwise) escapability holds, every restriction of the game to a binary state space has adequate learning.
\end{proposition}
\begin{proof}
  Fix $\{\omega_1,\omega_2\}\subseteq\Omega$. If the same action is optimal at both states, adequate learning is immediate. Otherwise, let $a_1\neq a_2$ be uniquely optimal at $\omega_1$ and $\omega_2$, respectively. Since $A$ is finite, $\omega_i\in\Omega_{a_i}^{n_i}$ for some $n_i\in\mathbb{N}$, $i=1,2$. Moreover, $\omega_1\in\Omega_{a_1,a_2}$ and $\omega_2\in\Omega_{a_2,a_1}$. (Pairwise) escapability makes the two states distinguishable from each other. Thus the restricted game has unbounded beliefs, implying adequate learning.
\end{proof}

The following table compares the implications of excludability, escapability, and pairwise escapability for learning and herding. Excludability implies escapability, which in turn implies pairwise escapability.

\begin{table}[htbp]
\centering
\begin{tabular}{@{}c|c|c@{}}
\textbf{Condition}
&
\begin{tabular}{@{}c@{}}
\textbf{Adequate learning}\\[0.2ex]
(if the condition holds)
\end{tabular}
&
\begin{tabular}{@{}c@{}}
\textbf{Herding}\\[0.2ex]
(if the condition fails)
\end{tabular}
\\[1.5mm]
\hline
\rule{0pt}{4ex}
Excludability
&
\begin{tabular}{@{}c@{}}
\textbf{all priors}\\[0.2ex]
fixed state space
\end{tabular}
&
\begin{tabular}{@{}c@{}}
some prior\\[0.2ex]
some binary action set
\end{tabular}
\\[2mm]
\hline
\rule{0pt}{3ex}
Escapability
& \multirow{2}{*}[-0.5mm]{all binary state restrictions}
& some prior
\\[1mm]
Pairwise escapability
&
& \textbf{every full-support prior}
\end{tabular}
\caption{Comparison of excludability, escapability, and pairwise escapability.}
\label{tab:learning_herding_comparison}

\end{table}

\subsection{Proof Sketch of \texorpdfstring{\Cref{thm1}}{Theorem \ref*{thm1}}}

We outline the main ideas of the proof of \Cref{thm1}. By \Cref{lem: martingale}, public beliefs converge almost surely to stationary beliefs. The role of escapability, like that of excludability, is therefore to restrict the set $\Delta^S$. For the first statement, it suffices to find a stationary belief without adequate knowledge and take it as the initial prior. If $\cup_{a'}\Omega_{a',a}$ is not distinguishable from $\Omega_a^n$, we can choose a belief $\mu'$ with $\mu'(\cup_{a'}\Omega_{a',a})>0$ and a ratio $\mu'(\cup_{a'}\Omega_{a',a})/\mu'(\Omega_a^n)$ small enough that $a$ is uniquely optimal under both $\mu'$ and $\mu'_s$ for $\mu'$-almost every signal $s$. This belief is stationary without adequate knowledge. 

For the second statement, we show that when pairwise escapability fails, the following lemma rules out convergence to a particular set of beliefs with adequate knowledge. For any $E \in \mathcal{B}(\Omega)$, denote $\Delta(E)$ by $\Delta^E$. We define $\Omega_a^+ := \{ \omega \in \Omega: u(a, \omega) \geq u(a', \omega), \forall a' \in A\}$.

\begin{lemma}\label{lem: ball around AK}
  Suppose that there exist $a \in A$ and $n \in \mathbb{N}$ such that some $\{\omega\} \subseteq \cup_{a'} \Omega_{a',a}$ is not distinguishable from some $\{\omega'\} \subseteq \Omega_a^n$. Then for every full-support prior and every equilibrium, $\mathbb{P}(\tilde{\mu}_\infty \in \Delta^*) = 0$, where $\Delta^* := \{\mu \in \Delta^{\Omega_a^+}: \mu(\omega') > 0\}$. 
\end{lemma}

When $A$ is finite and $\Omega_a=\Omega_a^+=\{\omega'\}$, we have $\Delta^*=\{\delta_{\omega'}\}$ as in \Cref{fig:scenario_exact_cascade}. In general, $\Delta^*$ is the set of adequate-knowledge beliefs that assign positive probability to $\omega'$. 

We first show that on the event $\{\tilde\mu_\infty\in\Delta^*\}$, agents eventually choose $a$ almost surely. On this event $\tilde\mu_t(\cup_{a'}\Omega_{a',a})\to 0$ and $\tilde\mu_t(\omega')\to\tilde\mu_\infty(\omega')>0$. This implies that for sufficiently large $t$, $a$ is uniquely optimal under the public belief. The overturning principle \citep{smith_pathological_2000} holds: if $a_t$ is observed by agent $t+1$, then she must find $a_t$ optimal under the updated public belief. Hence $a_t=a$ eventually, almost surely on this event. This agreement on $a$ need not be a cascade: private signals may still induce other actions, and public beliefs may continue to change.

We then argue that if $\{\omega\}$ is not distinguishable from $\{\omega'\}$, then repeated observations of $a$ cannot eliminate $\omega$ relative to $\omega'$. By indistinguishability, at any public history, the probability of choosing an action other than $a$ in state $\omega$, $\mathbb{P}(a_t \neq a \mid \mathcal{F}_t, \omega)$, is at most a constant $C$ times the corresponding probability in state $\omega'$, $\mathbb{P}(a_t \neq a \mid \mathcal{F}_t, \omega')$. Almost surely on the event $\{\tilde{\mu}_\infty \in \Delta^*\}$, actions different from $a$ occur only finitely often, so the conditional Borel--Cantelli lemma and probability bounds give
\[
\sum_t\mathbb{P}(a_t \neq a \mid \mathcal{F}_t, \omega) < \infty, \text{ and} \quad \sum_t\mathbb{P}(a_t \neq a \mid \mathcal{F}_t, \omega') < \infty.
\]

Consequently, the subsequent $a$ observations provide only a finite cumulative log-likelihood advantage to $\omega'$ over $\omega$. We thus have $\lim_{t\to\infty} \tilde{\mu}_t(h)(\omega)/\tilde{\mu}_t(h)(\omega') \in (0, \infty)$. Together with $\tilde{\mu}_\infty(h)(\omega')>0$, this implies $\tilde{\mu}_\infty(h)(\omega)>0$. However, $\omega\in\cup_{a'}\Omega_{a',a}$ and $\tilde{\mu}_\infty\in\Delta^{\Omega_a^+}$ require $\tilde{\mu}_\infty(h)(\omega)=0$, a contradiction. Thus, $\mathbb{P}(\tilde{\mu}_\infty\in\Delta^*)=0$.

Finally, the martingale property gives $\mathbb E[\tilde\mu_\infty(\omega')]=\mu_0(\omega')>0$, so $\tilde\mu_\infty(\omega')>0$ with positive probability. Since $\mathbb{P}(\tilde{\mu}_\infty\in\Delta^*)=0$, the public belief must converge to a stationary belief without adequate knowledge with positive probability. 

After identifying $\cup_{a'} \Omega_{a',a}$ and $\Omega_a^n$ as the relevant sets of states, we adapt the technique used in the proofs of Theorem 3 of \citet{arieli_general_2021} and Lemma SA.1 of \citet{kartik_beyond_arxiv_v5_2024} to show that indistinguishability prevents observations of $a$ from eliminating $\omega$ relative to $\omega'$. Rather than analyzing the probability of a particular infinite history, however, we use it to rule out a class of limiting public beliefs. This accommodates general action sets in which no single action history has positive probability.

The argument establishes herding without requiring public beliefs to become stationary in finite time. Indeed, failure of pairwise escapability need not imply the existence of any full-support stationary beliefs. The existence of stationary beliefs without adequate knowledge and the possibility of reaching or converging to them from a given initial prior are separate questions. As we show in the next section, a seemingly large set of stationary beliefs need not be reachable from every full-support initial prior.

\section{Initial-Prior Dependence}\label{sec:prior_dependence}

\Cref{thm1} leaves open whether herding occurs for every full-support prior when escapability fails but pairwise escapability holds. In the working-paper version, KLLR's Conjecture SA.1, through their Proposition SA.3, implies the following prediction for two-action environments.
\begin{conjecture}[\citealp{kartik_beyond_arxiv_v5_2024}]\label{conj:KLLR}
Suppose that $A = \{a_1, a_2\}$ and preferences are nontrivial. If $\Omega_{a_2,a_1}$ is not distinguishable from $\Omega_{a_1,a_2}$, then there is herding for every full-support prior.
\end{conjecture}

The following example, presented in the introduction, disproves this conjecture.

\begin{example}\label{ex:counter}
  Suppose that  $\Omega=\{1,2,3\}$ and $A=\{a_1,a_2\}$. Agents' utility function and the density of the information structure $F_1$ are given by
\[
u(a,\omega)=
\begin{array}{c|ccc}
 & 1 & 2 & 3\\ \hline a_1 & 1 & 0 & 1\\
a_2 & 0 & 1 & 0\\
\end{array}
\qquad\text{and}\qquad f_1(s\mid\omega)=
\begin{array}{c|ccc}
 & \omega=1 & \omega=2 & \omega=3\\ \cline{1-4}
s<0 & \tfrac12(-s)e^s & \tfrac12e^s & \tfrac12e^s\\
s>0 & \tfrac12e^{-s} & \tfrac12e^{-s} & \tfrac12se^{-s}
\end{array}.
\]
\end{example}

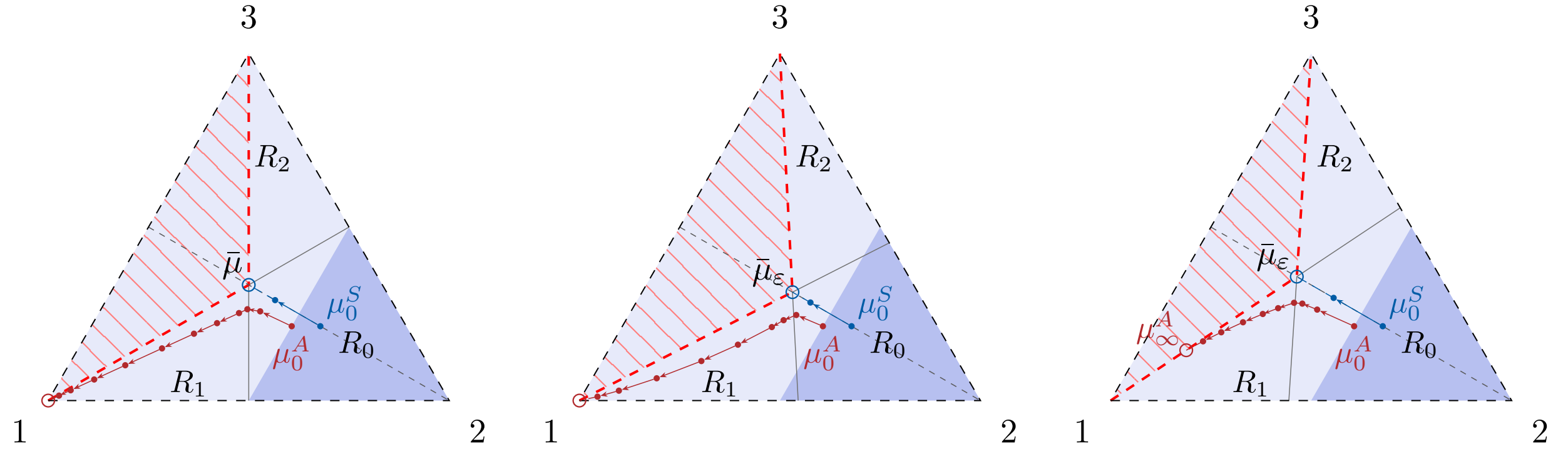
\begin{figure}[htbp]
\centering
\begin{subfigure}[b]{0.32\textwidth}
\centering
\resizebox{\linewidth}{!}{%
\begin{tikzpicture}[x=4.5cm,y=4.5cm,font=\small,
  path arrow/.style={-{Stealth[length=2.2pt,width=1.7pt]},line width=0.3pt,
    shorten <=1pt,shorten >=1pt},
  region boundary/.style={black!55,line width=0.3pt},
  belief label/.style={inner sep=1pt,font=\scriptsize}]
  \definecolor{learningred}{RGB}{178,42,45}
  \definecolor{herdingblue}{RGB}{0,92,166}
  \coordinate (S1) at (0,0);
  \coordinate (S2) at (1,0);
  \coordinate (S3) at (0.5,{sqrt(3)/2});
  \coordinate (B) at ($(S1)!0.5!(S2)$);
  \coordinate (T) at ($(S2)!0.5!(S3)$);
  \coordinate (M) at ($(S1)!0.5!(S3)$);
  \coordinate (C) at (0.5,{sqrt(3)/6});

  \fill[hazyblue!20] (S1)--(S3)--(T)--(B)--cycle;
  \fill[hazyblue!60] (B)--(T)--(S2)--cycle;
  \fill[pattern={Lines[angle=135,distance=6pt]},pattern color=red!50]
    (S1)--(C)--(S3)--cycle;
  \draw[region boundary] (B)--(C);
  \draw[region boundary] (C)--(T);
  \draw[thick,dashed,red] (S1)--(C)--(S3);
  \draw[black!65,dash pattern=on 2pt off 2pt,line width=0.3pt] (M)--(S2);
  \path (M)--(C);

  \node[font=\footnotesize] at (0.77,0.14) {$R_0$};
  \node[font=\footnotesize] at (0.35,0.042) {$R_1$};
  \node[font=\footnotesize] at (0.56,0.605) {$R_2$};

  \draw[dashed] (S1)--(S2)--(S3)--cycle;

  \coordinate (A0) at (0.607142857143,0.185576872240); 
  \coordinate (A1) at (0.528571428571,0.222692246687); 
  \coordinate (A2) at (0.496640465602,0.227982591499); 
  \coordinate (A3) at (0.475727387029,0.218382451802); 
  \coordinate (A4) at (0.420987238791,0.193254010366); 
  \coordinate (A5) at (0.363445825351,0.166839649348); 
  \coordinate (A6) at (0.283740470286,0.130250940488); 
  \coordinate (A7) at (0.192087756518,0.088177801769); 
  \coordinate (A8) at (0.114620211181,0.052616358499); 
  \coordinate (A9) at (0.055984818889,0.025699806961); 
  \coordinate (A10) at (0.026847107356,0.012324153051); 
  \coordinate (Alimit) at (0.000000000000,0.000000000000); 
  \draw[path arrow,learningred] (A0)--(A1);
  \draw[path arrow,learningred,shorten <=0.5pt,shorten >=0.5pt,-{Stealth[length=1.8pt,width=1.4pt]}] (A1)--(A2);
  \draw[learningred,line width=0.3pt] (A2)--(A3);
  \draw[path arrow,learningred] (A3)--(A4);
  \draw[path arrow,learningred] (A4)--(A5);
  \draw[path arrow,learningred] (A5)--(A6);
  \draw[path arrow,learningred] (A6)--(A7);
  \draw[path arrow,learningred] (A7)--(A8);
  \draw[path arrow,learningred] (A8)--(A9);
  \draw[path arrow,learningred,shorten <=0.5pt,shorten >=0.5pt,-{Stealth[length=1.8pt,width=1.4pt]}] (A9)--(A10);
  \draw[learningred,line width=0.3pt,densely dashed] (A10)--(Alimit);
  \foreach \i in {0,...,10} \fill[learningred] (A\i) circle[radius=1pt];
  \node[below=2pt,text=learningred,belief label,font=\footnotesize] at (A0) {$\mu_0^{A}$};
  \coordinate (H0) at (0.678571428571,0.185576872240); 
  \coordinate (H1) at (0.565789473684,0.250691564253); 
  \draw[path arrow,herdingblue] (H0)--(H1);
  \draw[herdingblue,line width=0.3pt,densely dashed] (H1)--(C);
  \foreach \i in {0,1} \fill[herdingblue] (H\i) circle[radius=1pt];
  \node[above right=1pt,text=herdingblue,belief label,font=\footnotesize] at (H0) {$\mu_0^{S}$};
  \draw[learningred,fill=none,line width=0.45pt] (Alimit) circle[radius=2pt];
  \draw[herdingblue,fill=none,line width=0.45pt] (C) circle[radius=2pt];
  \node[above left=1pt,belief label,font=\small] at (C) {$\bar\mu$};
  \node[below left=3pt] at (S1) {$1$};
  \node[below right=3pt] at (S2) {$2$};
  \node[above=4pt] at (S3) {$3$};
\end{tikzpicture}%
}
\caption{$\varepsilon=0$}
\label{fig:counterexample_baseline}
\end{subfigure}
\hfill
\begin{subfigure}[b]{0.32\textwidth}
\centering
\resizebox{\linewidth}{!}{%
\begin{tikzpicture}[x=4.5cm,y=4.5cm,font=\small,
  path arrow/.style={-{Stealth[length=2.2pt,width=1.7pt]},line width=0.3pt,
    shorten <=1pt,shorten >=1pt},
  region boundary/.style={black!55,line width=0.3pt},
  belief label/.style={inner sep=1pt,font=\scriptsize}]
  \definecolor{learningred}{RGB}{178,42,45}
  \definecolor{herdingblue}{RGB}{0,92,166}
  \coordinate (S1) at (0,0);
  \coordinate (S2) at (1,0);
  \coordinate (S3) at (0.5,{sqrt(3)/2});
  \coordinate (B) at ($(S1)!0.5!(S2)$);
  \coordinate (T) at ($(S2)!0.5!(S3)$);
  \coordinate (M) at ($(S1)!0.5!(S3)$);
  \coordinate (C) at (0.531250000000,0.270632938683);
  \coordinate (Be) at (0.545454545455,0.000000000000);
  \coordinate (Te) at (0.772727272727,0.393647910811);

  \fill[hazyblue!20] (S1)--(S3)--(T)--(B)--cycle;
  \fill[hazyblue!60] (B)--(T)--(S2)--cycle;
  \fill[pattern={Lines[angle=135,distance=6pt]},pattern color=red!50]
    (S1)--(C)--(S3)--cycle;
  \draw[region boundary] (Be)--(C)--(Te);
  \draw[thick,dashed,red] (S1)--(C)--(S3);
  \draw[black!65,dash pattern=on 2pt off 2pt,line width=0.3pt] (M)--(S2);
  \node[font=\footnotesize] at (0.77,0.14) {$R_0$};
  \node[font=\footnotesize] at (0.35,0.042) {$R_1$};
  \node[font=\footnotesize] at (0.585,0.605) {$R_2$};
  \draw[dashed] (S1)--(S2)--(S3)--cycle;

  \coordinate (A0) at (0.607142857143,0.185576872240); 
  \coordinate (A1) at (0.540710700355,0.213597682981); 
  \coordinate (A2) at (0.509186056789,0.200637566223); 
  \coordinate (A3) at (0.479762307354,0.183448558338); 
  \coordinate (A4) at (0.395265485930,0.139497611950); 
  \coordinate (A5) at (0.304099546422,0.098654401421); 
  \coordinate (A6) at (0.192671423230,0.055657866804); 
  \coordinate (A7) at (0.098427294031,0.024684905412); 
  \coordinate (A8) at (0.045117296119,0.009760278075); 
  \coordinate (Alimit) at (0.000000000000,0.000000000000); 
  \draw[path arrow,learningred] (A0)--(A1);
  \draw[path arrow,learningred,shorten <=0.5pt,shorten >=0.5pt,-{Stealth[length=1.8pt,width=1.4pt]}] (A1)--(A2);
  \draw[path arrow,learningred,shorten <=0.5pt,shorten >=0.5pt,-{Stealth[length=1.8pt,width=1.4pt]}] (A2)--(A3);
  \draw[path arrow,learningred] (A3)--(A4);
  \draw[path arrow,learningred] (A4)--(A5);
  \draw[path arrow,learningred] (A5)--(A6);
  \draw[path arrow,learningred] (A6)--(A7);
  \draw[path arrow,learningred] (A7)--(A8);
  \draw[learningred,line width=0.3pt,densely dashed] (A8)--(Alimit);
  \foreach \i in {0,...,8} \fill[learningred] (A\i) circle[radius=1pt];
  \node[below=2pt,text=learningred,belief label,font=\footnotesize] at (A0) {$\mu_0^{A}$};
  \coordinate (H0) at (0.678571428571,0.185576872240); 
  \coordinate (H1) at (0.576086956522,0.244746309765); 
  \draw[path arrow,herdingblue] (H0)--(H1);
  \draw[herdingblue,line width=0.3pt,densely dashed] (H1)--(C);
  \foreach \i in {0,1} \fill[herdingblue] (H\i) circle[radius=1pt];
  \node[above right=1pt,text=herdingblue,belief label,font=\footnotesize] at (H0) {$\mu_0^{S}$};
  \draw[learningred,fill=none,line width=0.45pt] (Alimit) circle[radius=2pt];
  \draw[herdingblue,fill=none,line width=0.45pt] (C) circle[radius=2pt];
  \node[above left=1pt,belief label,font=\small] at (C) {$\bar\mu_\varepsilon$};
  \node[below left=3pt] at (S1) {$1$};
  \node[below right=3pt] at (S2) {$2$};
  \node[above=4pt] at (S3) {$3$};
\end{tikzpicture}%
}
\caption{$\varepsilon=1/5$}
\label{fig:counterexample_positive}
\end{subfigure}
\hfill
\begin{subfigure}[b]{0.32\textwidth}
\centering
\resizebox{\linewidth}{!}{%
\begin{tikzpicture}[x=4.5cm,y=4.5cm,font=\small,
  path arrow/.style={-{Stealth[length=2.2pt,width=1.7pt]},line width=0.3pt,
    shorten <=1pt,shorten >=1pt},
  region boundary/.style={black!55,line width=0.3pt},
  belief label/.style={inner sep=1pt,font=\scriptsize}]
  \definecolor{learningred}{RGB}{178,42,45}
  \definecolor{herdingblue}{RGB}{0,92,166}
  \coordinate (S1) at (0,0);
  \coordinate (S2) at (1,0);
  \coordinate (S3) at (0.5,{sqrt(3)/2});
  \coordinate (B) at ($(S1)!0.5!(S2)$);
  \coordinate (T) at ($(S2)!0.5!(S3)$);
  \coordinate (M) at ($(S1)!0.5!(S3)$);
  \coordinate (C) at (0.464285714286,0.309294787066);
  \coordinate (Be) at (0.444444444444,0.000000000000);
  \coordinate (Te) at (0.722222222222,0.481125224325);

  \fill[hazyblue!20] (S1)--(S3)--(T)--(B)--cycle;
  \fill[hazyblue!60] (B)--(T)--(S2)--cycle;
  \fill[pattern={Lines[angle=135,distance=6pt]},pattern color=red!50]
    (S1)--(C)--(S3)--cycle;
  \draw[region boundary] (Be)--(C)--(Te);
  \draw[thick,dashed,red] (S1)--(C)--(S3);
  \draw[black!65,dash pattern=on 2pt off 2pt,line width=0.3pt] (M)--(S2);
  \node[font=\footnotesize] at (0.77,0.14) {$R_0$};
  \node[font=\footnotesize] at (0.35,0.042) {$R_1$};
  \node[font=\footnotesize] at (0.56,0.605) {$R_2$};
  \draw[dashed] (S1)--(S2)--(S3)--cycle;

  \coordinate (A0) at (0.607142857143,0.185576872240); 
  \coordinate (A1) at (0.515069938311,0.227018955063); 
  \coordinate (A2) at (0.478006641956,0.242003175511); 
  \coordinate (A3) at (0.457377518251,0.244030910445); 
  \coordinate (A4) at (0.417706250177,0.230608292963); 
  \coordinate (A5) at (0.379472526635,0.216113460954); 
  \coordinate (A6) at (0.330136481609,0.195463985968); 
  \coordinate (A7) at (0.275864571240,0.170484165603); 
  \coordinate (A8) at (0.230502732961,0.147892107758); 
  \coordinate (Alimit) at (0.188566021929,0.125617665601); 
  \draw[path arrow,learningred] (A0)--(A1);
  \draw[path arrow,learningred,shorten <=0.5pt,shorten >=0.5pt,-{Stealth[length=1.8pt,width=1.4pt]}] (A1)--(A2);
  \draw[learningred,line width=0.3pt] (A2)--(A3);
  \draw[path arrow,learningred,shorten <=0.5pt,shorten >=0.5pt,-{Stealth[length=1.8pt,width=1.4pt]}] (A3)--(A4);
  \draw[path arrow,learningred,shorten <=0.5pt,shorten >=0.5pt,-{Stealth[length=1.8pt,width=1.4pt]}] (A4)--(A5);
  \draw[path arrow,learningred] (A5)--(A6);
  \draw[path arrow,learningred] (A6)--(A7);
  \draw[path arrow,learningred] (A7)--(A8);
  \draw[learningred,line width=0.3pt,densely dashed] (A8)--(Alimit);
  \foreach \i in {0,...,8} \fill[learningred] (A\i) circle[radius=1pt];
  \node[below=2pt,text=learningred,belief label,font=\footnotesize] at (A0) {$\mu_0^{A}$};
  \coordinate (H0) at (0.678571428571,0.185576872240); 
  \coordinate (H1) at (0.556122448980,0.256272823569); 
  \draw[path arrow,herdingblue] (H0)--(H1);
  \draw[herdingblue,line width=0.3pt,densely dashed] (H1)--(C);
  \foreach \i in {0,1} \fill[herdingblue] (H\i) circle[radius=1pt];
  \node[above right=1pt,text=herdingblue,belief label,font=\footnotesize] at (H0) {$\mu_0^{S}$};
  \draw[learningred,fill=none,line width=0.45pt] (Alimit) circle[radius=2pt];
  \draw[herdingblue,fill=none,line width=0.45pt] (C) circle[radius=2pt];
  \node[above left=1pt,belief label,font=\small] at (C) {$\bar\mu_\varepsilon$};
  \node[above left=0.5pt,text=learningred,belief label,font=\footnotesize] at (Alimit) {$\mu_\infty^{A}$};
  \node[below left=3pt] at (S1) {$1$};
  \node[below right=3pt] at (S2) {$2$};
  \node[above=4pt] at (S3) {$3$};
\end{tikzpicture}%
}
\caption{$\varepsilon=-1/5$}
\label{fig:counterexample_negative}
\end{subfigure}
\caption{Belief paths under repeated $a_1$ observations for the original and perturbed information structures. The light and dark blue regions indicate where $a_1$ and $a_2$ are optimal, respectively. Red hatching marks $C_1^\varepsilon$. All panels use the same priors: $\mu_0^A=(4,7,3)/14$ for the red path and $\mu_0^S=(3,8,3)/14$ for the blue path. Open circles denote limits. In (a) and (b), the asymmetric path converges to $\delta_1$; in (c), both paths converge to full-support cascade beliefs.}
\label{fig:counterexample_regions}
\end{figure}

For a full-support belief $\mu \in \Delta(\Omega)$, define the two belief odds $\lambda_1(\mu) := \mu(1)/\mu(2)$ and $\lambda_2(\mu) := \mu(3)/\mu(2)$. Within the set of full-support beliefs, define the cascade belief set $C_1 = \{\mu \in \Delta(\Omega): \lambda_1(\mu) \geq 1, \lambda_2(\mu) \geq 1\}$. Preferences and $C_1$ are depicted in \Cref{fig:counterexample_baseline}.

Every pair of states is distinguishable in this example, so pairwise escapability holds. However, $\Omega_{a_2,a_1}=\{2\}$ is not distinguishable from $\Omega_{a_1,a_2}=\{1,3\}$, so escapability fails and herding occurs for some initial prior. \Cref{conj:KLLR} would imply herding for every full-support prior. The conjecture may seem plausible because $C_1$ contains full-support beliefs, whereas in \Cref{fig: figure_sa1_kllr} no full-support belief is stationary. However, the following proposition shows that learning occurs for generic full-support priors outside the cascade belief set.

\begin{proposition}\label{prop:prior_dependence}
  In \Cref{ex:counter}, there is adequate learning for generic full-support priors outside the cascade belief set. However, there is herding if the prior satisfies $0 < \lambda_1(\mu_0) = \lambda_2(\mu_0) < 1$.
  
\end{proposition} 

Here, ``generic'' means that, among full-support priors outside $C_1$, the exceptions form a meager set (a countable union of nowhere-dense sets) of two-dimensional Lebesgue measure zero. The proof has three steps. First, beliefs starting outside $C_1$ never enter it in finite time. Second, under repeated $a_1$ observations, their limit belongs to $C_1$ only when the two odds are equal. Finally, the priors from which a finite history makes these odds equal form a meager null set.

To describe the dynamics, label three regions outside $C_1$: $R_0 := \{\mu \in \Delta(\Omega): \lambda_1(\mu) < 1, \lambda_2(\mu) < 1\}$, $R_1 := \{\mu \in \Delta(\Omega): \lambda_1(\mu) > 1, \lambda_2(\mu) < 1\}$, and $R_2 := \{\mu \in \Delta(\Omega): \lambda_1(\mu) < 1, \lambda_2(\mu) > 1\}$, depicted in \Cref{fig:counterexample_baseline}. The following lemma summarizes the belief dynamics.

\begin{lemma}\label{lem:counter_dynamics_main}
In \Cref{ex:counter}, fix a full-support prior $\mu_0\notin C_1$. Public beliefs remain outside $C_1$ after every finite action history. Under repeated $a_1$ observations, the limit is $\bar\mu:=(1/3,1/3,1/3)$, $\delta_1$, or $\delta_3$ according as $\lambda_1(\mu_0)-\lambda_2(\mu_0)$ is zero, positive, or negative.
\end{lemma}

First, neither action can move a full-support belief from outside $C_1$ into $C_1$. For an $a_1$ observation, the conclusion follows from the updating rule under $F_1$. For an $a_2$ observation, this follows from the overturning principle: after $a_2$ is observed, that action is optimal under the public belief, whereas $a_1$ is strictly optimal throughout $C_1$. Herding hence requires that all sufficiently late observations be $a_1$; otherwise, the public belief cannot converge to $C_1$.

However, herding can occur for symmetric priors satisfying $0<\lambda_1(\mu_0)=\lambda_2(\mu_0)<1$. With positive probability, all agents choose $a_1$. Along this history, the two belief odds remain equal and converge to $1$, so the public belief converges to the cascade belief $\bar\mu=(1/3,1/3,1/3)$, as illustrated by the blue path from $\mu_0^S$ in \Cref{fig:counterexample_baseline}.

For priors outside $C_1$ satisfying $\lambda_1(\mu_0)\neq\lambda_2(\mu_0)$, repeated $a_1$ observations instead lead to the point mass on state $1$ or $3$, according to which ratio is larger. For example, consider $\mu_0^A\in R_0$ with $\lambda_1(\mu_0^A)>\lambda_2(\mu_0^A)$ in \Cref{fig:counterexample_baseline}. First, $a_1$ observations take the public belief into region $R_1$. Thereafter, action $a_2$ is chosen only after signals $s<0$ close to zero, where $f_1(s\mid 3)=f_1(s\mid 2)$. Observing $a_1$ thus increases $\lambda_1$ but leaves $\lambda_2$ unchanged. Hence $\lambda_1$ diverges and the public belief converges to the point mass on state $1$.

For priors outside $C_1$, herding thus requires the public belief to reach the diagonal after finitely many observations. This diagonal, $\{\mu:\lambda_1(\mu)=\lambda_2(\mu)<1\}$, has Lebesgue measure zero. On each $R_i$, the update maps are real analytic with Jacobian determinants not identically zero, so they are locally invertible outside a null set. The region boundaries are also null. Consequently, the priors leading to the diagonal under any fixed finite history form a null set. This set is also closed relative to the full-support beliefs outside $C_1$, since belief updates are continuous and the diagonal is relatively closed there; hence it is nowhere dense in that domain. The countable union of these sets over finite histories is therefore meager and null. This establishes learning for generic full-support priors outside $C_1$. The proofs are in \Cref{SA: ex_perturbed}, with supporting calculations in \Cref{SA: ex_counter}.

\Cref{ex:counter} thus disproves \Cref{conj:KLLR}. It also violates the following conjecture about infinite action histories, from which KLLR derive \Cref{conj:KLLR} in Proposition~SA.3 of their working paper.

\begin{conjecture}[Conjecture SA.1 in \citet{kartik_beyond_arxiv_v5_2024}]\label{conj:SA1}
  If learning is adequate at some full-support prior $\mu_0$ and some equilibrium with action set $\{a_1,a_2\}$, then there exist an infinite history $h^\infty$ and $\varepsilon>0$ such that $\mathbb P(h^\infty\mid\omega)>\varepsilon$ for every $\omega\in\Omega_{a_1,a_2}$.
\end{conjecture}

Intuitively, \Cref{conj:SA1} says that even though adequate learning requires learning the correct action, the public history should not \textit{always} distinguish states with the same optimal action. That is, there should be some infinite history that is consistent with all states in $\Omega_{a_1,a_2}$. However, this is not true. 

\begin{proposition}
  In \Cref{ex:counter}, fix any full-support prior and equilibrium at which learning is adequate. Then, for every infinite history $h^\infty$, $\mathbb P(h^\infty\mid 1) \mathbb P(h^\infty\mid 3)=0$.
\end{proposition}

\begin{proof}
Suppose that there exists an infinite history $h^\infty$ satisfying $\mathbb P(h^\infty\mid 1)>0$ and $\mathbb P(h^\infty\mid 3)>0$. Let $h^n$ denote the length-$(n-1)$ prefix of $h^\infty$. By Bayes' rule, $\tilde{\mu}_n(h^n)(\omega) = (\mu_0(\omega)\mathbb P(h^n\mid \omega))/ (\sum_j\mu_0(j)\mathbb P(h^n\mid j))$. Since the events $h^n$ decrease to $h^\infty$, continuity of probability from above implies $\mathbb P(h^n\mid \omega)\to\mathbb P(h^\infty\mid \omega)$. Since $\mu_0$ has full support, the denominator is bounded away from zero. Taking limits on both sides yields $\tilde{\mu}_\infty(h^\infty)(\omega) > 0$ for $\omega\in\{1,3\}$. Adequate learning requires $\tilde{\mu}_\infty(2)=0$. Consequently, both $\lambda_1(\tilde{\mu}_n(h^n))$ and $\lambda_2(\tilde{\mu}_n(h^n))$ eventually exceed $1$, so the public belief enters $C_1$, contradicting adequate learning.
\end{proof}

The geometry of the cascade region explains why adequate learning in \Cref{ex:counter} contradicts \Cref{conj:SA1}. The set $C_1\cup\Delta(\{1,3\})$ contains a neighborhood, relative to the belief simplex, of every nondegenerate belief in $\Delta(\{1,3\})$. Hence, full-support public beliefs outside $C_1$ cannot converge to such a belief. Adequate learning therefore requires the public history to distinguish states $1$ and $3$, even though they have identical payoff vectors, ruling out the common positive-probability history posited by the conjecture. To complete the counterexample, however, one must also establish that adequate learning occurs for some full-support prior. Geometry alone does not guarantee this: in \Cref{ex:perturbed} with $\varepsilon<0$ below, the cascade region has a similar shape, yet herding occurs for every full-support prior. \Cref{prop:prior_dependence} establishes the more surprising part: in \Cref{ex:counter}, adequate learning occurs for generic full-support priors outside $C_1$. From these priors, learning the correct action requires society to learn the true state almost surely. A cascade region with nonempty interior need not be ``attracting''.

\subsection{Perturbation}\label{subsec:perturbation}

In \Cref{ex:counter}, the public belief evades the cascade region $C_1$ along repeated $a_1$ observations. This feature relies on the fact that in $R_1$, for instance, $a_2$ is chosen only after signals $s<0$ close to zero, where $f_1(s\mid 3)=f_1(s\mid 2)$, and $\lambda_2$ is therefore unchanged by observing $a_1$. In this subsection we perturb the information structure so that $\lambda_2$ is affected by observing $a_1$ in $R_1$. The learning outcome then depends on the direction of the perturbation. 

\begin{example}\label{ex:perturbed}
  Preferences are as in \Cref{ex:counter}. For $\varepsilon\in(-1/4,1/4)$, the information structure $F_1^\varepsilon$ has density
\[
f_1^\varepsilon(s\mid\omega)=
\begin{array}{c|ccc}
 & \omega=1 & \omega=2 & \omega=3\\ \cline{1-4}
s<0 & \tfrac{1-\varepsilon}{2}(-s)e^s & \tfrac12e^s & \tfrac{1+\varepsilon}{2}e^s\\
s>0 & \tfrac{1+\varepsilon}{2}e^{-s} & \tfrac12e^{-s} & \tfrac{1-\varepsilon}{2}se^{-s}
\end{array}.
\]
\end{example}

At $\varepsilon=0$, $F_1^\varepsilon=F_1$. The perturbation is small in relative terms: $|f_1^\varepsilon(s\mid\omega)/f_1(s\mid\omega)-1|\le|\varepsilon|$ for every $\omega$ and $s\neq0$. As in \Cref{ex:counter}, every pair of states is distinguishable, whereas $f_1^\varepsilon(s\mid1)+f_1^\varepsilon(s\mid3)\geq(1+\varepsilon)f_1^\varepsilon(s\mid2)$ implies that $\{2\}$ is not distinguishable from $\{1,3\}$. The full-support cascade belief set is $C_1^\varepsilon:=\{\mu:\lambda_1(\mu)\geq b_{\varepsilon},\ \lambda_2(\mu)\geq b_{\varepsilon}\}$, where $b_{\varepsilon}:=1/(1+\varepsilon)$.

\begin{proposition}\label{prop:perturbed}
  In \Cref{ex:perturbed}, public beliefs starting from a full-support prior outside $C_1^\varepsilon$ remain outside $C_1^\varepsilon$ after every finite action history.
  \begin{enumerate}[label=(\roman*)]
    \item If $\varepsilon\geq0$, there is adequate learning for generic full-support priors outside $C_1^\varepsilon$. However, there is herding if the prior satisfies $0<\lambda_1(\mu_0)=\lambda_2(\mu_0)<b_{\varepsilon}$.
    \item If $\varepsilon<0$, there is herding for every full-support prior.
  \end{enumerate}
\end{proposition}

The sign of $\varepsilon$ determines what overturning signals reveal about the less likely extreme state. Consider beliefs with $\lambda_1\geq b_\varepsilon>\lambda_2$. Agents choose $a_2$ only after signals in an interval $(-\alpha,0)$ with $\alpha>0$, on which $f_1^\varepsilon(s\mid3)/f_1^\varepsilon(s\mid2)=1+\varepsilon$. If $\varepsilon>0$, overturning signals favor state $3$ over state $2$, so observing $a_1$ lowers $\lambda_2$. The asymmetry between the extreme states grows, and repeated $a_1$ observations lead to $\delta_1$, as shown in \Cref{fig:counterexample_positive}. If $\varepsilon<0$, overturning signals favor state $2$, so repeated $a_1$ observations raise $\lambda_2$ toward $b_\varepsilon$. In fact, $a_1$ observations make the public belief converge to the cascade region $C_1^\varepsilon$, and herding occurs for every full-support prior.

The perturbation makes it clear that identifying the cascade region and identifying herding are distinct problems. The cascade regions in all panels of \Cref{fig:counterexample_regions} are similar, but because observing $a_1$ drives the public belief toward or away from the cascade region, the learning outcome varies.

Prior dependence provides a new perspective on the release of public information in social learning \citep[Section III.A]{BHW1992}. In the standard binary-state setting, changing the public belief while preserving full support does not change the asymptotic learning outcome. With multiple states, this need not be true. In \Cref{ex:counter}, a public disclosure may move a belief from the cascade region $C_1$ to a full-support belief outside $C_1$, where learning occurs for generic full-support priors.

\section{Application}\label{sec: application}
\Cref{sec:prior_dependence} shows that the learning outcome can depend on the initial prior even when preferences and information are fixed. We now identify a class of one-dimensional environments in which escapability provides a complete characterization independent of the full-support initial prior.

\subsection{Interval Optimality}\label{subsec: scd}

In this subsection, we assume totally ordered one-dimensional state and signal spaces. For simplicity, let $\Omega \subseteq \mathbb{R}$ and $S \subseteq \mathbb{R}$. Recall the \textit{monotone likelihood ratio property} (MLRP): for any $s' > s$ and $\omega' > \omega$, $f(s' \mid \omega')/f(s' \mid \omega) \geq f(s \mid \omega')/f(s \mid \omega)$. We consider the following condition on preferences:

\begin{definition}
  Preferences satisfy \textit{interval optimality} if for every $a \in A$, the set of states where $a$ is uniquely optimal, $\Omega_a$, is an interval: if $\omega, \omega' \in \Omega_a$ and $\omega'' \in \Omega$ satisfies $\omega < \omega'' < \omega'$, then $\omega'' \in \Omega_a$.
\end{definition}

Interval optimality is a natural assumption in economics. Notably, it is implied by single-crossing differences \citep{kartik_single-crossing_2024} when each state has a unique optimal action. It is also closely related to convex choice \citep{kartik_convex_2025}. Though the discrete state space is not convex, if the utility function is defined on a convex set containing $\Omega$, then interval optimality is implied by convex choice. Under interval optimality and MLRP information, escapability is not only necessary but also sufficient for adequate learning.

\begin{theorem}\label{thm: escapability_necc_suff}
    Suppose that $A$ is finite, each state has a unique optimal action, interval optimality holds, and the information structure satisfies MLRP. There is adequate learning if (pairwise) escapability holds. If (pairwise) escapability fails, there is herding for every full-support prior.
\end{theorem}

\begin{proof}
  By \Cref{lem:escapability_equiv_pairwise_escapability}, escapability and pairwise escapability are equivalent under the stated assumptions. It only remains to show that pairwise escapability implies adequate learning. Suppose that pairwise escapability holds. By \Cref{lem:learning_limit}, it suffices to show that every stationary belief has adequate knowledge. 
  
  Take a belief $\mu$ without adequate knowledge and define $A':=\{a\in A:\mu(\Omega_a)>0\}$. Unique optimality implies that $\{\Omega_a\cap\mathrm{supp}(\mu)\}_{a\in A'}$ partitions $\mathrm{supp}(\mu)$, with $|A'|>1$. By interval optimality and finiteness of $A$, this partition has distinct lowest and highest regions, corresponding to $a^-$ and $a^+$. Countability of $\Omega$ lets us choose $\omega^-\in\Omega_{a^-}$ with $\mu(\omega^-)>0$. Since $A$ is finite and optimal actions are unique, $\omega^-\in\Omega_{a^-}^n$ for some $n\in\mathbb N$; every $\omega\in\mathrm{supp}(\mu)\setminus\Omega_{a^-}$ lies above $\omega^-$ and belongs to $\Omega_b^m$ for its unique optimal action $b\neq a^-$ and some $m\in\mathbb N$.

Pairwise escapability therefore implies that $\{\omega^-\}$ is distinguishable from $\{\omega\}$. By MLRP, the likelihood ratio $f(s\mid\omega)/f(s\mid\omega^-)$ is nondecreasing in $s$ and converges to 0 as $s\downarrow\inf S$.

Now consider the posterior ratio between $\Omega\setminus\Omega_{a^-}$ and $\{\omega^-\}$:
\begin{align*}
    \frac{\mu_s(\Omega\setminus\Omega_{a^-})}{\mu_s(\omega^-)} &= \frac{1}{\mu(\omega^-)} \sum_{\omega\in\mathrm{supp}(\mu)\setminus\Omega_{a^-}} \mu(\omega)\frac{f(s\mid\omega)}{f(s\mid\omega^-)}.
\end{align*}
Choose $s_0$ such that $0<f(s_0\mid\omega^-)<\infty$, and $\sum_{\omega\in\Omega}\mu(\omega)f(s_0\mid\omega)<\infty$. Such a signal exists because the signal distributions are mutually absolutely continuous and $\Pr_{\mu}(\cdot) < \infty$. Hence by the dominated convergence theorem, $\mu_s(\Omega\setminus\Omega_{a^-})/\mu_s(\omega^-)$ converges to 0 as $s\downarrow\inf S$. This implies that with positive probability, action $a^-$ is uniquely optimal after observing a sufficiently low signal. The symmetric argument, using a state $\omega^+\in\Omega_{a^+}$ with $\mu(\omega^+)>0$, shows that action $a^+$ is uniquely optimal after observing a sufficiently high signal with positive probability. This means that $\mu$ is not stationary, and hence every stationary belief has adequate knowledge.
\end{proof}

As a corollary, under the assumptions of \Cref{thm: escapability_necc_suff}, \Cref{conj:KLLR} is true, which is consistent with Claim~SA.1 of KLLR's working paper. Note that our counterexample violates both MLRP and interval optimality. In fact, by interchanging the labels of states $2$ and $3$ in the utility function and the information structure, the counterexample can be modified to satisfy interval optimality (in fact, it satisfies SCD defined later) but violate MLRP without changing the dynamics. Moreover, in \Cref{SA: ex_mlrp}, we provide an adaptation that satisfies strict MLRP but violates interval optimality, in which prior dependence similar to that in \Cref{prop:prior_dependence} arises. In this sense, MLRP and interval optimality are a minimal pair of conditions for prior-independence. 

Under the assumptions of \Cref{thm: escapability_necc_suff}, preferences and information jointly characterize learning and herding, independently of the full-support initial prior. We next ask whether information alone can characterize learning within this class. The next proposition gives a positive answer when the information structure is location-shift. An information structure is location-shift if $S = \mathbb{R}$ and there exists some density $g: \mathbb{R} \to \mathbb{R}_{++}$ such that $f(s \mid \omega) = g(s - \omega)$ for all $\omega \in \Omega$ and $s \in S$.  For location-shift information structures, log-concavity of $g$ implies MLRP.

\begin{definition}\label{def: pairwise_dist}
    An information structure satisfies \textit{pairwise distinguishability} if for every pair of states $\omega, \omega' \in \Omega$, $\{\omega\}$ is distinguishable from $\{\omega'\}$.
\end{definition}

\begin{proposition}\label{prop: pairwise}
In the setting of \Cref{thm: escapability_necc_suff} with nontrivial preferences, if the information structure is also location-shift, then (pairwise) escapability is equivalent to pairwise distinguishability.
\end{proposition}

\begin{proof}
 It suffices to show pairwise escapability implies pairwise distinguishability. Suppose that pairwise distinguishability fails. Consider first the case where there exist $\omega<\omega'$ such that $\{\omega'\}$ is not distinguishable from $\{\omega\}$. Let $\delta:=\omega'-\omega>0$. Then there exists some $1\leq C<\infty$ such that $g(s-\delta)\leq Cg(s)$ for a.e.\ $s\in\mathbb{R}$. Let $\overline G(s):=\int_s^\infty g(t)\,\mathrm{d}t$. Integrating the preceding inequality gives $\overline G(s-\delta)\leq C\overline G(s)$. Iterating, we obtain $\overline G(s-m\delta)\leq C^m\overline G(s)$ for every $m\in\mathbb{N}$.

  Take any $\xi<\xi'$ in $\Omega$, and let $\delta':=\xi'-\xi$ and $m:=\lceil\delta'/\delta\rceil$. By MLRP, $g(s-\delta')/g(s)$ is nondecreasing in $s$. Hence, for a.e.\ $s$,
  \begin{align*}
    \frac{g(s-\delta')}{g(s)} \leq \frac{\int_s^\infty g(t-\delta')\,\mathrm{d}t} {\int_s^\infty g(t)\,\mathrm{d}t} = \frac{\overline G(s-\delta')}{\overline G(s)} \leq\frac{\overline G(s-m\delta)}{\overline G(s)} \leq C^m<\infty.
  \end{align*}
  The first inequality follows by integrating $g(s-\delta')/g(s)\leq g(t-\delta')/g(t)$ over $t\geq s$ after multiplying both sides by $g(t)$; the second uses $\delta'\leq m\delta$. Thus, $\{\xi'\}$ is not distinguishable from $\{\xi\}$ for every $\xi<\xi'$ in $\Omega$.

  By assumption, we can choose $\xi<\xi'$ with different optimal actions. Let $a$ be the unique optimal action at $\xi$. Finiteness of $A$ implies that $\xi\in\Omega_a^n$ for some $n\in\mathbb{N}$, whereas $\xi'\in\cup_{a'}\Omega_{a',a}$. Since $\{\xi'\}$ is not distinguishable from $\{\xi\}$, pairwise escapability fails. The case where $\{\omega\}$ is not distinguishable from $\{\omega'\}$ for some $\omega<\omega'$ is symmetric. This completes the proof.
\end{proof}

For location-shift information, KLLR show that if the tails are ``subexponential'' in the sense that there exist some $p > 1$ and $M > 0$ such that $g(s) < \exp(-|s|^p)$ for all $|s| > M$, then there is adequate learning for SCD preferences. Subexponential tails imply pairwise distinguishability \citep[Lemma 2]{kartik_beyond_2024}. The converse is not true: pairwise distinguishability is weaker than subexponential tails even with MLRP. In fact, if $g$ is log-concave, then pairwise distinguishability is equivalent to $\lim_{|s| \to \infty} -\log g(s) / |s| = \infty$. For example, the density $g(s) = (1/Z)\exp(-|s|\log(1 +|s|))$ is log-concave and satisfies pairwise distinguishability but does not have subexponential tails. 

Pairwise distinguishability can be viewed as the pairwise analogue of the unbounded-beliefs condition in \citet{smith_pathological_2000}. \Cref{prop: pairwise} thus shows that, although learning depends on preferences, information, and initial priors, the familiar binary-state dichotomy, which depends only on the information structure, extends to a class of multi-state environments. 

The only previous complete characterizations we know of are Theorem 3 of \citet{arieli_general_2021} and the binary-action characterization implied by Claim SA.1 of \citet{kartik_beyond_arxiv_v5_2024}. The former covers finite $\Omega=A$ with matching utility $\mathbb{1}_{\{\omega = a\}}$. In this environment, pairwise distinguishability is equivalent to excludability and pairwise escapability, so the result is immediate. The latter considers binary action spaces with SCD preferences and strict MLRP information on a discrete ordered state space. Together with their Proposition SA.3, KLLR's Claim SA.1 establishes \Cref{conj:KLLR}, making excludability necessary and sufficient for learning at every full-support prior. When each state has a unique optimal action, \Cref{thm: escapability_necc_suff} yields the corresponding characterization under weak MLRP. For binary actions, the same conclusion also holds when some states admit both actions as optimal.\footnote{Let $A=\{a_1,a_2\}$. Tie states belong to neither $\Omega_{a_1,a_2}$ nor $\Omega_{a_2,a_1}$, and excludability implies adequate learning \citep{kartik_beyond_2024}. Conversely, suppose that pairwise escapability holds. Since $A$ is finite, each state of $\Omega_{a_1,a_2}=\Omega_{a_1}$ lies in some $\Omega_{a_1}^n$, and similarly for $a_2$, so each state of $\Omega_{a_2,a_1}$ is distinguishable from each state of $\Omega_{a_1,a_2}$, and vice versa. Under SCD, all states of one set lie below all states of the other. The dominated-convergence argument in the proof of \Cref{lem:escapability_equiv_pairwise_escapability} then shows that excludability holds. Hence, if excludability fails, so does pairwise escapability, and \Cref{thm1} gives herding for every full-support prior.}

\subsection{Single-Crossing Differences (SCD)}

Interval optimality is closely connected to another property of preferences called single-crossing differences (SCD) \citep{kartik_beyond_2024,kartik_single-crossing_2024}. A function $h$ is \textit{single-crossing} if either (i) for all $\omega < \omega'$, $h(\omega) \geq (>) 0 \implies h(\omega') \geq (>) 0$ or (ii) for all $\omega < \omega'$, $h(\omega) \leq (<) 0 \implies h(\omega') \leq (<) 0$.\footnote{We adopt the definition of single-crossing from \citet{kartik_single-crossing_2024}, which is slightly different from that in KLLR. Specifically, KLLR define single-crossing as follows: a function $h$ is single-crossing if either (i) for all $\omega < \omega', h(\omega) > 0 \implies h(\omega') \geq 0$ or (ii) for all $\omega < \omega', h(\omega) < 0 \implies h(\omega') \leq 0$. This distinction does not affect our results; we use the definition of \citet{kartik_single-crossing_2024} for ease of exposition.}

\begin{definition}[\citealp{kartik_beyond_2024}]\label{def:SCD}
  Preferences represented by $u : A \times \Omega \to \mathbb{R}$ have \textit{single-crossing differences}  (SCD) if for all $a, a' \in A$, the difference $u(a, \cdot) - u(a', \cdot)$ is single-crossing.
\end{definition}

To relate SCD to interval optimality, consider the interval-choice property of \citet{kartik_single-crossing_2024}. Interval choice requires that the weak choice region $\{\omega:a\in\arg\max_{a'\in B}u(a',\omega)\} $ be an interval for every action set $B\subseteq A$, which is implied by SCD \citep[Theorem 1]{kartik_single-crossing_2024}. We require that the strict optimality region $\Omega_a$ be an interval under the fixed action set $A$. When each state has a unique optimal action, the weak and strict optimality regions coincide for $B=A$; hence interval optimality is implied by interval choice and SCD. The converse is not true: in \Cref{fig:scenario_triangle_a3}, interval optimality holds but SCD and interval choice fail.

KLLR show that if preferences satisfy SCD, then there is adequate learning if the information structure satisfies \textit{Directionally Unbounded Beliefs (DUB)}, which requires that for any state $\bar{\omega}$, $\{\omega: \omega \geq \bar{\omega}\}$ and $\{\omega: \omega < \bar{\omega}\}$ are distinguishable from each other, and $\{\omega: \omega > \bar{\omega} \}$ and $\{\omega: \omega \leq \bar{\omega} \}$ are distinguishable from each other. The idea is that under SCD preferences, the sets $\Omega_{a_1, a_2}$ and $\Omega_{a_2, a_1}$ satisfy $\inf \Omega_{a_1, a_2} \geq \sup \Omega_{a_2, a_1}$ or $\sup \Omega_{a_1, a_2} \leq \inf \Omega_{a_2, a_1}$, so DUB implies excludability and hence adequate learning. 

As remarked by KLLR, DUB is equivalent to pairwise distinguishability under MLRP. Hence pairwise distinguishability implies excludability. At the same time, excludability implies escapability, which is equivalent to pairwise distinguishability in the setting of \Cref{prop: pairwise}. We thus have the following equivalence.

\begin{corollary}\label{prop:equiv_excludability_necess_MLRP}
  Suppose that $A$ is finite, preferences are nontrivial and satisfy SCD, and each state has a unique optimal action. If the information structure is location-shift and satisfies MLRP, then excludability, (pairwise) escapability, and pairwise distinguishability are all equivalent. 
\end{corollary}
\begin{proof}
  The proof follows from the discussion above. 
\end{proof}
\subsection{Necessity of Interval Optimality}

With interval optimality, the results above show that pairwise escapability (or pairwise distinguishability under stronger assumptions) can completely characterize adequate learning. Without interval optimality, however, escapability naturally fails with MLRP even when every pair of states is distinguishable. Combining this observation with \Cref{prop: pairwise} yields the following proposition.

\begin{proposition}\label{prop: non-interval}
  Suppose that $A$ is finite, each state has a unique optimal action, and the information structure is location-shift and satisfies MLRP. With nontrivial preferences, there is adequate learning if and only if interval optimality and pairwise distinguishability hold.
\end{proposition}

\begin{proof}
  It only remains to show that if interval optimality fails, learning is inadequate for some initial prior. The idea is simple: if interval optimality fails, there exist some action $a$ and states $\omega_1 < \omega_2 < \omega_3$ such that $\omega_1, \omega_3 \in \Omega_a$ but $\omega_2 \notin \Omega_a$. Since each state has a unique optimal action, we have $\omega_2 \in \cup_{a' \neq a} \Omega_{a',a}$. Since $A$ is finite, there exists $n$ such that $\{\omega_1, \omega_3\} \subset \Omega_a^n$. By MLRP, the middle state $\omega_2$ is not distinguishable from the set of states $\{\omega_1, \omega_3\}$; see \Cref{lem:middle_mlrp} for a proof.  By the subset property of distinguishability, $\cup_{a' \neq a} \Omega_{a',a}$ is not distinguishable from $\Omega_a^n$, which implies that there is herding for some initial prior by \Cref{thm1}. 
\end{proof}

\section{Discussion}\label{sec: conclusion}

When pairwise escapability holds but escapability fails, it remains open which full-support priors lead to learning and which lead to herding in general. One such case is the failure of interval optimality: \Cref{prop: non-interval} guarantees herding for some prior, but whether herding occurs for every full-support prior with location-shift information satisfying MLRP remains open. We discuss a simple Gaussian example in the hope that it is useful for further research.

\begin{figure}[htbp]
\centering
\begin{subfigure}[t]{0.48\textwidth}
\centering
\begin{tikzpicture}[x=4.5cm,y=4.5cm,font=\small,
  path arrow/.style={-{Stealth[length=2.2pt,width=1.7pt]},line width=0.3pt,
    shorten <=1pt,shorten >=1pt},
  belief label/.style={inner sep=1pt,font=\small}]
  \path[use as bounding box] (-0.15,-0.17) rectangle (1.17,1.02);
  \definecolor{learningred}{RGB}{178,42,45}
  \definecolor{herdingblue}{RGB}{0,92,166}
  \coordinate (S1) at (0,0);
  \coordinate (S2) at (1,0);
  \coordinate (S3) at (0.5,{sqrt(3)/2});
  \coordinate (B) at ($(S1)!0.5!(S2)$);
  \coordinate (T) at ($(S2)!0.5!(S3)$);
  \coordinate (M) at ($(S1)!0.5!(S3)$);
  \fill[hazyblue!20] (S1)--(S3)--(T)--(B)--cycle;
  \fill[hazyblue!60] (B)--(T)--(S2)--cycle;

  \tikzset{declare function={
    gaussianMuTwo(\t)=(-\t+sqrt(\t*\t-exp(1)*(\t*\t-\t)))/(exp(1)/2);
  }}
  \fill[pattern={Lines[angle=135,distance=4pt]},pattern color=red!60]
    (S1)--plot[variable=\t,domain=0:1,samples=180]
    ({gaussianMuTwo(\t)+0.5*\t},{sqrt(3)*\t/2})--cycle;
  \draw[thick,dashed,red] plot[variable=\t,domain=0:1,samples=180]
    ({gaussianMuTwo(\t)+0.5*\t},{sqrt(3)*\t/2});
  \draw[black!65,dash pattern=on 2pt off 2pt,line width=0.3pt] (M)--(S2);
  \coordinate (A0) at (0.450000000000,0.086602540378); 
  \coordinate (A1) at (0.334713936928,0.048620833758); 
  \coordinate (A2) at (0.269727684035,0.025725383229); 
  \coordinate (A3) at (0.224938057150,0.012804399704); 
  \coordinate (A4) at (0.170612486408,0.003436923311); 
  \coordinate (A5) at (0.115648082703,0.000366510660); 
  \coordinate (A6) at (0.081041861055,0.000028639830); 
  \coordinate (A7) at (0.041003690443,0.000000023205); 
  \draw[path arrow,herdingblue] (A0)--(A1);
  \draw[path arrow,herdingblue] (A1)--(A2);
  \draw[path arrow,herdingblue] (A2)--(A3);
  \draw[path arrow,herdingblue] (A3)--(A4);
  \draw[path arrow,herdingblue] (A4)--(A5);
  \draw[path arrow,herdingblue] (A5)--(A6);
  \draw[path arrow,herdingblue] (A6)--(A7);
  \draw[herdingblue,line width=0.3pt,densely dashed] (A7)--(S1);
  \foreach \i in {0,...,7} \fill[herdingblue] (A\i) circle[radius=1pt];
  \node[below right=2pt,belief label] at (A0) {$\mu_0$};
  \draw[dashed] (S1)--(S2)--(S3)--cycle;
  \draw[herdingblue,fill=white,line width=0.45pt] (S1) circle[radius=2pt];
  \node[below left=3pt] at (S1) {$1$};
  \node[below right=3pt] at (S2) {$2$};
  \node[above=4pt] at (S3) {$3$};
  \node[font=\small] at (0.4,0.43) {$a_1$};
  \node[font=\small] at (0.75,0.25) {$a_2$};
  \node[font=\footnotesize] at (0.29,0.29) {$C_2$};
\end{tikzpicture}
\caption{Repeated $a_1$ observations.}
\label{fig:counterexample_gaussian_a1}
\end{subfigure}
\begin{subfigure}[t]{0.48\textwidth}
\centering
\begin{tikzpicture}[x=4.5cm,y=4.5cm,font=\small,
  path arrow/.style={-{Stealth[length=2.2pt,width=1.7pt]},line width=0.3pt,
    shorten <=1pt,shorten >=1pt},
  belief label/.style={inner sep=1pt,font=\small}]
  \path[use as bounding box] (-0.15,-0.17) rectangle (1.17,1.02);
  \definecolor{learningred}{RGB}{178,42,45}
  \definecolor{herdingblue}{RGB}{0,92,166}
  \coordinate (S1) at (0,0);
  \coordinate (S2) at (1,0);
  \coordinate (S3) at (0.5,{sqrt(3)/2});
  \coordinate (B) at ($(S1)!0.5!(S2)$);
  \coordinate (T) at ($(S2)!0.5!(S3)$);
  \coordinate (M) at ($(S1)!0.5!(S3)$);
  \fill[hazyblue!20] (S1)--(S3)--(T)--(B)--cycle;
  \fill[hazyblue!60] (B)--(T)--(S2)--cycle;

  \tikzset{declare function={
    gaussianMuTwo(\t)=(-\t+sqrt(\t*\t-exp(1)*(\t*\t-\t)))/(exp(1)/2);
  }}
  \fill[pattern={Lines[angle=135,distance=4pt]},pattern color=red!60]
    (S1)--plot[variable=\t,domain=0:1,samples=180]
    ({gaussianMuTwo(\t)+0.5*\t},{sqrt(3)*\t/2})--cycle;
  \draw[thick,dashed,red] plot[variable=\t,domain=0:1,samples=180]
    ({gaussianMuTwo(\t)+0.5*\t},{sqrt(3)*\t/2});
  \draw[black!65,dash pattern=on 2pt off 2pt,line width=0.3pt] (M)--(S2);
  \coordinate (P0) at (0.450000000,0.086602540); 
  \coordinate (P1) at (0.650795257,0.152755780); 
  \coordinate (P2) at (0.712937458,0.146159945); 
  \coordinate (P3) at (0.749961300,0.134067262); 
  \coordinate (P4) at (0.775668434,0.123224280); 
  \coordinate (P5) at (0.794832098,0.114218317); 
  \coordinate (P6) at (0.538284897,0.241893910); 
  \coordinate (P7) at (0.504149593,0.253393591); 
  \draw[path arrow,learningred] (P0)--(P1);
  \draw[path arrow,learningred] (P1)--(P2);
  \draw[path arrow,learningred] (P2)--(P3);
  \draw[path arrow,learningred] (P3)--(P4);
  \draw[path arrow,learningred] (P4)--(P5);
  \draw[path arrow,herdingblue] (P5)--(P6);
  \draw[path arrow,herdingblue] (P6)--(P7);
  \foreach \i in {0,...,4} \fill[learningred] (P\i) circle[radius=1pt];
  \foreach \i in {6,7} \fill[herdingblue] (P\i) circle[radius=1pt];
  \fill[herdingblue] (P5) circle[radius=1pt];
  \draw[learningred,line width=0.45pt] (P5) circle[radius=2pt];
  \fill[herdingblue] (P7) circle[radius=2pt];
  \node[below right=2pt,belief label] at (P0) {$\mu_0$};
  \node[right=3pt,belief label] at (P5) {$\mu_5$};
  \node[above left=2pt,text=herdingblue,belief label] at (P7) {$\mu_7$};
  \draw[dashed] (S1)--(S2)--(S3)--cycle;
  \node[below left=3pt] at (S1) {$1$};
  \node[below right=3pt] at (S2) {$2$};
  \node[above=4pt] at (S3) {$3$};
    \node[font=\small] at (0.4,0.43) {$a_1$};
  \node[font=\small] at (0.75,0.25) {$a_2$};
  \node[font=\footnotesize] at (0.29,0.29) {$C_2$};
\end{tikzpicture}
\caption{The history $a_2^5a_1^2$.}
\label{fig:counterexample_gaussian_entry}
\end{subfigure}
\caption{Two belief paths from the same prior $\mu_0=(0.5,0.4,0.1)$ in the Gaussian investor example. Blue denotes $a_1$ observations and red denotes $a_2$ observations; hatching marks the cascade region $C_2$. In (a), beliefs converge to $\delta_1$ (open circle). In (b), beliefs enter $C_2$ at $\mu_7$ (filled dot) and remain there; \Cref{tab:gaussian_path} reports the beliefs along this path.}
\label{fig:counterexample_gaussian_path}
\label{fig:counterexample_gaussian_regions}
\end{figure}
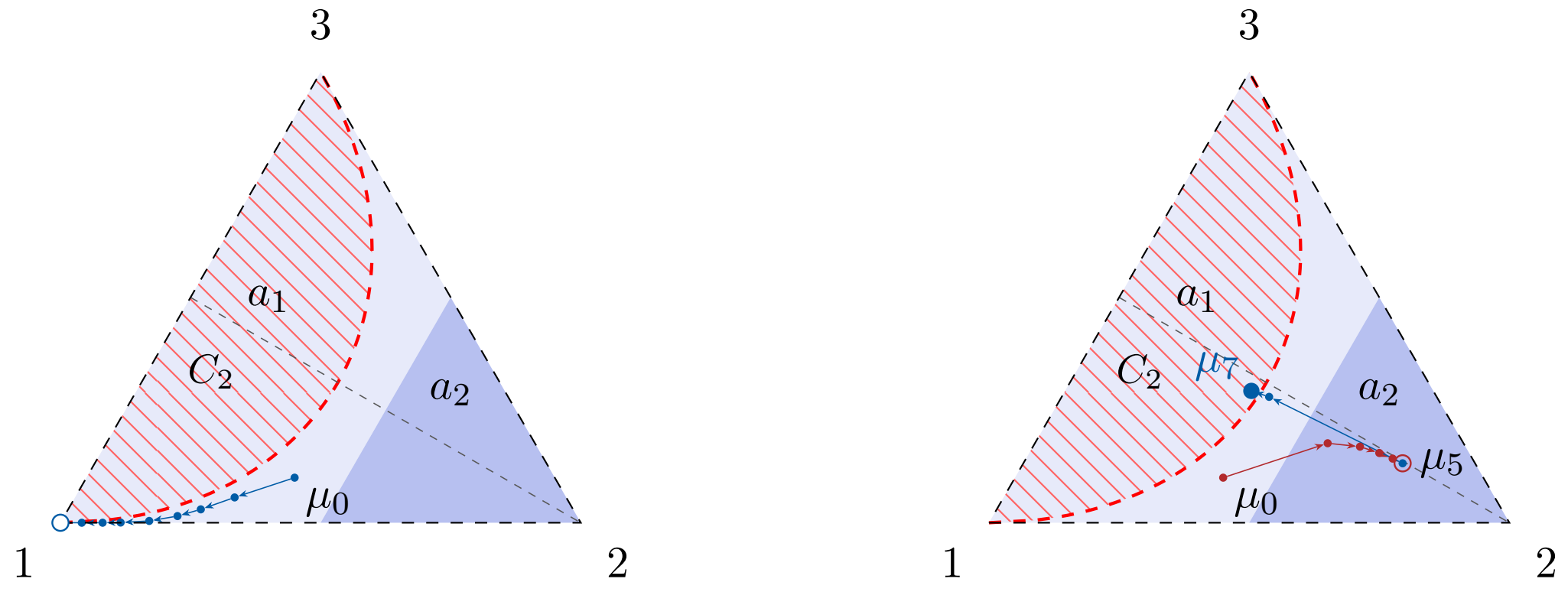

In the Gaussian investor example of \Cref{fig:scenario_2}, herding occurs for every full-support prior, through a mechanism different from that in \Cref{ex:perturbed} with $\varepsilon < 0$. Starting from $\mu_0 = (0.5, 0.4, 0.1)$, repeated $a_1$ observations drive the public belief to $\delta_1$ without entering the cascade region $C_2$ (\Cref{lem:gaussian_a1_learning}). However, a different history from the same prior can lead to herding. From every full-support prior outside $C_2$, a sufficiently long block of $a_2$ observations reduces public beliefs' asymmetry between states $1$ and $3$ enough for subsequent $a_1$ observations to move the public belief into $C_2$ with positive probability (\Cref{fig:counterexample_gaussian_path,prop:pairwise_no_learning}). 

The technical difficulty is that $a_2$ observations reduce asymmetry but also move beliefs farther from the cascade region, while subsequent $a_1$ observations can amplify the remaining asymmetry. Because the required lengths of both blocks depend on the initial prior, the proof must choose them jointly. The asymmetry that the $a_1$ block can tolerate shrinks as beliefs move away from the cascade region, so the proof must show that the $a_2$ block reduces asymmetry even faster. Establishing these rates requires delicate estimates of Gaussian tail probabilities (\Cref{SA:gaussian_a1,SA:gaussian_a2}).

These findings invite further study of how insights from binary-state models extend to multiple states. The interaction among information, preferences, and the initial prior may have implications for the comparison of information structures \citep{sato2025valueinformationsociallearning}, the efficiency of social learning \citep{rosenberg_efficiency_2019}, and the effects of misspecification \citep{frick_misinterpreting_2020,bohren_learning_2021}.

\appendix

\setcounter{lemma}{0}
\setcounter{definition}{0}
\setcounter{theorem}{0}
\setcounter{proposition}{0}

\renewcommand{\thedefinition}{\thesection.\arabic{definition}}
\renewcommand{\theHdefinition}{appendix.\thesection.\arabic{definition}}

\renewcommand{\thelemma}{\thesection.\arabic{lemma}}
\renewcommand{\theHlemma}{appendix.\thesection.\arabic{lemma}}

\renewcommand{\thetheorem}{\thesection.\arabic{theorem}}
\renewcommand{\theHtheorem}{appendix.\thesection.\arabic{theorem}}

\renewcommand{\theproposition}{\thesection.\arabic{proposition}}
\renewcommand{\theHproposition}{appendix.\thesection.\arabic{proposition}}

\section{General Model}\label{app:general_model}

We extend the model in \Cref{sec: model} to a Polish state space $(\Omega,d)$, with beliefs given by Borel probability measures on $\Omega$. The action and signal spaces remain standard Borel, and the information structure $F(\cdot\mid\omega)$ is a Markov kernel with mutually absolutely continuous conditional laws. We retain conditional independence of private signals given the state and complete observation of predecessors' actions. Payoffs satisfy $|u(a,\omega)|\leq\bar u$, the family $\{u(a,\cdot):a\in A\}$ is pointwise equicontinuous, and an optimal action exists at every belief. We use the notation of \Cref{subsec: public belief}, with $\mathcal F_\infty:=\sigma(\cup_n\mathcal F_n)$. The public beliefs $\mathbb{P}(\omega\in\cdot\mid\mathcal F_n)$ and private beliefs $\mathbb{P}(\omega\in\cdot\mid\mathcal F_n,s_n)$ are well defined as regular conditional probabilities since $\Omega$ is Polish \citep[Theorem 4.1.17]{durrett2019probability}. 

Following KLLR, write $u(\mu):=\sup_{a\in A}\int_\Omega u(a,\omega)\,\mu(\mathrm{d}\omega)$ for the optimal payoff at belief $\mu$. Let $u^*(\omega):=\sup_{a\in A}u(a,\omega)$, so $u^*(\mu)=\int_\Omega u^*(\omega)\,\mu(\mathrm{d}\omega)$. The payoff-continuity results in \citet[Lemma 5 and its proof]{kartik_beyond_2024} imply that $u(\mu)$ and $u^*(\mu)$ are weakly continuous (i.e., we consider the weak topology on the space of probability measures) and $u^*(\omega)$ is continuous. All three functions above are bounded. 

\section{Omitted Proofs}\label{app:proof}\label{appendix: proofs}\label{sec: appendix}

\subsection{Public Belief Process}

\begin{lemma}\label{lem:mtg_general}
For every prior and equilibrium, $\tilde\mu_n\Rightarrow\tilde\mu_\infty$ and $\tilde\mu_\infty\in\Delta^S$ almost surely. Moreover, $\lim_{n\to\infty}\mathbb E[u(a_n,\omega)]=\mathbb E[u(\tilde\mu_\infty)]$.
\end{lemma}

\begin{proof}
\emph{We first show weak convergence}. Since $\Omega$ is Polish, there is a countable family $\mathcal C\subset C_b(\Omega)$ that determines weak convergence \citep[Corollary 2.2.6]{bogachev2018weak}. For each $\chi\in\mathcal C$, L\'evy's upward theorem gives, almost surely, $\int_\Omega \chi(\omega)\,\tilde\mu_n(\mathrm{d}\omega) =\mathbb E[\chi(\omega)\mid\mathcal F_n] \to \mathbb E[\chi(\omega)\mid\mathcal F_\infty] =\int_\Omega \chi(\omega)\,\tilde\mu_\infty(\mathrm{d}\omega)$. Since $\mathcal C$ is countable, these convergences hold simultaneously on an event of probability one. Hence $\tilde\mu_n\Rightarrow\tilde\mu_\infty$ almost surely.

\emph{We prove the utility limit}. Note that the following holds for every equilibrium:
\[
\mathbb{E}[u(\tilde\mu_n)]\leq \mathbb{E}[u(a_n,\omega)] \leq \mathbb{E}[u(\tilde\mu_{n+1})].
\]
The first inequality holds because agent $n$ can ignore her private signal. For the second inequality, note that $a_n$ is $\mathcal F_{n+1}$-measurable, so $\mathbb E[u(a_n,\omega)\mid\mathcal F_{n+1}]\leq u(\tilde\mu_{n+1})$. The law of iterated expectations gives the second inequality. 

By bounded continuity of $u(\mu)$, we interchange the limit and expectation to obtain $\mathbb{E}[u(\tilde\mu_n)]\to \mathbb{E}[u(\tilde\mu_\infty)]$ and $\mathbb{E}[u(\tilde\mu_{n+1})]\to \mathbb{E}[u(\tilde\mu_\infty)]$. Thus, we have $\lim_{n\to\infty}\mathbb{E}[u(a_n,\omega)] = \mathbb{E}[u(\tilde\mu_\infty)]$.

\emph{It remains to prove stationarity}. Define $u^F(\mu):=\int_S u(\mu_s)\Pr_\mu(\mathrm{d}s)$, which is the expected private utility given public belief $\mu$. Suppose that at time $0$, a signal $\hat{s}$ is drawn according to the information structure. Consider an observer who at each time $n$ sees the public history and signal $\hat s$. Then there is an extended probability space $(X_\infty\times S,\mathcal X_\infty\otimes\mathcal S,\bar{\mathbb P})$, where $\bar{\mathbb P}(\mathrm{d}x, \mathrm{d}s) = \mathbb P(\mathrm{d}x)F(\mathrm{d}s\mid\omega(x))$. Write $\bar{\mathbb E}$ for expectation under $\bar{\mathbb P}$. At time $n$, the observer's information is given by $\mathcal F_n\vee\sigma(\hat s)$. Define the observer's belief $\tilde\nu_n:=\bar{\mathbb P}(\omega\in\cdot\mid\mathcal F_n\vee\sigma(\hat s))$ for $n\in\mathbb N\cup\{\infty\}$. Note that the realization of $\hat{s}$ does not influence the original game.

For each $n$, the distribution of $(\omega,h_n,\hat s)$ under $\bar{\mathbb P}$ is the same as that of $(\omega,h_n,s_n)$ under $\mathbb P$. Hence the observer, by choosing an optimal action, has the same ex ante expected utility as agent $n$. That is, $\mathbb E[u(a_n,\omega)] = \bar{\mathbb E}[u(\tilde\nu_n)]$. We claim that the following holds:
\[
\lim_{n\to\infty}\mathbb E[u(a_n,\omega)] = \lim_{n\to\infty}\bar{\mathbb E}[u(\tilde\nu_n)] = \bar{\mathbb E}[u(\tilde\nu_\infty)] = \bar{\mathbb E}[u^F(\tilde\mu_\infty)].
\]
The second equality follows from the same martingale and utility limit argument used for $\tilde{\mu}_n$. The last equality comes from $\bar{\mathbb E}[u(\tilde\nu_\infty)] = \bar{\mathbb E}[\bar{\mathbb E}[u(\tilde\nu_\infty) \mid \mathcal{F}_\infty] ] = \bar{\mathbb E}[u^F(\tilde\mu_\infty)]$.

Moreover, since $a_n$ is $\mathcal F_\infty$-measurable, $\bar{\mathbb E}[u(a_n,\omega)\mid\mathcal F_\infty] \leq u(\tilde\mu_\infty)$. Since $\bar{\mathbb E}[u(a_n,\omega)]=\mathbb E[u(a_n,\omega)]$, taking expectations and passing to the limit yields $\bar{\mathbb E}[u^F(\tilde\mu_\infty)]\leq \bar{\mathbb E}[u(\tilde\mu_\infty)]$. Since $u^F(\mu)\geq u(\mu)$ for every $\mu$, equality must hold. Hence $u^F(\tilde\mu_\infty)=u(\tilde\mu_\infty)$ $\bar{\mathbb P}$-almost surely.

Finally, for any belief $\mu$ and any $a\in c(\mu)$, Bayes plausibility gives
 \[
 u^F(\mu)-u(\mu) = \int_S 
\left[
    u(\mu_s)
    -\int_\Omega u(a,\omega)\,\mu_s(\mathrm{d}\omega)
\right]\mathrm{Pr}_\mu(\mathrm{d}s).
\] 
The integrand is nonnegative, so the difference vanishes if and only if $a\in c(\mu_s)$ for $\Pr_\mu$-almost every signal $s$. Thus $u^F(\mu)=u(\mu)$ if and only if $\mu\in\Delta^S$. Consequently, $\tilde\mu_\infty\in\Delta^S$ $\bar{\mathbb P}$-almost surely. Therefore, $\mathbb P(\tilde\mu_\infty\in\Delta^S) = \bar{\mathbb P}(\{\tilde\mu_\infty\in\Delta^S\} \times S) = 1$.
\end{proof}

\begin{proof}[Proof of \Cref{lem: martingale}]
The countable-state model satisfies the assumptions above under the discrete metric. Hence the result follows from \Cref{lem:mtg_general}.
\end{proof}

\begin{proof}[Proof of \Cref{lem:learning_limit}]
We prove this lemma in the general model. Fix an equilibrium. We have 
\[u^*(\mu_0)-\lim_{n\to\infty}\mathbb E[u(a_n,\omega)] = \mathbb E[u^*(\tilde\mu_\infty)]- \mathbb{E}[u(\tilde\mu_\infty)] \geq 0.\]

For the first equality, \Cref{lem:mtg_general} implies that  $\lim_{n\to\infty}\mathbb{E}[u(a_n,\omega)] = \mathbb E[u(\tilde\mu_\infty)]$. Since $\tilde\mu_\infty$ is the conditional distribution given $\mathcal F_\infty$, Bayes plausibility (or the law of iterated expectations) gives $\mathbb E[u^*(\tilde\mu_\infty)]=u^*(\mu_0)$. The inequality holds because for any belief $\mu$, $u^*(\mu)-u(\mu) \geq 0$, with equality if and only if $\mu\in\Delta^{AK}$.\footnote{Formally, $u^*(\mu) - u(\mu) = 0$ if and only if $\mu(\Omega_a^+) = 1$ for some $a$. Since $\Omega_a^+ = \{\omega: u(a,\omega) = u^*(\omega)\}$ is closed by the equicontinuity assumption, we have $\mu(\Omega_a^+) = 1$ if and only if $\mathrm{supp} \mu \subseteq \Omega_a^+$, i.e., $\mu \in \Delta^{AK}$.} Thus, there is adequate learning for $\mu_0$ if and only if $\tilde\mu_\infty\in\Delta^{AK}$ a.s.\ in every equilibrium.

For (ii), note that $u(\tilde\mu_\infty) \leq u^*(\tilde\mu_\infty)$. The former is strictly below the latter exactly when $\tilde\mu_\infty\notin\Delta^{AK}$. Since $\tilde\mu_\infty\in\Delta^S$ almost surely by \Cref{lem:mtg_general}, herding occurs precisely when $\tilde\mu_\infty\in\Delta^S\setminus\Delta^{AK}$ with positive probability in some equilibrium.
\end{proof}

\subsection{\texorpdfstring{\Cref{thm1}}{Theorem \ref*{thm1}}}

\begin{lemma}\label{lem: Delta^*_prob0}
Suppose that there exist some $a\in A$, $n\in\mathbb{N}$, a nonempty open set $V\subseteq\Omega_a^n$, a Borel set $W\subseteq\cup_{a'}\Omega_{a',a}$, and $C<\infty$ such that for every $\omega\in W$ and $\omega'\in V$, $f(s\mid\omega)\le C f(s\mid\omega')$ for a.e.\ $s$. Then, for any full-support initial prior $\mu_0$ satisfying $\mu_0(W)>0$ and every equilibrium $\sigma$, we have $\mathbb{P}(\tilde{\mu}_\infty\in\Delta^*)=0$, where $\Delta^*:= \{\mu\in\Delta^{\Omega_a^+}:\mu(V)>0\}.$
\end{lemma}

\begin{proof}
Fix $a\in A$, $n\in\mathbb{N}$, $V$, and $W$ satisfying the condition in the lemma, and take any full-support prior $\mu_0$ with $\mu_0(W)>0$ and any equilibrium $\sigma$. Since $V$ is nonempty and open, $\mu_0(V)>0$. Mutual absolute continuity of the signal distributions implies that $\tilde{\mu}_t(V)>0$ and $\tilde{\mu}_t(W)>0$ at every finite time $t$, almost surely.

Recall that for a random history $\tilde{h}_{t+1} = (a_1, \ldots, a_t)$, the public belief filtration is $\mathcal{F}_{t+1} = \sigma(\tilde{h}_{t+1})$. At the public belief $\tilde{\mu}_{t+1}(h_{t+1})$, $a_t$ is optimal almost surely, since agent $t+1$, who sees only the public history, knows less than agent $t$, who also sees her private signal, i.e., $\mathcal{F}_{t+1} \subseteq \mathcal{F}_t \vee \sigma(a_t, s_t)$.  This is the overturning principle \citep{smith_pathological_2000}. Conditional on the event $\{a_t \neq a\}$, almost surely, the expected utility difference between $a_t$ and $a$ at $\tilde{\mu}_{t+1}$ satisfies
\begin{align*}
    0 \leq \int_\Omega [u(a_t,\omega) - u(a,\omega)] \tilde{\mu}_{t+1}(\mathrm{d}\omega) \leq 2\bar u\tilde{\mu}_{t+1}(\cup_{a'}\Omega_{a',a}) -\frac{1}{n}\tilde{\mu}_{t+1}(V), 
\end{align*}
where the first inequality follows from the overturning principle. The second inequality follows from evaluating the integral over the three disjoint sets $\cup_{a'}\Omega_{a',a}$, $V$, and $\Omega_a^+\setminus V$. The first part is bounded above by $2\bar u\tilde{\mu}_{t+1}(\cup_{a'}\Omega_{a',a})$ because $ |u| \leq \bar u$; the second is bounded above by $-\tilde{\mu}_{t+1}(V)/n$ because $V\subseteq\Omega_a^n$; and the last is nonpositive because $a$ is weakly optimal on $\Omega_a^+$.

Almost surely on the event $\tilde{\mu}_\infty\in\Delta^*$, martingale convergence implies that $\tilde{\mu}_t(\cup_{a'}\Omega_{a',a})$ converges to $0$ and $\tilde{\mu}_t(V)$ converges to $\tilde{\mu}_\infty(V)>0$. That is, if $a_t \neq a$, for sufficiently large $t$, the above integral is negative, contradicting the overturning principle. Hence $\mathbb{P}(\{a_t\neq a \text{ i.o.}\} \cap \{\tilde{\mu}_\infty\in\Delta^*\}) = 0$. In other words, $a_t = a$ for all sufficiently large $t$ almost surely on $\{\tilde{\mu}_\infty\in\Delta^*\}$.

For $E\in\{V,W\}$, define $r_t^E := \mathbb{P}(a_t\neq a\mid\mathcal{F}_t,\omega\in E)$ as the probability that the agent at time $t$ does not choose $a$ conditional on the public history and the state being in $E$. Fix a public history $h_t$, and let $\mathbb{P}(a_t\neq a\mid h_t, \omega) =  \int_S \sigma_t(h_t,s)(A\setminus\{a\}) F(\mathrm{d}s\mid\omega)$ be the probability that the agent at time $t$ does not choose $a$ conditional on a realized public history $h_t$ and the state being $\omega$. Hence, 
\begin{align*}
    r_t^W(h_t) &= \frac{1}{\tilde{\mu}_t(h_t)(V)}\int_V r_t^W(h_t) \tilde{\mu}_{t}(h_t)(\mathrm{d}\omega') \\
    &= \frac{1}{\tilde{\mu}_t(h_t)(V)\tilde{\mu}_t(h_t)(W)}\int_V\int_W \mathbb{P}(a_t\neq a\mid h_t, \omega) \tilde{\mu}_{t}(h_t)(\mathrm{d}\omega) \tilde{\mu}_{t}(h_t)(\mathrm{d}\omega') \\
    &\leq \frac{C}{\tilde{\mu}_t(h_t)(V)\tilde{\mu}_t(h_t)(W)}\int_V \int_W \mathbb{P}(a_t\neq a\mid h_t, \omega') \tilde{\mu}_{t}(h_t)(\mathrm{d}\omega) \tilde{\mu}_{t}(h_t)(\mathrm{d}\omega') \\
    &= \frac{C}{\tilde{\mu}_t(h_t)(V)}\int_V \mathbb{P}(a_t\neq a\mid h_t, \omega') \tilde{\mu}_{t}(h_t)(\mathrm{d}\omega') = C r_t^V(h_t).
\end{align*}
The first equality holds since $r_t^W(h_t)$ is constant for $\omega' \in V$. The second equality comes from $r_t^E(h_t) = \frac{1}{\tilde{\mu}_t(h_t)(E)}\int_E \mathbb{P}(a_t\neq a\mid h_t, \omega) \tilde{\mu}_{t}(h_t)(\mathrm{d}\omega)$. For the first inequality, since $f(s \mid \omega) \leq C f(s \mid \omega')$ for any $\omega\in W$ and $\omega'\in V$, we have $\mathbb{P}(a_t\neq a\mid h_t, \omega) \leq C \mathbb{P}(a_t\neq a\mid h_t, \omega')$ for any $\omega\in W$ and $\omega'\in V$. Thus, $r_t^W \leq C r_t^V$ almost surely.  Moreover,
\begin{align*}
    \mathbb{P}(a_t\neq a\mid h_t) &= \int_\Omega \int_S \sigma_t(h_t,s)(A\setminus\{a\}) F(\mathrm{d}s\mid\omega) \tilde{\mu}_{t}(h_t)(\mathrm{d}\omega)\\
    & \geq \int_V \int_S \sigma_t(h_t,s)(A\setminus\{a\}) F(\mathrm{d}s\mid\omega') \tilde{\mu}_{t}(h_t)(\mathrm{d}\omega') = \tilde{\mu}_t(h_t)(V) r_t^V(h_t), 
\end{align*}
which implies $\tilde{\mu}_t(V)r_t^V \leq \mathbb{P}(a_t\neq a\mid\mathcal{F}_t)$ almost surely.

Since $\{a_t\neq a\}\in\mathcal{F}_{t+1}$, the conditional Borel--Cantelli lemma \citep[Theorem 4.3.4]{durrett2019probability} implies $\{a_t\neq a\text{ finitely often}\}= \left\{ \sum_{t=1}^{\infty} \mathbb{P}(a_t\neq a\mid\mathcal{F}_t)<\infty \right\}$ up to a probability-zero event. Since $a_t=a$ for all sufficiently large $t$ on $\{\tilde{\mu}_\infty\in\Delta^*\}$, we have $\sum_{t=1}^{\infty} \mathbb{P}(a_t\neq a\mid\mathcal{F}_t)<\infty$ almost surely on $\{\tilde{\mu}_\infty\in\Delta^*\}$. Since $\tilde{\mu}_t(V)$ is bounded away from zero eventually on $\{\tilde{\mu}_\infty\in\Delta^*\}$, it follows that, almost surely on this event, $\sum_{t=1}^{\infty}r_t^V<\infty$ and $\sum_{t=1}^{\infty}r_t^W<\infty$. 

Suppose, toward a contradiction, that $\mathbb{P}(\tilde{\mu}_\infty\in\Delta^*)>0$. Choose a history $h$ in this event and outside the probability-zero sets on which the preceding statements may fail, and write $\nu:=\tilde{\mu}_\infty(h)$. There exists some finite $N$ such that, for all $t\geq N$, $a_t(h)=a$ and $r_t^V(h),r_t^W(h)\leq 1/2$. Along this history, Bayes' rule gives 
$\frac{\tilde{\mu}_{t+1}(h)(W)} {\tilde{\mu}_{t+1}(h)(V)} = \frac{\tilde{\mu}_t(h)(W)} {\tilde{\mu}_t(h)(V)} \frac{1-r_t^W(h)}{1-r_t^V(h)}$. Notice that
\begin{align*}
    \sum_{t= N}^{\infty} \left|\log \frac{1 - r_t^W(h)}{1 - r_t^V(h)}\right| &\leq \sum_{t=N}^{\infty}
\bigl[|\log(1-r_t^W(h))|+|\log(1-r_t^V(h))|\bigr]\\
    &\leq 2\sum_{t=N}^{\infty}r_t^W(h)+2\sum_{t=N}^{\infty}r_t^V(h) < \infty.
\end{align*}
The second inequality is due to $|\log(1-x)|\leq 2|x|$ for $|x|\leq 1/2$. This implies that $0 < \prod_{t=N}^{\infty} \frac{1-r_t^W(h)}{1-r_t^V(h)} < \infty$. Hence,
\begin{align*}
    \lim_{T\to\infty} \frac{\tilde{\mu}_{T+1}(h)(W)} {\tilde{\mu}_{T+1}(h)(V)} &\;= \frac{\tilde{\mu}_N(h)(W)} {\tilde{\mu}_N(h)(V)} \prod_{t=N}^{\infty} \frac{1-r_t^W(h)}{1-r_t^V(h)} \in(0,\infty).
\end{align*}
Since $\tilde{\mu}_t(h)(V)\to\nu(V)>0$, this implies $\nu(W)>0$. However, $W\subseteq\cup_{a'}\Omega_{a',a}$ and $\nu\in\Delta^*$ imply $\nu(W)=0$, a contradiction. Therefore, $\mathbb{P}(\tilde{\mu}_\infty\in\Delta^*)=0$.
\end{proof}

\begin{theorem}\label{thmA1}
Suppose that there exist some $a\in A$, $n\in\mathbb{N}$, nonempty open sets $V\subseteq\Omega_a^n$ and $W\subseteq\cup_{a'}\Omega_{a',a}$, and $C<\infty$ such that for every $\omega\in W$ and $\omega'\in V$, $f(s\mid\omega)\le C f(s\mid\omega')$ for a.e.\ $s$. Then learning is inadequate in every equilibrium for every full-support prior.
\end{theorem}

\begin{proof}
Fix any full-support prior $\mu_0$ and any equilibrium $\sigma$. Since $V$ and $W$ are nonempty and open, $\mu_0(V)>0$ and $\mu_0(W)>0$. Hence \Cref{lem: Delta^*_prob0} applies.

We show that $\tilde{\mu}_\infty\in\Delta^S\setminus\Delta^{AK}$ with positive probability. Suppose not. Since $\tilde{\mu}_\infty\in\Delta^S$ almost surely, we would have $\tilde{\mu}_\infty\in\Delta^{AK}$ almost surely. By \Cref{lem: Delta^*_prob0}, we also have $\tilde{\mu}_\infty\notin\Delta^*$ $\mathbb{P}$-a.s. We claim that for any $\nu\in\Delta^{AK}\setminus\Delta^*$, we have $\nu(V)=0$. Suppose instead that $\nu(V)>0$. Since $\nu$ has adequate knowledge, some action is optimal at every state in its support. Because $V\subseteq\Omega_a^n$ and $\nu(V)>0$, this action must be $a$. Thus $\nu\in\Delta^{\Omega_a^+}$, and hence $\nu\in\Delta^*$, a contradiction. Therefore, $\nu(V)=0$. It follows that $\tilde{\mu}_\infty(V)=0$ almost surely, and hence $\mathbb{E}[\tilde{\mu}_\infty(V)]=0$. However, $\tilde{\mu}_t(V)$ is a bounded martingale converging to $\tilde{\mu}_\infty(V)$, so $\mathbb{E}[\tilde{\mu}_\infty(V)] =\mu_0(V)>0$, a contradiction.

Therefore, $\tilde{\mu}_\infty\in\Delta^S\setminus\Delta^{AK}$ with positive probability. By \Cref{lem:learning_limit}, learning is inadequate in the fixed equilibrium. Since $\mu_0$ and $\sigma$ were arbitrary, the conclusion holds in every equilibrium for every full-support prior.
\end{proof}

\begin{proof}[Proof of \Cref{lem: ball around AK}]
  When $\Omega$ is countable with the discrete metric, every singleton is open. The result follows from \Cref{lem: Delta^*_prob0} by taking $V = \{\omega'\}$ and $W = \{\omega\}$.
\end{proof}

\begin{proof}[Proof of \Cref{thm1}]
  For the first statement, failure of escapability yields an action $a$, $n\in\mathbb{N}$, $\varepsilon\in(0,1)$, and a belief $\mu \in \Delta(\cup_{a'} \Omega_{a',a} \cup \Omega_a^n)$ with $\mu(\cup_{a'} \Omega_{a',a})>0$ such that $\mu_s(\cup_{a'} \Omega_{a',a})\leq1-\varepsilon$ for $\mu$-a.e.\ signal $s$. In particular, $\mu(\Omega_a^n)\geq\varepsilon>0$.

  Take $\mu' \in \Delta(\Omega)$ assigning positive probability to both sets, with $\mu'(\cdot \mid \cup_{a'}\Omega_{a',a}) = \mu(\cdot\mid \cup_{a'}\Omega_{a',a})$ and $\mu'(\cdot\mid \Omega_a^n) = \mu(\cdot \mid \Omega_a^n)$. Bayes' rule gives, for $\mu'$-a.e.\ signal $s$,
  \begin{align*}
    \frac{\mu'_s(\cup_{a'} \Omega_{a',a})}{\mu'_s(\Omega_a^n)} = \frac{\mu_s(\cup_{a'} \Omega_{a',a}) / \mu(\cup_{a'} \Omega_{a',a})}{\mu_s(\Omega_a^n) / \mu(\Omega_a^n)} \frac{\mu'(\cup_{a'} \Omega_{a',a})}{\mu'(\Omega_a^n)} \leq \frac{1 - \varepsilon}{\varepsilon} \frac{\mu(\Omega_a^n)}{\mu(\cup_{a'} \Omega_{a',a})} \frac{\mu'(\cup_{a'} \Omega_{a',a})}{\mu'(\Omega_a^n)}.
  \end{align*}
  As shown in the proof of \Cref{lem: Delta^*_prob0}, the expected gain from switching from $a$ to any $a'\neq a$ under $\mu'_s$ is at most $2\bar{u} \mu'_s(\cup_{a'} \Omega_{a',a}) - \frac{1}{n} \mu'_s(\Omega_a^n)$. Hence $a$ is uniquely optimal whenever $\mu'_s(\cup_{a'} \Omega_{a',a}) / \mu'_s(\Omega_a^n) < 1/(2n\bar{u})$. This holds if $\mu'(\cup_{a'}\Omega_{a',a})/\mu'(\Omega_a^n)
< [\varepsilon/(2n\bar{u}(1-\varepsilon))]
[\mu(\cup_{a'}\Omega_{a',a})/\mu(\Omega_a^n)]$.

  With $\mu$ and $\varepsilon$ fixed, the RHS is positive and independent of $\mu'$. Choose $\mu'$ satisfying this inequality and $\mu'(\cup_{a'} \Omega_{a',a})>0$. Then $a$ is uniquely optimal after almost every signal and, by averaging, under $\mu'$ itself. Thus $\mu'\in\Delta^S\setminus\Delta^{AK}$. Starting from $\mu_0=\mu'$, every agent chooses $a$ almost surely, so $\tilde\mu_n=\mu'$ for all $n$. Herding follows from \Cref{lem:learning_limit}.

  For the second statement, failure of pairwise escapability yields $a\in A$, $n\in\mathbb N$, $\omega'\in\Omega_a^n$, and $\omega\in\cup_{a'}\Omega_{a',a}$ such that $\{\omega\}$ is not distinguishable from $\{\omega'\}$. Thus there exist a belief $\mu\in\Delta(\{\omega,\omega'\})$ assigning positive probability to both states and $\varepsilon\in(0,1)$ such that $\mu_s(\omega)\leq1-\varepsilon$ for $\mu$-a.e.\ signal $s$. Equivalently, $f(s\mid \omega)/f(s\mid \omega') \leq [(1-\varepsilon)/\varepsilon]\,[\mu(\omega')/\mu(\omega)] < \infty$ for $\mu$-a.e.\ signal $s$ and hence a.e.\ $s$. In the countable discrete state space of \Cref{thm1}, the singletons $V=\{\omega'\}$ and $W=\{\omega\}$ are open. Hence \Cref{thmA1} and \Cref{lem:learning_limit} imply herding for every full-support prior.
\end{proof}

For the following proofs, we consider the same environment as in the main text: the state space $\Omega$ is a countable set with the discrete metric. 

\subsection{\texorpdfstring{\Cref{thm: escapability_necc_suff}}{Theorem \ref*{thm: escapability_necc_suff}}}

\begin{lemma}\label{lem:escapability_equiv_pairwise_escapability}
   Suppose that $A$ is finite, preferences satisfy interval optimality, each state has a unique optimal action, and the information structure satisfies MLRP. Then escapability is equivalent to pairwise escapability.
\end{lemma}

\begin{proof}
  By the subset property of distinguishability, escapability implies pairwise escapability regardless of the assumptions. We prove the converse.

  Fix any $a\in A$ and $n\in\mathbb{N}$. Take any belief $\mu\in\Delta({(\cup_{a'}\Omega_{a',a})\cup\Omega_a^n})$ with $\mu(\cup_{a'}\Omega_{a',a})>0$. If $\mu(\Omega_a^n)=0$, the claim is immediate. Hence suppose that $\mu(\Omega_a^n)>0$.

  Since $\Omega_a$ is an interval, each state has a unique optimal action, and $\Omega_a^n\subseteq\Omega_a$, every state in $\cup_{a'}\Omega_{a',a}$ lies either strictly below every state in $\Omega_a^n$ or strictly above every state in $\Omega_a^n$. At least one of these two sets has positive probability under $\mu$. Suppose that $L:= \left\{\omega\in\cup_{a'}\Omega_{a',a}: \omega<\omega', \forall \omega'\in\Omega_a^n\right\}$ satisfies $\mu(L)>0$. The other case is symmetric.
  
  For any $\omega\in L$ and $\omega'\in\Omega_a^n$, pairwise escapability implies that $\{\omega\}$ is distinguishable from $\{\omega'\}$. That is, for any $\nu \in \Delta(\{\omega, \omega'\})$ with $\nu(\omega)>0$, for any $\varepsilon>0$, $\mathrm{Pr}_\nu(s: \nu_s(\omega) > 1-\varepsilon) > 0$. Equivalently, for any $\varepsilon \in (0,1)$ and any of these signals $s$, we have $\nu_s(\omega')/\nu_s(\omega) = [f(s\mid\omega')\nu(\omega')]/[f(s\mid\omega)\nu(\omega)] < \varepsilon/(1-\varepsilon)$. Since $\omega<\omega'$, MLRP implies that $f(s\mid\omega')/f(s\mid\omega)$ is nondecreasing in $s$. Therefore, distinguishability implies that $f(s\mid\omega')/f(s\mid\omega)$ converges to $0$ as $s \downarrow \inf S$. Choose $\omega \in L$ with $\mu(\omega) > 0$. Fix a signal $s_0$ where $0 < f(s_0 \mid \omega) < \infty$ and $\sum_{\omega'\in\Omega_a^n} \mu(\omega') f(s_0 \mid \omega') < \infty$. For $s\leq s_0$, MLRP gives
  \[
    0\leq \frac{f(s\mid\omega')}{f(s\mid\omega)} \leq \frac{f(s_0\mid\omega')}{f(s_0\mid\omega)}, \text{ and}\quad      \sum_{\omega'\in\Omega_a^n} \mu(\omega') \frac{f(s_0\mid\omega')}{f(s_0\mid\omega)} < \infty.
  \]
 By the dominated convergence theorem,
  \[
    \frac{\mu_s(\Omega_a^n)}{\mu_s(\omega)} = \frac{1}{\mu(\omega)} \sum_{\omega'\in\Omega_a^n} \mu(\omega') \frac{f(s\mid\omega')}{f(s\mid\omega)} \to 0 \qquad\text{as }s\downarrow\inf S.
  \]
  Since $\mu_s(\omega)\leq 1$, it follows that $\mu_s(\Omega_a^n)\to 0$, and hence $\mu_s(\cup_{a'}\Omega_{a',a})\to 1$. Therefore, $\cup_{a'}\Omega_{a',a}$ is distinguishable from $\Omega_a^n$. When $\omega$ lies above every state in $\Omega_a^n$, the same argument applies as $s$ approaches the upper endpoint of the signal support. Since $\mu$ is arbitrary, $\cup_{a'}\Omega_{a',a}$ is distinguishable from $\Omega_a^n$. Since $a$ and $n$ are arbitrary, escapability holds.
\end{proof}

\subsection{\texorpdfstring{\Cref{prop: non-interval}}{Proposition \ref*{prop: non-interval}}}

\begin{lemma}\label{lem:middle_mlrp}
Suppose that the information structure satisfies MLRP. For any
$\omega_L,\omega,\omega_H\in\Omega$ with $\omega_L<\omega<\omega_H$, $\{\omega\}$ is not distinguishable from $\{\omega_L,\omega_H\}$.
\end{lemma}

\begin{proof}
Fix any full-support belief
$\mu\in\Delta(\{\omega_L,\omega,\omega_H\})$. Fix a signal $s$. The posterior ratio is
\[
\frac{\mu_s(\omega_L)+\mu_s(\omega_H)}{\mu_s(\omega)}
=
\frac{\mu(\omega_L)}{\mu(\omega)}
\frac{f(s\mid\omega_L)}{f(s\mid\omega)}
+
\frac{\mu(\omega_H)}{\mu(\omega)}
\frac{f(s\mid\omega_H)}{f(s\mid\omega)}.
\]
By MLRP, the first term on the right-hand side is nonincreasing in $s$, while the second is nondecreasing in $s$.

Choose $s^*$ at which all three densities are positive and finite. For $s\leq s^*$,
\[
\frac{\mu_s(\omega_L)+\mu_s(\omega_H)}{\mu_s(\omega)}
\geq
\frac{\mu(\omega_L)}{\mu(\omega)}
\frac{f(s^*\mid\omega_L)}{f(s^*\mid\omega)},
\]
whereas for $s\geq s^*$,
\[
\frac{\mu_s(\omega_L)+\mu_s(\omega_H)}{\mu_s(\omega)}
\geq
\frac{\mu(\omega_H)}{\mu(\omega)}
\frac{f(s^*\mid\omega_H)}{f(s^*\mid\omega)}.
\]
Hence, there exists $k > 0$ such that for $\mu$-a.e.\ signal $s$,
\[
\frac{\mu_s(\omega_L)+\mu_s(\omega_H)}{\mu_s(\omega)}
\geq \min\left\{
\frac{\mu(\omega_L)}{\mu(\omega)}
\frac{f(s^*\mid\omega_L)}{f(s^*\mid\omega)},
\frac{\mu(\omega_H)}{\mu(\omega)}
\frac{f(s^*\mid\omega_H)}{f(s^*\mid\omega)}
\right\}
> k >0.
\]
Therefore,
\[
\mu_s(\omega)
=
\frac{1}{
1+\dfrac{\mu_s(\omega_L)+\mu_s(\omega_H)}{\mu_s(\omega)}
}
\leq
\frac{1}{1+k}
<1
\]
for $\mu$-a.e.\ $s$. Thus $\{\omega\}$ is not distinguishable from
$\{\omega_L,\omega_H\}$.
\end{proof}

\subsection{Details for \texorpdfstring{\Cref{sec:prior_dependence}}{Section \ref*{sec:prior_dependence}}}\label{SA: ex_perturbed}

Throughout this subsection, fix $\varepsilon\in(-1/4,1/4)$ and consider the information structure $F_1^\varepsilon$ of \Cref{ex:perturbed}. \Cref{ex:counter} is the case $\varepsilon=0$.

\subsubsection{Preliminaries}\label{app:counter_prelim}

Represent a full-support belief by $(\lambda_1,\lambda_2)=(\mu(1)/\mu(2),\mu(3)/\mu(2))$, and let $b_\varepsilon:=1/(1+\varepsilon)$. In these coordinates, the cascade region is $C_1^\varepsilon = [b_\varepsilon,\infty)^2$. Its complement is $\mathcal{W} = (0,\infty)^2\setminus C_1^\varepsilon$, and the symmetric non-cascade beliefs are $\mathcal{D} = \{(r,r):0<r<b_\varepsilon\}$. Define $R_0=(0,b_\varepsilon)^2$, $R_1=(b_\varepsilon,\infty)\times(0,b_\varepsilon)$, and $R_2=(0,b_\varepsilon)\times(b_\varepsilon,\infty)$. They cover $\mathcal{W}$ except for the boundaries $\lambda_1=b_\varepsilon$ and $\lambda_2=b_\varepsilon$, and they are the regions of \Cref{sec:prior_dependence} when $\varepsilon=0$.

Let $T_a$ be the public-belief update after action $a$, with coordinates $T_{a,1}$ and $T_{a,2}$, and let $\psi_\omega^a$ be the probability of that action in state $\omega$. Write
\[
\alpha=\left(\frac{1-(1+\varepsilon)\lambda_2}{(1-\varepsilon)\lambda_1}\right)_+,\quad
\beta=\left(\frac{1-(1+\varepsilon)\lambda_1}{(1-\varepsilon)\lambda_2}\right)_+.
\]
The calculations in \Cref{SA:counter_updates} give
\begin{equation}\label{eq:counter_Ta1}
\begin{aligned}
T_{a_1,1}(\lambda_1,\lambda_2)&=\lambda_1+\varepsilon\lambda_1\tanh\frac{\alpha-\beta}{2}+\frac{(1-(1+\varepsilon)\lambda_2)_+e^{-\alpha}}{e^{-\alpha}+e^{-\beta}},\\
T_{a_1,2}(\lambda_1,\lambda_2)&=\lambda_2-\varepsilon\lambda_2\tanh\frac{\alpha-\beta}{2}+\frac{(1-(1+\varepsilon)\lambda_1)_+e^{-\beta}}{e^{-\alpha}+e^{-\beta}},
\end{aligned}
\end{equation}
and $\psi_2^{a_1}(\lambda_1,\lambda_2)=(e^{-\alpha}+e^{-\beta})/2$. Ties have probability zero in every state, so these quantities do not depend on tie-breaking. Exchanging states $1$ and $3$ while replacing $s$ with $-s$ leaves the densities unchanged. Hence, with $P(\lambda_1,\lambda_2):=(\lambda_2,\lambda_1)$, we have $T_a\circ P=P\circ T_a$ for each action $a$.

For a finite history $h=(a^1,\ldots,a^n)$, write $T_h=T_{a^n}\circ\cdots\circ T_{a^1}$. Define the set of priors from which $\mathcal{D}$ is reachable by $\mathcal{N}=\bigcup_{n\geq0}\ \bigcup_{h\in\{a_1,a_2\}^n}T_h^{-1}(\mathcal{D})$. Here $T_{\varnothing}$ is the identity, so $\mathcal{D}\subseteq \mathcal{N}$, and all preimages are taken within $\mathcal{W}$.

  \subsubsection{Characterization}

  \begin{lemma}\label{lem:counter_W_invariant}
    For any $(\lambda_1,\lambda_2) \in \mathcal{W}$, we have $T_{a_1}(\lambda_1,\lambda_2) \in \mathcal{W}$ and $T_{a_2}(\lambda_1,\lambda_2) \in \mathcal{W}$. That is, the set of cascade beliefs cannot be reached from any non-cascade belief in finite time.
  \end{lemma}

  \begin{proof}
    We first show that $T_{a_2}(\lambda_1,\lambda_2) \in \mathcal{W}$ for any $(\lambda_1,\lambda_2) \in \mathcal{W}$. This is an immediate consequence of the overturning principle: after observing the previous agent choose $a_2$, the current agent finds $a_2$ optimal under the updated public belief. Since $\lambda_1+\lambda_2\geq2b_\varepsilon>1$ on $C_1^\varepsilon$, action $a_1$ is strictly optimal under every public belief in $C_1^\varepsilon$. Thus, $T_{a_2}(\lambda_1,\lambda_2) \notin C_1^\varepsilon$ for any $(\lambda_1,\lambda_2) \in \mathcal{W}$.

    We then consider $T_{a_1}(\lambda_1,\lambda_2)$. On $R_0$, both positive parts in \eqref{eq:counter_Ta1} are positive. Thus
    \begin{align*}
      T_{a_1,1}(\lambda_1,\lambda_2) + T_{a_1,2}(\lambda_1,\lambda_2) &= \lambda_1+\lambda_2 +\varepsilon(\lambda_1-\lambda_2)\tanh\frac{\alpha-\beta}{2}\\
      &\quad+ \frac{(1-(1+\varepsilon)\lambda_2)e^{-\alpha} + (1-(1+\varepsilon)\lambda_1)e^{-\beta}}{e^{-\alpha}+e^{-\beta}} \\
      &= 1 + (1-\varepsilon)\frac{\lambda_1 e^{-\alpha} + \lambda_2 e^{-\beta}}{e^{-\alpha}+e^{-\beta}} < 1+(1-\varepsilon)b_\varepsilon=2b_\varepsilon.
    \end{align*}
    This implies that $T_{a_1}(\lambda_1,\lambda_2) \in \mathcal{W}$.

    On $R_1$, and also on the boundary $\lambda_1=b_\varepsilon$ within $\mathcal{W}$, we have $\beta = 0$ and $\alpha = (1-(1+\varepsilon)\lambda_2)/((1-\varepsilon)\lambda_1)>0$. Since $1-(1+\varepsilon)\lambda_2=(1-\varepsilon)\lambda_1\alpha$, \eqref{eq:counter_Ta1} becomes
    \begin{equation}\label{eq:counter_R1}
    T_{a_1}(\lambda_1,\lambda_2) = \left(\lambda_1 + \lambda_1\frac{(1-\varepsilon)\alpha+\varepsilon(e^\alpha-1)}{1+e^{\alpha}},\; \lambda_2\Bigl(1-\varepsilon\tanh\frac\alpha2\Bigr) \right).
    \end{equation}
    If $\varepsilon\geq0$, then $T_{a_1,1}(\lambda_1,\lambda_2) > \lambda_1 \geq b_\varepsilon$ and $T_{a_1,2}(\lambda_1,\lambda_2) \leq \lambda_2 < b_\varepsilon$. If $\varepsilon<0$, then $\alpha<1/((1-\varepsilon)b_\varepsilon)=(1+\varepsilon)/(1-\varepsilon)<1$, so $e^\alpha-1\leq\alpha e^\alpha<e\alpha$ and $|\varepsilon|(e^\alpha-1)<e\alpha/4<(1-\varepsilon)\alpha$. Hence again $T_{a_1,1}(\lambda_1,\lambda_2)>\lambda_1\geq b_\varepsilon$. Moreover, $\tanh(\alpha/2)<\alpha/2$ and $(1+\varepsilon)\lambda_2<1<(1-\varepsilon)\lambda_1$ give
    \begin{align*}
    1-(1+\varepsilon)T_{a_1,2}(\lambda_1,\lambda_2)&=\bigl(1-(1+\varepsilon)\lambda_2\bigr)-|\varepsilon|(1+\varepsilon)\lambda_2\tanh\frac\alpha2\\
    &>\bigl(1-(1+\varepsilon)\lambda_2\bigr)\left(1-\frac{|\varepsilon|(1+\varepsilon)\lambda_2}{2(1-\varepsilon)\lambda_1}\right)>0,
    \end{align*}
    so $T_{a_1,2}(\lambda_1,\lambda_2)<b_\varepsilon$. In either case, the updated belief lies in $R_1\subset \mathcal{W}$. The argument for $R_2$ and the boundary $\lambda_2=b_\varepsilon$ within $\mathcal{W}$ is symmetric.

    Hence we have $T_h(\lambda_1,\lambda_2)\in \mathcal{W}$ for every $(\lambda_1,\lambda_2)\in \mathcal{W}$ and every finite action history $h$. Thus, no finite history can take a full-support non-cascade belief into the cascade region.
  \end{proof}

  \begin{lemma}\label{lem:counter_gap_expansion}
    If $(\lambda_1,\lambda_2)\in R_0\setminus \mathcal{D}$, then $\mathrm{sgn}(T_{a_1,1}(\lambda_1,\lambda_2) - T_{a_1,2}(\lambda_1,\lambda_2)) = \mathrm{sgn}(\lambda_1-\lambda_2)$ and the following uniform bound holds: $|T_{a_1,1}(\lambda_1,\lambda_2) - T_{a_1,2}(\lambda_1,\lambda_2)| > \frac{5+\varepsilon}{4}|\lambda_1-\lambda_2|$.
  \end{lemma}

\begin{proof}
On $R_0$, both positive parts in \eqref{eq:counter_Ta1} are positive, so
\begin{align*}
    T_{a_1,1}(\lambda_1,\lambda_2)-T_{a_1,2}(\lambda_1,\lambda_2)
    &=\lambda_1-\lambda_2+\varepsilon(\lambda_1+\lambda_2)\tanh\frac{\alpha-\beta}{2}\\
    &\quad+\frac{(1-(1+\varepsilon)\lambda_2)e^{-\alpha}-(1-(1+\varepsilon)\lambda_1)e^{-\beta}}{e^{-\alpha}+e^{-\beta}}.
\end{align*}
Substituting $1-(1+\varepsilon)\lambda_2=1-(1+\varepsilon)\frac{\lambda_1+\lambda_2}{2}+(1+\varepsilon)\frac{\lambda_1-\lambda_2}{2}$, $1-(1+\varepsilon)\lambda_1=1-(1+\varepsilon)\frac{\lambda_1+\lambda_2}{2}-(1+\varepsilon)\frac{\lambda_1-\lambda_2}{2}$, and
\[
    \tanh\left(\frac{\alpha-\beta}{2}\right)
    =\frac{e^{\frac{\alpha-\beta}{2}}-e^{-\frac{\alpha-\beta}{2}}}{e^{\frac{\alpha-\beta}{2}}+e^{-\frac{\alpha-\beta}{2}}}
    =\frac{e^{-\beta}-e^{-\alpha}}{e^{-\alpha}+e^{-\beta}}
\]
gives
\begin{equation}\label{eq:difference-update}
    T_{a_1,1}(\lambda_1,\lambda_2)-T_{a_1,2}(\lambda_1,\lambda_2)
    =\frac{3+\varepsilon}2(\lambda_1-\lambda_2)+\left(1-(1+3\varepsilon)\frac{\lambda_1+\lambda_2}{2}\right)\frac{e^{-\alpha}-e^{-\beta}}{e^{-\alpha}+e^{-\beta}}.
\end{equation}
Also,
\begin{equation}\label{eq:alpha-beta}
    \alpha-\beta=\frac{1-(1+\varepsilon)\lambda_2}{(1-\varepsilon)\lambda_1}-\frac{1-(1+\varepsilon)\lambda_1}{(1-\varepsilon)\lambda_2}
    =\frac{(\lambda_1-\lambda_2)\bigl((1+\varepsilon)(\lambda_1+\lambda_2)-1\bigr)}{(1-\varepsilon)\lambda_1\lambda_2}.
\end{equation}
By symmetry, assume $\lambda_1>\lambda_2$ and consider two cases.

\noindent\textbf{Case 1: $(1+\varepsilon)(\lambda_1+\lambda_2)\leq1$.}
By \eqref{eq:alpha-beta}, $\alpha\leq\beta$, so $e^{-\alpha}\geq e^{-\beta}$. Since $1+3\varepsilon>0$, the coefficient of the last fraction in \eqref{eq:difference-update} is at least $1-(1+3\varepsilon)/(2(1+\varepsilon))=(1-\varepsilon)/(2(1+\varepsilon))>0$. Hence
\[
    T_{a_1,1}(\lambda_1,\lambda_2)-T_{a_1,2}(\lambda_1,\lambda_2)
    \geq\frac{3+\varepsilon}2(\lambda_1-\lambda_2)>\frac{5+\varepsilon}4(\lambda_1-\lambda_2).
\]

\noindent\textbf{Case 2: $(1+\varepsilon)(\lambda_1+\lambda_2)>1$.}
Now \eqref{eq:alpha-beta} gives $\alpha>\beta$ and $e^{-\beta}>e^{-\alpha}$. If $1-(1+3\varepsilon)(\lambda_1+\lambda_2)/2\leq0$, the second term in \eqref{eq:difference-update} is nonnegative, and the bound of Case 1 applies. Otherwise,
\begin{equation}\label{eq:difference-negative}
    0<1-(1+3\varepsilon)\frac{\lambda_1+\lambda_2}{2}<1-\frac{1+3\varepsilon}{2(1+\varepsilon)}=\frac{1-\varepsilon}{2(1+\varepsilon)}.
\end{equation}
Since $\tanh(x)<x$ for $x>0$, we obtain
\begin{equation}\label{eq:ratio-bound}
    \frac{e^{-\beta}-e^{-\alpha}}{e^{-\alpha}+e^{-\beta}}
    <\frac{\alpha-\beta}{2}
    =\frac{(\lambda_1-\lambda_2)\bigl((1+\varepsilon)(\lambda_1+\lambda_2)-1\bigr)}{2(1-\varepsilon)\lambda_1\lambda_2}.
\end{equation}
Since $0<(1+\varepsilon)\lambda_1,(1+\varepsilon)\lambda_2<1$, expanding $(1-(1+\varepsilon)\lambda_1)(1-(1+\varepsilon)\lambda_2)>0$ yields $(1+\varepsilon)^2\lambda_1\lambda_2>(1+\varepsilon)(\lambda_1+\lambda_2)-1$, so
\begin{equation}\label{eq:ratio-simple}
    \frac{e^{-\beta}-e^{-\alpha}}{e^{-\alpha}+e^{-\beta}}<\frac{(1+\varepsilon)^2(\lambda_1-\lambda_2)}{2(1-\varepsilon)}.
\end{equation}
Combining \eqref{eq:difference-update}, \eqref{eq:difference-negative}, and \eqref{eq:ratio-simple} yields
\[
    T_{a_1,1}(\lambda_1,\lambda_2)-T_{a_1,2}(\lambda_1,\lambda_2)
    >\frac{3+\varepsilon}2(\lambda_1-\lambda_2)-\frac{1-\varepsilon}{2(1+\varepsilon)}\cdot\frac{(1+\varepsilon)^2(\lambda_1-\lambda_2)}{2(1-\varepsilon)}
    =\frac{5+\varepsilon}4(\lambda_1-\lambda_2).
\]
Both cases give the claimed sign and bound; symmetry covers $\lambda_1<\lambda_2$.
\end{proof}

\begin{lemma}\label{lem:counter_symmetric_herding}
  For any initial prior in $\mathcal{D}$, herding occurs.
\end{lemma}

\begin{proof}
  We show that starting from any initial prior with $\lambda_1 = \lambda_2 = r \in (0,b_\varepsilon)$, the history of actions $(a_1, a_1, \ldots)$ occurs with positive probability and the public belief converges to $\bar\mu_\varepsilon:=(1,1+\varepsilon,1)/(3+\varepsilon)$ along this history. Note that $\bar\mu_0=\bar\mu=(1/3,1/3,1/3)$.

  When $\lambda_1 = \lambda_2 = r \in (0,b_\varepsilon)$, we have $\alpha = \beta = (1-(1+\varepsilon)r)/((1-\varepsilon)r)$, so the $\tanh$ terms in \eqref{eq:counter_Ta1} vanish. By \eqref{eq:counter_Ta1}, $T_{a_1}(r,r) = (r',r')$ with $r'=(1+(1-\varepsilon)r)/2$. Since $b_\varepsilon=(1+(1-\varepsilon)b_\varepsilon)/2$, we have $b_\varepsilon-r'=(1-\varepsilon)(b_\varepsilon-r)/2$. Thus, along the history $(a_1, \ldots, a_1)$ of length $n$, the first belief-odds coordinate satisfies $r_n:= b_\varepsilon-(b_\varepsilon-r)((1-\varepsilon)/2)^n$, which converges to $b_\varepsilon$ as $n \to \infty$. Thus, $T_{a_1}^n(r,r) \to (b_\varepsilon,b_\varepsilon)$, which represents $\bar\mu_\varepsilon$.

  It remains to show that the infinite history $(a_1, a_1, \ldots)$ occurs with positive probability. Recall that conditional on state $\omega = 2$, $\psi_2^{a_1}(r,r) = e^{-\alpha} = e^{-(1-(1+\varepsilon)r)/((1-\varepsilon)r)}$. Moreover, $1-(1+\varepsilon)r_n=(1+\varepsilon)(b_\varepsilon-r_n)=(1-(1+\varepsilon)r)((1-\varepsilon)/2)^n$ and $r_n\geq r$. Thus, the probability of observing the infinite history $(a_1, a_1, \ldots)$ conditional on state $\omega = 2$, denoted by $\mathrm{Pr}((a_1, a_1, \ldots)\mid \omega = 2)$, satisfies
  \[\prod_{n = 0}^{\infty} \psi_2^{a_1}(r_n,r_n) = \prod_{n = 0}^{\infty} \exp\left(-\frac{1-(1+\varepsilon)r_n}{(1-\varepsilon)r_n}\right) \geq \prod_{n = 0}^{\infty} \exp\left(-\frac{1-(1+\varepsilon)r}{(1-\varepsilon)r}\left(\frac{1-\varepsilon}{2}\right)^n\right).\]
  The logarithm of the right-hand side equals
\[-\frac{1-(1+\varepsilon)r}{(1-\varepsilon)r}\sum_{n=0}^{\infty}\left(\frac{1-\varepsilon}{2}\right)^n = -\frac{2(1-(1+\varepsilon)r)}{(1-\varepsilon^2)r}>-\infty.\]
Thus, $\mathrm{Pr}((a_1,a_1,\ldots)\mid\omega=2)\geq\exp(-2(1-(1+\varepsilon)r)/((1-\varepsilon^2)r))>0$. Since $\mu_0(2)>0$, the public belief converges to $\bar\mu_\varepsilon\in\Delta^S\setminus\Delta^{AK}$ with positive probability. Herding follows from \Cref{lem:learning_limit}.
\end{proof}

\begin{lemma}\label{lem:counter_asymmetric_paths}
  Fix $(\lambda_{1,0},\lambda_{2,0})\in\mathcal{W}$ with $\lambda_{1,0}>\lambda_{2,0}$, and write $(\lambda_{1,t},\lambda_{2,t}):=T_{a_1}^t(\lambda_{1,0},\lambda_{2,0})$.
  \begin{enumerate}[label=(\roman*)]
    \item If $\varepsilon\geq0$, then $\lambda_{1,t}\to\infty$ while $\lambda_{2,t}<b_\varepsilon$ for all $t$. Hence the public belief converges to $\delta_1$.
    \item If $\varepsilon<0$, then $\lambda_{1,t}$ converges to some $\lambda_{1,\infty}\in(b_\varepsilon,\infty)$ and $\lambda_{2,t}\to b_\varepsilon$. Hence the public belief converges to the full-support belief represented by $(\lambda_{1,\infty},b_\varepsilon)\in C_1^\varepsilon$. Moreover, the history $(a_1,a_1,\ldots)$ has positive probability conditional on state $2$.
  \end{enumerate}
  The case $\lambda_{1,0}<\lambda_{2,0}$ is symmetric, with $\delta_3$ in place of $\delta_1$.
\end{lemma}

\begin{proof}
Since $\lambda_{1,0}>\lambda_{2,0}$, the initial belief lies in $R_0$ or in $\bar R_1:=\{(\lambda_1,\lambda_2)\in\mathcal{W}:\lambda_1\geq b_\varepsilon>\lambda_2\}$. While the belief is in $R_0$, \Cref{lem:counter_gap_expansion} shows that $\lambda_{1,t}-\lambda_{2,t}$ remains positive and increases by a factor greater than $(5+\varepsilon)/4>1$ at each update. Since this difference is less than $b_\varepsilon$ in $R_0$, the belief leaves $R_0$ at some finite time $t_0$; by \Cref{lem:counter_W_invariant}, it then lies in $\bar R_1$. The proof of \Cref{lem:counter_W_invariant} shows that $\bar R_1$ is invariant under $T_{a_1}$. Hence, for $t\geq t_0$, \eqref{eq:counter_R1} gives
\begin{equation}\label{eq:counter_R1_path}
\lambda_{1,t+1}-\lambda_{1,t}=\lambda_{1,t}\frac{(1-\varepsilon)\alpha_t+\varepsilon(e^{\alpha_t}-1)}{1+e^{\alpha_t}}>0,\qquad
\lambda_{2,t+1}=\lambda_{2,t}\left(1-\varepsilon\tanh\frac{\alpha_t}{2}\right),
\end{equation}
where $\alpha_t:=(1-(1+\varepsilon)\lambda_{2,t})/((1-\varepsilon)\lambda_{1,t})\leq(1+\varepsilon)/(1-\varepsilon)$.

(i) Suppose $\varepsilon\geq0$. By \eqref{eq:counter_R1_path}, $\lambda_{2,t}$ is nonincreasing for $t\geq t_0$, so $1-(1+\varepsilon)\lambda_{2,t}\geq 1-(1+\varepsilon)\lambda_{2,t_0}>0$. Since $\lambda_{1,t}(1-\varepsilon)\alpha_t=1-(1+\varepsilon)\lambda_{2,t}$, \eqref{eq:counter_R1_path} gives
\[
\lambda_{1,t+1}-\lambda_{1,t}\geq\frac{1-(1+\varepsilon)\lambda_{2,t}}{1+e^{\alpha_t}}\geq\frac{1-(1+\varepsilon)\lambda_{2,t_0}}{1+e^{(1+\varepsilon)/(1-\varepsilon)}}>0.
\]
Thus $\lambda_{1,t}\to\infty$, while $\lambda_{2,t}<b_\varepsilon$. The public belief is $(\lambda_{1,t},1,\lambda_{2,t})/(1+\lambda_{1,t}+\lambda_{2,t})$, which converges to $\delta_1$ as $t\to\infty$.

(ii) Suppose $\varepsilon<0$. By \eqref{eq:counter_R1_path}, $\lambda_{2,t+1}=\lambda_{2,t}(1+|\varepsilon|\tanh(\alpha_t/2))$ for $t\geq t_0$, and $\lambda_{2,t}<b_\varepsilon$ for all $t$. Hence $\prod_{t\geq t_0}(1+|\varepsilon|\tanh(\alpha_t/2))\leq b_\varepsilon/\lambda_{2,t_0}$. Since $\log(1+x)\geq x/2$ for $x\in[0,1]$, we get $\sum_{t\geq t_0}\tanh(\alpha_t/2)<\infty$. The proof of \Cref{lem:counter_W_invariant} shows that $\alpha_t<1$. By concavity of $\tanh$ on $[0,\infty)$, $\tanh(\alpha_t/2)\geq\alpha_t\tanh(1/2)\geq\alpha_t/3$, so $\sum_{t\geq t_0}\alpha_t<\infty$.

By \eqref{eq:counter_R1_path}, $1<\lambda_{1,t+1}/\lambda_{1,t}\leq1+(1-\varepsilon)\alpha_t/2$ for $t\geq t_0$. Summability of $\alpha_t$ implies that $\lambda_{1,t}$ increases to a finite limit $\lambda_{1,\infty}>b_\varepsilon$. Since $\alpha_t\to0$, we have $1-(1+\varepsilon)\lambda_{2,t}=(1-\varepsilon)\lambda_{1,t}\alpha_t\leq(1-\varepsilon)\lambda_{1,\infty}\alpha_t\to0$, so $\lambda_{2,t}\to b_\varepsilon$.

Finally, for $t\geq t_0$, the probability of $a_2$ in state $2$ is $1-\psi_2^{a_1}(\lambda_{1,t},\lambda_{2,t})=(1-e^{-\alpha_t})/2\leq\alpha_t/2\leq1/2$. Since $\log(1-x)\geq-2x$ for $x\in[0,1/2]$,
\[
\prod_{t\geq t_0}\psi_2^{a_1}(\lambda_{1,t},\lambda_{2,t})\geq\exp\left(-\sum_{t\geq t_0}\alpha_t\right)>0.
\]
For $t<t_0$, $\psi_2^{a_1}(\lambda_{1,t},\lambda_{2,t})>0$. Hence $\mathrm{Pr}((a_1,a_1,\ldots)\mid\omega=2)>0$.
\end{proof}

The following lemma on preimages of null sets under real analytic maps is standard; we include a proof for completeness.

\begin{lemma}\label{lem:analytic_null_preimages}
Let $U\subseteq\mathbb R^d$ be connected and open, and let $T:U\to\mathbb R^d$ be real analytic. If $\det DT(x_0)\neq0$ for some $x_0\in U$, then $T^{-1}(E)$ is Lebesgue-null for every Lebesgue-null set $E\subseteq\mathbb R^d$.
\end{lemma}

\begin{proof}
Let $Z:=\{x\in U:\det DT(x)=0\}$. The function $\det DT$ is real analytic and not identically zero on the connected open set $U$, so $Z$ is Lebesgue-null \citep{mityagin2020zero}. By the inverse function theorem, every $x\in U\setminus Z$ has an open neighborhood on which $T$ is a $C^1$ diffeomorphism onto an open set. Countably many such neighborhoods $V_1,V_2,\ldots$ cover $U\setminus Z$. For each $k$, $T^{-1}(E)\cap V_k$ is the image of $E\cap T(V_k)$ under the $C^1$ inverse of $T|_{V_k}$. Since $C^1$ maps send null sets to null sets, $T^{-1}(E)\cap V_k$ is null. Hence $T^{-1}(E)\subseteq Z\cup\bigcup_k\bigl(T^{-1}(E)\cap V_k\bigr)$ is null.
\end{proof}

\begin{lemma}\label{lem:counter_exceptional_null}
  $\mathcal{N}$ is meager and has Lebesgue measure zero in $\mathbb{R}^2$.
\end{lemma}

\begin{proof}
Apart from the null boundaries $\lambda_1=b_\varepsilon$ and $\lambda_2=b_\varepsilon$, $\mathcal W$ is the union of the connected open sets $R_0,R_1,R_2$. On each region, $\alpha$, $\beta$, and both update maps are real analytic, with positive action probabilities in every state. By \Cref{lem:counter_jacobian}, $\det DT_a$ is not identically zero on any of these regions. Applying \Cref{lem:analytic_null_preimages} on each region, $T_a^{-1}(E)$ is null for every null set $E\subseteq\mathbb R^2$ and each action $a$.

Both update maps preserve $\mathcal W$ by \Cref{lem:counter_W_invariant}, and $\mathcal D$ is null. Applying the previous step repeatedly, $E_h:=T_h^{-1}(\mathcal D)$ is null for every finite history $h$. Since $\mathcal D$ is closed relative to $\mathcal W$ and $T_h$ is continuous, $E_h$ is closed relative to the open set $\mathcal W$. Being null, it has empty interior, so it is nowhere dense in $\mathbb R^2$. There are countably many finite histories, so $\mathcal N=\bigcup_hE_h$ is null and meager.
\end{proof}

\begin{proof}[Proof of \Cref{lem:counter_dynamics_main}]
Set $\varepsilon=0$, so that $C_1^\varepsilon=C_1$ and $\bar\mu_\varepsilon=\bar\mu$. The first claim follows from \Cref{lem:counter_W_invariant}. Under repeated $a_1$ observations, if $\lambda_1(\mu_0)=\lambda_2(\mu_0)$, then the proof of \Cref{lem:counter_symmetric_herding} shows that the public belief converges to $\bar\mu$. Otherwise, \Cref{lem:counter_asymmetric_paths}(i) shows that it converges to $\delta_1$ if $\lambda_1(\mu_0)>\lambda_2(\mu_0)$ and to $\delta_3$ if $\lambda_1(\mu_0)<\lambda_2(\mu_0)$.
\end{proof}

\begin{proof}[Proof of \Cref{prop:perturbed}]
The first claim follows from \Cref{lem:counter_W_invariant}.

(i) Suppose $\varepsilon\geq0$, and take a full-support prior in $\mathcal{W}\setminus \mathcal{N}$. By \Cref{lem: martingale}, the public belief converges almost surely to a stationary belief. The possible limits are full-support beliefs represented by $C_1^\varepsilon$ and the three point masses: a nondegenerate limit supported on $\{1,3\}$ would force both ratios eventually to exceed $b_\varepsilon$, contradicting \Cref{lem:counter_W_invariant}, while nondegenerate beliefs on the other two edges are not stationary.

Suppose that $\mathbb P(\tilde\mu_\infty\in C_1^\varepsilon)>0$. Since $a_1$ is strictly optimal throughout $C_1^\varepsilon$, convergence and the overturning principle imply that, almost surely on this event, only finitely many $a_2$ observations occur. Thus, there is a history $(h_{t_0},a_1,a_1,\ldots)$ along which the public belief has a limit in $C_1^\varepsilon$. By invariance of $\mathcal{W}$ and the definition of $\mathcal{N}$, the belief after $h_{t_0}$ lies in $\mathcal{W}\setminus \mathcal{D}$. By \Cref{lem:counter_asymmetric_paths}(i), the belief along $(a_1,a_1,\ldots)$ converges to $\delta_1$ or $\delta_3$, a contradiction.

Hence $\tilde\mu_\infty\in\{\delta_1,\delta_2,\delta_3\}\subseteq\Delta^{AK}$ almost surely. Ties have probability zero, so this holds in every equilibrium; \Cref{lem:learning_limit} therefore gives adequate learning. By \Cref{lem:counter_exceptional_null}, $\mathcal{N}$ is meager and null. The coordinate map $(\lambda_1,\lambda_2)\mapsto(\lambda_1,1,\lambda_2)/(1+\lambda_1+\lambda_2)$ is a diffeomorphism from $(0,\infty)^2$ onto the full-support simplex, so the corresponding exceptional priors are also meager and null. This establishes learning for generic full-support priors outside $C_1^\varepsilon$. Finally, \Cref{lem:counter_symmetric_herding} proves herding for priors with $\lambda_1(\mu_0)=\lambda_2(\mu_0)\in(0,b_\varepsilon)$.

(ii) Suppose $\varepsilon<0$. For a full-support prior in $C_1^\varepsilon$, every agent chooses $a_1$ almost surely, so the public belief remains at the prior, which lies in $\Delta^S\setminus\Delta^{AK}$; herding follows from \Cref{lem:learning_limit}. For a prior in $\mathcal{D}$, herding follows from \Cref{lem:counter_symmetric_herding}. For any other full-support prior, \Cref{lem:counter_asymmetric_paths}(ii) shows that, with positive probability conditional on state $2$, all agents choose $a_1$ and the public belief converges to a full-support belief in $C_1^\varepsilon$. This limit lies in $\Delta^S\setminus\Delta^{AK}$. Since $\mu_0(2)>0$, herding follows from \Cref{lem:learning_limit}.
\end{proof}

\begin{proof}[Proof of \Cref{prop:prior_dependence}]
This is \Cref{prop:perturbed}(i) with $\varepsilon=0$, for which $F_1^\varepsilon=F_1$ and $C_1^\varepsilon=C_1$.
\end{proof}

\bibliographystyle{ecta}
\bibliography{Learning}

\begin{thebibliography}{27}
\newcommand{\enquote}[1]{``#1''}
\expandafter\ifx\csname natexlab\endcsname\relax\def\natexlab#1{#1}\fi

\bibitem[\protect\citeauthoryear{Acemoglu, Dahleh, Lobel, and
  Ozdaglar}{Acemoglu et~al.}{2011}]{acemoglu_bayesian_2011}
\textsc{Acemoglu, D., M.~A. Dahleh, I.~Lobel, and A.~Ozdaglar} (2011):
  \enquote{Bayesian {Learning} in {Social} {Networks},} \emph{The Review of
  Economic Studies}, 78, 1201--1236.

\bibitem[\protect\citeauthoryear{Aghion, Bolton, Harris, and Jullien}{Aghion
  et~al.}{1991}]{aghion1991}
\textsc{Aghion, P., P.~Bolton, C.~Harris, and B.~Jullien} (1991):
  \enquote{Optimal Learning by Experimentation,} \emph{The Review of Economic
  Studies}, 58, 621--654.

\bibitem[\protect\citeauthoryear{Ali}{Ali}{2018}]{ALI201879}
\textsc{Ali, S.~N.} (2018): \enquote{On the Role of Responsiveness in Rational
  Herds,} \emph{Economics Letters}, 163, 79--82.

\bibitem[\protect\citeauthoryear{Arieli and Mueller-Frank}{Arieli and
  Mueller-Frank}{2021}]{arieli_general_2021}
\textsc{Arieli, I. and M.~Mueller-Frank} (2021): \enquote{A {General}
  {Analysis} of {Sequential} {Social} {Learning},} \emph{Mathematics of
  Operations Research}, 46, 1235--1249.

\bibitem[\protect\citeauthoryear{Awaya and Krishna}{Awaya and
  Krishna}{2026}]{awaya_krishna_2026_social_learning_correlated_information}
\textsc{Awaya, Y. and V.~Krishna} (2026): \enquote{Social Learning with
  Correlated Information,} UTMD Working Paper UTMD-113, The University of
  Tokyo.

\bibitem[\protect\citeauthoryear{Banerjee}{Banerjee}{1992}]{banerjee_simple_1992}
\textsc{Banerjee, A.~V.} (1992): \enquote{A {Simple} {Model} of {Herd}
  {Behavior},} \emph{The Quarterly Journal of Economics}, 107, 797--817.

\bibitem[\protect\citeauthoryear{Bikhchandani, Hirshleifer, Tamuz, and
  Welch}{Bikhchandani et~al.}{2024}]{bikhchandani_information_2024}
\textsc{Bikhchandani, S., D.~Hirshleifer, O.~Tamuz, and I.~Welch} (2024):
  \enquote{Information {Cascades} and {Social} {Learning},} \emph{Journal of
  Economic Literature}, 62, 1040--1093.

\bibitem[\protect\citeauthoryear{Bikhchandani, Hirshleifer, and
  Welch}{Bikhchandani et~al.}{1992}]{BHW1992}
\textsc{Bikhchandani, S., D.~Hirshleifer, and I.~Welch} (1992): \enquote{A
  Theory of Fads, Fashion, Custom, and Cultural Change as Informational
  Cascades,} \emph{Journal of Political Economy}, 100, 992--1026.

\bibitem[\protect\citeauthoryear{Bogachev}{Bogachev}{2018}]{bogachev2018weak}
\textsc{Bogachev, V.~I.} (2018): \emph{Weak Convergence of Measures}, vol. 234
  of \emph{Mathematical Surveys and Monographs}, American Mathematical Society.

\bibitem[\protect\citeauthoryear{Bohren and Hauser}{Bohren and
  Hauser}{2021}]{bohren_learning_2021}
\textsc{Bohren, J.~A. and D.~N. Hauser} (2021): \enquote{Learning {With}
  {Heterogeneous} {Misspecified} {Models}: {Characterization} and
  {Robustness},} \emph{Econometrica}, 89, 3025--3077.

\bibitem[\protect\citeauthoryear{Brunnermeier and Nagel}{Brunnermeier and
  Nagel}{2004}]{k_brunnermeier_hedge_2004}
\textsc{Brunnermeier, M.~K. and S.~Nagel} (2004): \enquote{Hedge {Funds} and
  the {Technology} {Bubble},} \emph{The Journal of Finance}, 59, 2013--2040.

\bibitem[\protect\citeauthoryear{{\c{C}}elen and Kariv}{{\c{C}}elen and
  Kariv}{2004}]{CELEN200472}
\textsc{{\c{C}}elen, B. and S.~Kariv} (2004): \enquote{Observational Learning
  under Imperfect Information,} \emph{Games and Economic Behavior}, 47, 72--86.

\bibitem[\protect\citeauthoryear{Danenberg and Wolitzky}{Danenberg and
  Wolitzky}{2026}]{danenberg2026sequential}
\textsc{Danenberg, T. and A.~Wolitzky} (2026): \enquote{Sequential Social
  Learning from Outcomes,} Working Paper, Massachusetts Institute of
  Technology.

\bibitem[\protect\citeauthoryear{Durrett}{Durrett}{2019}]{durrett2019probability}
\textsc{Durrett, R.} (2019): \emph{Probability: Theory and Examples}, Cambridge
  University Press, 5th ed.

\bibitem[\protect\citeauthoryear{Frick, Iijima, and Ishii}{Frick
  et~al.}{2020}]{frick_misinterpreting_2020}
\textsc{Frick, M., R.~Iijima, and Y.~Ishii} (2020): \enquote{Misinterpreting
  {Others} and the {Fragility} of {Social} {Learning},} \emph{Econometrica},
  88, 2281--2328.

\bibitem[\protect\citeauthoryear{Hann-Caruthers, Martynov, and
  Tamuz}{Hann-Caruthers et~al.}{2018}]{tamuz_speed2018383}
\textsc{Hann-Caruthers, W., V.~V. Martynov, and O.~Tamuz} (2018): \enquote{The
  Speed of Sequential Asymptotic Learning,} \emph{Journal of Economic Theory},
  173, 383--409.

\bibitem[\protect\citeauthoryear{Kartik and Kleiner}{Kartik and
  Kleiner}{2025}]{kartik_convex_2025}
\textsc{Kartik, N. and A.~Kleiner} (2025): \enquote{Convex Choice,} CRC TR 224
  Discussion Paper No.~676, arXiv:2406.19063.

\bibitem[\protect\citeauthoryear{Kartik, Lee, Liu, and Rappoport}{Kartik
  et~al.}{2024{\natexlab{a}}}]{kartik_beyond_2024}
\textsc{Kartik, N., S.~Lee, T.~Liu, and D.~Rappoport} (2024{\natexlab{a}}):
  \enquote{Beyond {Unbounded} {Beliefs}: {How} {Preferences} and {Information}
  {Interplay} in {Social} {Learning},} \emph{Econometrica}, 92, 1033--1062.

\bibitem[\protect\citeauthoryear{Kartik, Lee, Liu, and Rappoport}{Kartik
  et~al.}{2024{\natexlab{b}}}]{kartik_beyond_arxiv_v5_2024}
---\hspace{-.1pt}---\hspace{-.1pt}--- (2024{\natexlab{b}}): \enquote{Beyond
  Unbounded Beliefs: How Preferences and Information Interplay in Social
  Learning,} Working Paper, April 2024 version, arXiv:2103.02754v5.

\bibitem[\protect\citeauthoryear{Kartik, Lee, and Rappoport}{Kartik
  et~al.}{2024{\natexlab{c}}}]{kartik_single-crossing_2024}
\textsc{Kartik, N., S.~Lee, and D.~Rappoport} (2024{\natexlab{c}}):
  \enquote{Single-{Crossing} {Differences} in {Convex} {Environments},}
  \emph{The Review of Economic Studies}, 91, 2981--3012.

\bibitem[\protect\citeauthoryear{Lakonishok, Shleifer, and Vishny}{Lakonishok
  et~al.}{1994}]{lakonishok_contrarian_1994}
\textsc{Lakonishok, J., A.~Shleifer, and R.~W. Vishny} (1994):
  \enquote{Contrarian {Investment}, {Extrapolation}, and {Risk},} \emph{The
  Journal of Finance}, 49, 1541--1578.

\bibitem[\protect\citeauthoryear{Lee}{Lee}{1993}]{LEE1993395}
\textsc{Lee, I.~H.} (1993): \enquote{On the Convergence of Informational
  Cascades,} \emph{Journal of Economic Theory}, 61, 395--411.

\bibitem[\protect\citeauthoryear{Lobel and Sadler}{Lobel and
  Sadler}{2015}]{Lobel_Sadler_2015}
\textsc{Lobel, I. and E.~Sadler} (2015): \enquote{Information Diffusion in
  Networks through Social Learning,} \emph{Theoretical Economics}, 10,
  807--851.

\bibitem[\protect\citeauthoryear{Mityagin}{Mityagin}{2020}]{mityagin2020zero}
\textsc{Mityagin, B.~S.} (2020): \enquote{The Zero Set of a Real Analytic
  Function,} \emph{Mathematical Notes}, 107, 529--530.

\bibitem[\protect\citeauthoryear{Rosenberg and Vieille}{Rosenberg and
  Vieille}{2019}]{rosenberg_efficiency_2019}
\textsc{Rosenberg, D. and N.~Vieille} (2019): \enquote{On the {Efficiency} of
  {Social} {Learning},} \emph{Econometrica}, 87, 2141--2168.

\bibitem[\protect\citeauthoryear{Sato and Shimizu}{Sato and
  Shimizu}{2026}]{sato2025valueinformationsociallearning}
\textsc{Sato, H. and K.~Shimizu} (2026): \enquote{Value of Information in
  Social Learning,} \emph{Journal of Economic Theory}, 237, 106232.

\bibitem[\protect\citeauthoryear{Smith and Sørensen}{Smith and
  Sørensen}{2000}]{smith_pathological_2000}
\textsc{Smith, L. and P.~Sørensen} (2000): \enquote{Pathological {Outcomes} of
  {Observational} {Learning},} \emph{Econometrica}, 68, 371--398.

\end{thebibliography}

\newpage

\section{Supplementary Appendix}\label{sec: SA}
\setcounter{equation}{0}
\renewcommand{\theequation}{\thesection.\arabic{equation}}
\renewcommand{\theHequation}{appendix.\thesection.\arabic{equation}}
\setcounter{lemma}{0}
\makeatletter \let\c@proposition\c@lemma \makeatother
\renewcommand{\thelemma}{\thesection.\arabic{lemma}}
\renewcommand{\theproposition}{\thesection.\arabic{lemma}}
\renewcommand{\theHlemma}{appendix.\thesection.\arabic{lemma}}
\renewcommand{\theHproposition}{appendix.\thesection.\arabic{lemma}}

\subsection{Supporting Calculations for \texorpdfstring{\Cref{ex:perturbed}}{Example \ref*{ex:perturbed}}}\label{SA: ex_counter}

\subsubsection{Action probabilities and updating}\label{SA:counter_updates}

Use the belief-odds coordinates and notation of \Cref{app:counter_prelim}, with $\varepsilon\in(-1/4,1/4)$ fixed; \Cref{ex:counter} is the case $\varepsilon=0$. At a private belief $\mu$, action $a_1$ is optimal if and only if $\mu(1)+\mu(3)-\mu(2)\geq0$. Thus, from a public belief $(\lambda_1,\lambda_2)$, an agent chooses $a_1$ after signal $s$ if and only if
\[
\lambda_1\frac{f(s\mid1)}{f(s\mid2)}+\lambda_2\frac{f(s\mid3)}{f(s\mid2)}\geq1,
\]
where, writing $f=f_1^\varepsilon$,
\[
\frac{f(s\mid1)}{f(s\mid2)}=\begin{cases}(1-\varepsilon)(-s),&s<0,\\1+\varepsilon,&s>0,\end{cases}
\qquad
\frac{f(s\mid3)}{f(s\mid2)}=\begin{cases}1+\varepsilon,&s<0,\\(1-\varepsilon)s,&s>0.\end{cases}
\]
Consequently, $a_1$ is chosen when $s\leq-\alpha$ or $s\geq\beta$, and $a_2$ when $-\alpha<s<\beta$, where
\[
\alpha=\left(\frac{1-(1+\varepsilon)\lambda_2}{(1-\varepsilon)\lambda_1}\right)_+,
\qquad
\beta=\left(\frac{1-(1+\varepsilon)\lambda_1}{(1-\varepsilon)\lambda_2}\right)_+.
\]
Ties have probability zero in every state, so tie-breaking does not affect action probabilities. The belief is an $a_1$-cascade belief exactly when $\alpha=\beta=0$, or equivalently $(\lambda_1,\lambda_2)\in C_1^\varepsilon$.

Integrating the signal densities gives
\[
\psi_\omega^{a_1}(\lambda_1,\lambda_2)
=\int_{-\infty}^{-\alpha}f(s\mid\omega)\,\mathrm{d}s
+\int_\beta^\infty f(s\mid\omega)\,\mathrm{d}s.
\]
Since $\int_{-\infty}^{-\alpha}(-s)e^s\,\mathrm{d}s=(1+\alpha)e^{-\alpha}$, we obtain
\begin{gather*}
\psi_1^{a_1}(\lambda_1,\lambda_2)=\frac{(1-\varepsilon)(1+\alpha)e^{-\alpha}+(1+\varepsilon)e^{-\beta}}{2},\qquad
\psi_2^{a_1}(\lambda_1,\lambda_2)=\frac{e^{-\alpha}+e^{-\beta}}{2},\\
\psi_3^{a_1}(\lambda_1,\lambda_2)=\frac{(1+\varepsilon)e^{-\alpha}+(1-\varepsilon)(1+\beta)e^{-\beta}}{2}.
\end{gather*}
For each state, $\psi_\omega^{a_2}=1-\psi_\omega^{a_1}$. Bayes' rule yields, whenever action $a$ has positive probability,
\[
T_a(\lambda_1,\lambda_2)
=\left(\lambda_1\frac{\psi_1^a(\lambda_1,\lambda_2)}{\psi_2^a(\lambda_1,\lambda_2)},\;
\lambda_2\frac{\psi_3^a(\lambda_1,\lambda_2)}{\psi_2^a(\lambda_1,\lambda_2)}\right).
\]
In particular,
\[
T_{a_1,1}(\lambda_1,\lambda_2)
=\lambda_1\frac{(1-\varepsilon)(1+\alpha)e^{-\alpha}+(1+\varepsilon)e^{-\beta}}{e^{-\alpha}+e^{-\beta}}
=\lambda_1+\varepsilon\lambda_1\frac{e^{-\beta}-e^{-\alpha}}{e^{-\alpha}+e^{-\beta}}+\frac{(1-\varepsilon)\lambda_1\alpha e^{-\alpha}}{e^{-\alpha}+e^{-\beta}}.
\]
Since $(1-\varepsilon)\lambda_1\alpha=(1-(1+\varepsilon)\lambda_2)_+$ and $(e^{-\beta}-e^{-\alpha})/(e^{-\alpha}+e^{-\beta})=\tanh((\alpha-\beta)/2)$, this is the first line of \eqref{eq:counter_Ta1}; the second is analogous. Similarly,
\begin{equation}\label{eq:counter_Ta2}
\begin{aligned}
T_{a_2,1}(\lambda_1,\lambda_2)&=\lambda_1+\frac{\varepsilon\lambda_1(e^{-\alpha}-e^{-\beta})-(1-(1+\varepsilon)\lambda_2)_+e^{-\alpha}}{2-e^{-\alpha}-e^{-\beta}},\\
T_{a_2,2}(\lambda_1,\lambda_2)&=\lambda_2+\frac{\varepsilon\lambda_2(e^{-\beta}-e^{-\alpha})-(1-(1+\varepsilon)\lambda_1)_+e^{-\beta}}{2-e^{-\alpha}-e^{-\beta}}.
\end{aligned}
\end{equation}

\subsubsection{Jacobian determinants}\label{SA:counter_jacobian}

\begin{lemma}\label{lem:counter_jacobian}
For each $a\in\{a_1,a_2\}$ and $i\in\{0,1,2\}$, $\det DT_a$ is not identically zero on $R_i$.
\end{lemma}

\begin{proof}
With gradients written as row vectors, Bayes' rule gives
\begin{equation}\label{eq:counter_update_jacobian}
 DT_a=
 \begin{pmatrix}
 \psi_1^a/\psi_2^a&0\\
 0&\psi_3^a/\psi_2^a
 \end{pmatrix}
 +\frac1{(\psi_2^a)^2}
 \begin{pmatrix}
 \lambda_1(\psi_2^a\nabla\psi_1^a-\psi_1^a\nabla\psi_2^a)\\
 \lambda_2(\psi_2^a\nabla\psi_3^a-\psi_3^a\nabla\psi_2^a)
 \end{pmatrix}.
\end{equation}
The cutoff gradients are
\begin{align*}
 \nabla\alpha&=-\frac1{\lambda_1}\left(\alpha,\frac{1+\varepsilon}{1-\varepsilon}\right)
 &&\text{on }R_0\cup R_1,\\
 \nabla\beta&=-\frac1{\lambda_2}\left(\frac{1+\varepsilon}{1-\varepsilon},\beta\right)
 &&\text{on }R_0,
\end{align*}
while $\nabla\beta=0$ on $R_1$.

On $R_0$, evaluate at $(\lambda_1,\lambda_2)=(1/2,1/2)$, where $\alpha=\beta=1$. Since this point is fixed by $P$ and $T_a\circ P=P\circ T_a$ (\Cref{app:counter_prelim}), at this point
\[
DT_a=\begin{pmatrix}p_a&q_a\\ q_a&p_a\end{pmatrix},
\qquad
\det DT_a=(p_a+q_a)(p_a-q_a).
\]
Substituting the action probabilities from \Cref{SA:counter_updates} and their derivatives into \eqref{eq:counter_update_jacobian} gives
\begin{align*}
p_{a_1}+q_{a_1}&=\frac{1-\varepsilon}{2}, &
p_{a_1}-q_{a_1}&=\frac{5\varepsilon^2-4\varepsilon+3}{2(1-\varepsilon)},\\
p_{a_2}+q_{a_2}&=\frac{2e^2-(5-\varepsilon)e+1-\varepsilon}{2(e-1)^2}, &
p_{a_2}-q_{a_2}&=\frac{2(1-\varepsilon)e-5\varepsilon^2+4\varepsilon-3}{2(1-\varepsilon)(e-1)}.
\end{align*}
All four are positive for $|\varepsilon|<1/4$: $5\varepsilon^2-4\varepsilon+3$ has negative discriminant; by $e>5/2$, the last numerator exceeds $2-\varepsilon-5\varepsilon^2>0$; and since $2z^2-(5-\varepsilon)z$ is increasing for $z\geq2$, the third numerator exceeds $1+3\varepsilon/2>0$.

On $R_1$, $\beta=0$, so the ratios $\zeta_{a,\omega}:=\psi_\omega^a/\psi_2^a$, $\omega\in\{1,3\}$, depend on $(\lambda_1,\lambda_2)$ only through $\alpha$, and $T_a=(\lambda_1\zeta_{a,1}(\alpha),\lambda_2\zeta_{a,3}(\alpha))$ with
\begin{align*}
\zeta_{a_1,1}(\alpha)&=\frac{(1-\varepsilon)(1+\alpha)e^{-\alpha}+1+\varepsilon}{1+e^{-\alpha}}, &
\zeta_{a_1,3}(\alpha)&=\frac{(1+\varepsilon)e^{-\alpha}+1-\varepsilon}{1+e^{-\alpha}},\\
\zeta_{a_2,1}(\alpha)&=\frac{(1-\varepsilon)\bigl(1-(1+\alpha)e^{-\alpha}\bigr)}{1-e^{-\alpha}}, &
\zeta_{a_2,3}(\alpha)&=1+\varepsilon.
\end{align*}
The cutoff gradient $\nabla\alpha$ above gives
\[
\det DT_a=(\zeta_{a,1}-\alpha\zeta_{a,1}')\zeta_{a,3}-\frac{(1+\varepsilon)\lambda_2}{(1-\varepsilon)\lambda_1}\,\zeta_{a,1}\zeta_{a,3}'.
\]
Since $\zeta_{a_2,3}'=0$,
\[
\det DT_{a_2}=(1-\varepsilon^2)\left[1-\left(\frac{\alpha}{2\sinh(\alpha/2)}\right)^2\right]>0
\]
throughout $R_1$, because $\alpha>0$ there and $\sinh z>z$ for $z>0$. For $a_1$, take $(\lambda_1,\lambda_2)=(2/(1-\varepsilon),\,b_\varepsilon/2)$, which lies in $R_1$ because $2(1+\varepsilon)>1-\varepsilon$. There $\alpha=1/4$ and $(1+\varepsilon)\lambda_2/((1-\varepsilon)\lambda_1)=\alpha$, so
\[
\det DT_{a_1}=1+\frac{e^{\alpha}\bigl[(1-\varepsilon)\alpha^2\bigl((1-\varepsilon)e^{\alpha}+1+3\varepsilon\bigr)+2\varepsilon^2(e^{\alpha}-1)(2\alpha-\sinh\alpha)\bigr]}{(e^{\alpha}+1)^3}>1,
\]
since $(1-\varepsilon)e^{\alpha}+1+3\varepsilon>2+2\varepsilon>0$ and $\sinh(1/4)<1/2$. Finally, $P$ maps $R_1$ onto $R_2$ and commutes with $T_a$, so $\det DT_a\circ P=\det DT_a$ and the same conclusion holds on $R_2$.
\end{proof}

\subsection{Strict MLRP Example}\label{SA: ex_mlrp}

This subsection shows that prior dependence does not rely on the failure of MLRP. We keep the preferences of \Cref{ex:counter} and use an information structure that satisfies strict MLRP.

\begin{example}\label{ex:counter_mlrp}
  Let $\Omega=\{1,2,3\}$ and $A=\{a_1,a_2\}$, with preferences as in \Cref{ex:counter}. The information structure $F_3$ has density
\[
f_3(s\mid\omega)=
\begin{array}{c|ccc}
 & \omega=1 & \omega=2 & \omega=3\\ \cline{1-4}
s<0 & \bigl(\tfrac34-\tfrac{9}{40}s\bigr)e^{s} & \tfrac12e^{s} & \tfrac1{20}e^{2s}\\[3pt]
s>0 & \tfrac1{20}e^{-2s} & \tfrac12e^{-s} & \bigl(\tfrac34+\tfrac{9}{40}s\bigr)e^{-s}
\end{array}.
\]
\end{example}

For $s\geq0$, let $\ell_+(s):=3/2+9s/20$ and $\ell_-(s):=e^{-s}/10$. Writing $f=f_3$, the likelihood ratios relative to state $2$ are
\[
\left(\frac{f(s\mid1)}{f(s\mid2)},\frac{f(s\mid3)}{f(s\mid2)}\right)=
\begin{cases}
(\ell_+(-s),\ell_-(-s)),&s<0,\\
(\ell_-(s),\ell_+(s)),&s>0.
\end{cases}
\]
The ratio $f(s\mid1)/f(s\mid2)$ is strictly decreasing on $\mathbb R\setminus\{0\}$, and $f(s\mid3)/f(s\mid2)$ is strictly increasing. This holds on each half-line, and across $s=0$ because $\ell_+(0)=3/2>1/10=\ell_-(0)$. Hence $F_3$ satisfies strict MLRP. Every pairwise likelihood ratio is unbounded, so pairwise distinguishability holds. However, $f(s\mid1)+f(s\mid3)=[\ell_+(|s|)+\ell_-(|s|)]f(s\mid2)\geq\frac85f(s\mid2)$, so $\{2\}$ is not distinguishable from $\{1,3\}$, and escapability fails.

\subsubsection{Preliminaries}\label{SA:mlrp_prelim}

Use the belief odds $(\lambda_1,\lambda_2)$ of \Cref{app:counter_prelim}. An agent chooses $a_1$ if and only if $\varrho(s)\geq1$, where
\[
\varrho(s):=\begin{cases}
\lambda_1\ell_+(-s)+\lambda_2\ell_-(-s),&s<0,\\
\lambda_1\ell_-(s)+\lambda_2\ell_+(s),&s>0.
\end{cases}
\]
By symmetry, suppose $\lambda_1\geq\lambda_2$. Then $\varrho(s)\geq\varrho(-s)$ for $s<0$, and $\varrho$ is strictly increasing in $|s|$ on that half-line. Minimizing on $s\geq0$ gives
\[
m(\lambda_1,\lambda_2):=\min_{s\geq0}\bigl[\lambda_1\ell_-(s)+\lambda_2\ell_+(s)\bigr]
=\begin{cases}
\dfrac{\lambda_1}{10}+\dfrac{3\lambda_2}{2}, & \lambda_1\leq \dfrac{9\lambda_2}2,\\[8pt]
\lambda_2\left[\dfrac32+\dfrac9{20}\left(1+\log\dfrac{2\lambda_1}{9\lambda_2}\right)\right], & \lambda_1>\dfrac{9\lambda_2}2,
\end{cases}
\]
with minimizer $\max\{0,\log(2\lambda_1/(9\lambda_2))\}$. For $\lambda_1<\lambda_2$, exchange the coordinates. In these coordinates, the cascade region is $C_3=\{m\geq1\}$. In this subsection, $\mathcal W:=(0,\infty)^2\setminus C_3$ and $\mathcal D:=\{(r,r):0<r<5/8\}$. Since $m(r,r)=8r/5$, $\mathcal D$ is the set of symmetric non-cascade beliefs.

Define $T_a$, $\psi_\omega^a$, $T_h$, $P$, and $\mathcal N$ as in \Cref{app:counter_prelim}, with $\mathcal W$ and $\mathcal D$ as above. For $(\lambda_1,\lambda_2)\in\mathcal W$, both actions have positive probability in every state, and $T_a$ is given by Bayes' rule as in \Cref{SA:counter_updates}.

Define the open sets
\begin{align*}
Z_0&:=\{(\lambda_1,\lambda_2):3\lambda_1/2+\lambda_2/10<1,\ \lambda_1/10+3\lambda_2/2<1\},\\
Z_1&:=\{(\lambda_1,\lambda_2):\lambda_1>\lambda_2,\ 3\lambda_1/2+\lambda_2/10>1>\lambda_1/10+3\lambda_2/2\},\\
Z_2&:=\{(\lambda_1,\lambda_2):\lambda_1>\lambda_2,\ \lambda_1/10+3\lambda_2/2>1>m(\lambda_1,\lambda_2)\},
\end{align*}
and let $Z_1'$ and $Z_2'$ be the images of $Z_1$ and $Z_2$ under $P$. These five sets lie in $\mathcal W$ and cover $\mathcal W$ except for the null lines $3\lambda_1/2+\lambda_2/10=1$ and $\lambda_1/10+3\lambda_2/2=1$. On $Z_0$, $a_2$ is chosen exactly when $-L<s<U$, where $L,U>0$ solve
\begin{equation}\label{eq:mlrp_cut}
\lambda_1\ell_+(L)+\lambda_2\ell_-(L)=1,\qquad \lambda_1\ell_-(U)+\lambda_2\ell_+(U)=1.
\end{equation}
If $\lambda_1\geq\lambda_2$, then $L\leq U$, because $\varrho(L)\leq\varrho(-L)=1$. On $Z_1$, $a_2$ is chosen exactly when $0<s<U$, with $U$ given by \eqref{eq:mlrp_cut}. On $Z_2$, it is chosen exactly when $L<s<U$, where $0<L<U$ are the two roots of $\lambda_1\ell_-(s)+\lambda_2\ell_+(s)=1$.

\begin{figure}[htbp]
\centering
\begin{tikzpicture}[x=5cm,y=5cm,font=\small,
  region boundary/.style={black!55,line width=0.3pt},
  cascade boundary/.style={thick,dashed,red,line join=round},
  region pointer/.style={black!65,line width=0.3pt}]
  \coordinate (S1) at (0,0);
  \coordinate (S2) at (1,0);
  \coordinate (S3) at (0.5,{sqrt(3)/2});
  \coordinate (B) at ($(S1)!0.5!(S2)$);
  \coordinate (T) at ($(S2)!0.5!(S3)$);
  \coordinate (Bp) at (0.6,0);
  \coordinate (Bm) at (0.8,{sqrt(3)/5});
  \coordinate (Ep) at ({1/11},0);
  \coordinate (Em) at ({6/11},{5*sqrt(3)/11});
  \coordinate (Q) at ({7/12},{5*sqrt(3)/36});
  \coordinate (Jp) at ({49/149},{10*sqrt(3)/149});
  \coordinate (Jm) at ({84/149},{45*sqrt(3)/149});

  \fill[hazyblue!20] (S1)--(S3)--(T)--(B)--cycle;
  \fill[hazyblue!60] (B)--(T)--(S2)--cycle;
  \fill[pattern={Lines[angle=135,distance=6pt]},pattern color=red!50]
    (S1)--(S3)
    --plot[domain=16:0,samples=161,variable=\t]
      ({(45+(39+9*\t)*exp(-\t))/(90+(59+9*\t)*exp(-\t))},
       {45*sqrt(3)/(90+(59+9*\t)*exp(-\t))})
    --(Q)--(Jp)
    --plot[domain=0:16,samples=161,variable=\t]
      ({(49+9*\t)*exp(-\t)/(90+(59+9*\t)*exp(-\t))},
       {10*sqrt(3)*exp(-\t)/(90+(59+9*\t)*exp(-\t))})
    --cycle;

  \draw[region boundary] (Ep)--(Q)--(Bm);
  \draw[region boundary] (Bp)--(Q)--(Em);
  \draw[cascade boundary] (Q)--(Jp)
    --plot[domain=0:16,samples=161,variable=\t]
      ({(49+9*\t)*exp(-\t)/(90+(59+9*\t)*exp(-\t))},
       {10*sqrt(3)*exp(-\t)/(90+(59+9*\t)*exp(-\t))})
    --(S1);
  \draw[cascade boundary] (Q)--(Jm)
    --plot[domain=0:16,samples=161,variable=\t]
      ({(45+(39+9*\t)*exp(-\t))/(90+(59+9*\t)*exp(-\t))},
       {45*sqrt(3)/(90+(59+9*\t)*exp(-\t))})
    --(S3);
  \draw[black!65,dash pattern=on 2pt off 2pt,line width=0.3pt] (S2)--(Q);
  \draw[dashed] (S1)--(S2)--(S3)--cycle;

  \node[font=\footnotesize] at (0.32,0.34) {$C_3$};
  \node[font=\footnotesize] at (0.78,0.20) {$Z_0$};
  \node[font=\footnotesize] at (0.415,0.060) {$Z_1$};
  \node[font=\footnotesize] at (0.650,0.455) {$Z_1'$};
  \node[font=\footnotesize,rotate=-30] at (0.815,0.080) {$\mathcal D$};
  \node[font=\footnotesize,anchor=north] (Z2label) at (0.13,-0.045) {$Z_2$};
  \draw[region pointer] (Z2label.north)--(0.13,-0.015)
    --({1/12},{sqrt(3)/132});
  \node[font=\footnotesize,anchor=west] (Z2plabel) at (0.65,0.81) {$Z_2'$};
  \draw[region pointer] (Z2plabel.west)--(0.61,0.81)
    --({35/66},{5*sqrt(3)/11});
  \node[below left=3pt] at (S1) {$1$};
  \node[below right=3pt] at (S2) {$2$};
  \node[above=4pt] at (S3) {$3$};
\end{tikzpicture}
\caption{The belief simplex for \Cref{ex:counter_mlrp}. Lighter and darker blue shading indicate where $a_1$ and $a_2$, respectively, are optimal at the public belief. Gray lines separate the signal-choice regions. The red-hatched region is $C_3$, and the dashed open segment represents $\mathcal D$. Shading depicts the closures of the full-support regions.}
\label{fig:mlrp_regions}
\end{figure}
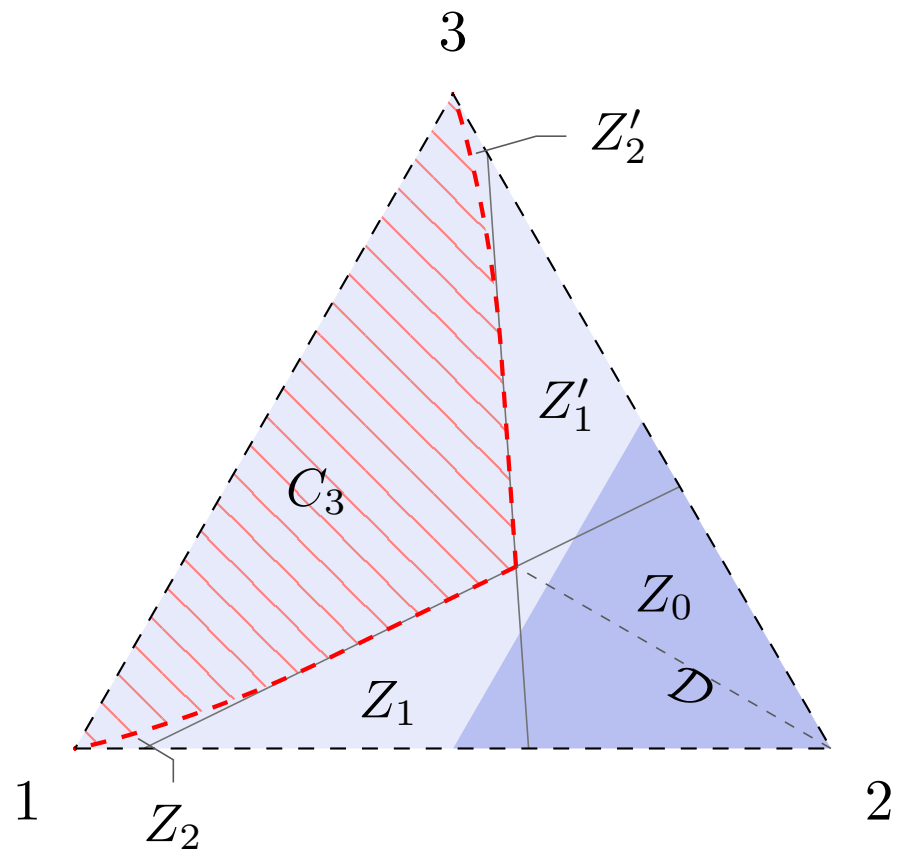

\Cref{fig:mlrp_regions} illustrates these regions in the belief simplex.

For $z\geq0$, write $\bar F_3(z\mid\omega):=\int_z^\infty f_3(s\mid\omega)\,\mathrm ds$. Direct integration gives
\[
\bigl(\bar F_3(z\mid1),\bar F_3(z\mid2),\bar F_3(z\mid3)\bigr)
=\frac{e^{-z}}2\left(\frac{\ell_-(z)}2,\;1,\;\ell_+(z)+\frac9{20}\right).
\]
Since $f_3(-s\mid\omega)=f_3(s\mid4-\omega)$, the probability of signals below $-z$ in state $\omega$ is $\bar F_3(z\mid4-\omega)$, so
\begin{equation}\label{eq:mlrp_tail_probabilities}
\psi_\omega^{a_2}=\begin{cases}
1-\bar F_3(L\mid4-\omega)-\bar F_3(U\mid\omega),&\text{on }Z_0,\\
\bar F_3(0\mid\omega)-\bar F_3(U\mid\omega),&\text{on }Z_1,\\
\bar F_3(L\mid\omega)-\bar F_3(U\mid\omega),&\text{on }Z_2.
\end{cases}
\end{equation}
For each state, $\psi_\omega^{a_1}=1-\psi_\omega^{a_2}$.
The same identity gives $T_a\circ P=P\circ T_a$, so the formulas on $Z_1'$ and $Z_2'$ follow by symmetry. At each cutoff, $\varrho$ crosses $1$ with nonzero slope, so the cutoffs and both update maps are real analytic on each region. Ties have probability zero in every state, so tie-breaking does not affect action probabilities, and the probabilities and update maps are continuous on $\mathcal W$.

\begin{samepage}
\begin{proposition}\label{prop:mlrp}
  In \Cref{ex:counter_mlrp}, public beliefs starting from a full-support prior outside the cascade belief set $C_3$ remain outside $C_3$ after every finite action history. There is adequate learning for generic full-support priors outside $C_3$. However, there is herding if the prior satisfies $0<\lambda_1(\mu_0)=\lambda_2(\mu_0)<5/8$.
\end{proposition}
\end{samepage}

\subsubsection{Characterization}

\begin{lemma}\label{lem:mlrp_positive}
Suppose $(\lambda_1,\lambda_2)\in\mathcal W\setminus Z_0$ and $\lambda_1>\lambda_2$. Then $T_{a_1}(\lambda_1,\lambda_2)\in\mathcal W\setminus Z_0$, $T_{a_1,1}(\lambda_1,\lambda_2)>\lambda_1$, $T_{a_1,2}(\lambda_1,\lambda_2)<\lambda_2$, and $m(T_{a_1}(\lambda_1,\lambda_2))<m(\lambda_1,\lambda_2)$. Moreover, along repeated $a_1$ observations from $(\lambda_1,\lambda_2)$, the public belief converges to $\delta_1$.
\end{lemma}

\begin{proof}
Since $\lambda_1>\lambda_2$ and $(\lambda_1,\lambda_2)\notin Z_0$, we have $3\lambda_1/2+\lambda_2/10\geq1$. Thus $a_2$ is chosen on a nonempty interval $J\subseteq(0,\infty)$. On $J$, $\ell_-<1/10$, $\ell_+>3/2$, and $\varrho<1$. By integration,
\[
 0<\psi_1^{a_2}<\psi_2^{a_2}/10,\qquad
 \psi_3^{a_2}>3\psi_2^{a_2}/2,\qquad
 \lambda_1\psi_1^{a_2}+\lambda_2\psi_3^{a_2}<\psi_2^{a_2}.
\]
Bayes' rule therefore gives $T_{a_1,1}>\lambda_1$ and $T_{a_1,2}<\lambda_2$. Moreover,
\begin{align*}
 \frac32(T_{a_1,1}-\lambda_1)+\frac1{10}(T_{a_1,2}-\lambda_2) =&\frac{30\lambda_1(\psi_2^{a_2}-\psi_1^{a_2})
 +2\lambda_2(\psi_2^{a_2}-\psi_3^{a_2})}{20(1-\psi_2^{a_2})}\\
 > & \frac{27\lambda_1\psi_2^{a_2}
 -2\lambda_2\psi_3^{a_2}}{20(1-\psi_2^{a_2})}\\
 > &\left(\frac{27\lambda_1}{20}-\frac1{10}\right)
 \frac{\psi_2^{a_2}}{1-\psi_2^{a_2}}\\
> &\left(\frac{27}{20}\cdot\frac58-\frac1{10}\right)
 \frac{\psi_2^{a_2}}{1-\psi_2^{a_2}}
 =\frac{119}{160}\frac{\psi_2^{a_2}}{1-\psi_2^{a_2}}>0.
\end{align*}
The first inequality uses $\psi_1^{a_2}<\psi_2^{a_2}/10$ and drops the positive term $2\lambda_2\psi_2^{a_2}$; the second uses $\lambda_2\psi_3^{a_2}<\psi_2^{a_2}$; and the third uses $\lambda_1>5/8$, which follows from $1\leq3\lambda_1/2+\lambda_2/10<8\lambda_1/5$. Thus $T_{a_1}\notin Z_0$.

\begin{samepage}
Let $s\geq0$ minimize $\lambda_1\ell_-(s)+\lambda_2\ell_+(s)$. Since $\ell_-'(s)=-\ell_-(s)$ and $\ell_+'(s)=9/20$, the derivative of this objective is $-\lambda_1\ell_-(s)+9\lambda_2/20$. At an interior minimizer it is zero; at $s=0$ the right derivative is nonnegative. Thus both cases give $\lambda_1\ell_-(s)\leq9\lambda_2/20$.
\begin{align*}
 &m(T_{a_1})-m(\lambda_1,\lambda_2)\\
 \leq &\ell_-(s)(T_{a_1,1}-\lambda_1)+\ell_+(s)(T_{a_1,2}-\lambda_2)\\
 =&\frac{\lambda_1\ell_-(s)(\psi_2^{a_2}-\psi_1^{a_2})
 +\lambda_2\ell_+(s)(\psi_2^{a_2}-\psi_3^{a_2})}{1-\psi_2^{a_2}}\\
 \leq &\frac{\lambda_2}{1-\psi_2^{a_2}}
 \left[\frac9{20}(\psi_2^{a_2}-\psi_1^{a_2})
 +\ell_+(s)(\psi_2^{a_2}-\psi_3^{a_2})\right]\\
 < &\left(\frac9{20}-\frac{\ell_+(s)}2\right)
 \frac{\lambda_2\psi_2^{a_2}}{1-\psi_2^{a_2}}\\
 \leq &\left(\frac9{20}-\frac34\right)
 \frac{\lambda_2\psi_2^{a_2}}{1-\psi_2^{a_2}}
 =-\frac{3\lambda_2\psi_2^{a_2}}{10(1-\psi_2^{a_2})}<0.
\end{align*}
The first inequality evaluates the updated minimum at the original minimizer $s$; the second uses $\lambda_1\ell_-(s)\leq9\lambda_2/20$ and $\psi_1^{a_2}<\psi_2^{a_2}$; the third uses $\psi_1^{a_2}>0$ and $\psi_3^{a_2}>3\psi_2^{a_2}/2$; and the fourth uses $\ell_+(s)\geq3/2$.
\end{samepage}
This proves all the one-step claims.

Along repeated $a_1$ observations, $\lambda_{1,t}$ increases, $\lambda_{2,t}$ decreases, and $m\leq m(\lambda_{1,0},\lambda_{2,0})<1$. If $\lambda_{1,t}$ had a finite limit, the limiting odds would admit some fixed $s>0$ with $\varrho(s)<1$: use the bound on $m$ if $\lambda_{2,\infty}>0$, and choose $s$ large if $\lambda_{2,\infty}=0$. A fixed interval around $s$ would then induce $a_2$ for all sufficiently large $t$, keeping $\psi_2^{a_2}$ bounded away from zero. This contradicts $\lambda_{1,t+1}/\lambda_{1,t}\geq1+9\psi_2^{a_2}/10$. Hence $\lambda_{1,t}\to\infty$. The formula for $m$ then forces $\lambda_{2,t}\to0$, proving convergence to $\delta_1$.
\end{proof}

\begin{lemma}\label{lem:mlrp_W_invariant}
For any $(\lambda_1,\lambda_2)\in\mathcal W$, we have $T_{a_1}(\lambda_1,\lambda_2)\in\mathcal W$ and $T_{a_2}(\lambda_1,\lambda_2)\in\mathcal W$.
\end{lemma}

\begin{proof}
Every cascade belief satisfies $3\lambda_1/2+\lambda_2/10\geq1$ and $\lambda_1/10+3\lambda_2/2\geq1$, by taking the two one-sided limits of $\varrho$ at zero. Adding gives $\lambda_1+\lambda_2\geq5/4$. After $a_2$, the overturning principle instead gives $T_{a_2,1}+T_{a_2,2}\leq1$, so $T_{a_2}\in\mathcal W$.

For $a_1$, \Cref{lem:mlrp_positive} and symmetry handle $\mathcal W\setminus Z_0$. On $Z_0$, action $a_2$ is chosen for $-L<s<U$. By symmetry, assume $\lambda_1\geq\lambda_2$, so $0<L\leq U$. Here $L$ and $U$ solve \eqref{eq:mlrp_cut}. To rule out a cascade after $a_1$, it suffices to show $T_{a_1,1}/10+3T_{a_1,2}/2<1$. 

Write $\ell_\pm=\ell_\pm(L)$, with $\ell_+\geq3/2$ and $0<\ell_-\leq1/10$. Then
\begin{align*}
 &(1+e^{L-U})\left(\frac{T_{a_1,1}}{10}+\frac{3T_{a_1,2}}2-1\right)\\
 &=\frac{\lambda_1}{10}\left[\ell_++\frac9{20}+\frac{\ell_-}{2}e^{2(L-U)}\right]
 +\frac{3\lambda_2}{2}\left[\frac{\ell_-}{2}+e^{L-U}\left(\ell_+(U)+\frac9{20}\right)\right]-1-e^{L-U}\\
 &=-\frac9{10}+\frac{9\lambda_1}{200}+\frac{13\lambda_2\ell_-}{20}
 +e^{L-U}\left(\frac12+\frac{27\lambda_2}{40}\right)
 -\frac{29\lambda_1\ell_-}{20}e^{2(L-U)}\\
 &\leq-\frac9{10}+\left(\frac9{200}+\frac{29\ell_-}{20}\right)\lambda_1
 +\frac{13\lambda_2\ell_-}{20}
 +e^{L-U}\left(\frac12+\frac{27\lambda_2}{40}-\frac{29\lambda_1\ell_-}{10}\right)\\
 &\leq-\frac25+\left(\frac9{200}-\frac{29\ell_-}{20}\right)\lambda_1
 +\left(\frac{27}{40}+\frac{13\ell_-}{20}\right)\lambda_2\\
 &=\left(\frac9{200}-\frac{29\ell_-}{20}-\frac{2\ell_+}{5}\right)\lambda_1
 +\left(\frac{27}{40}+\frac{\ell_-}{4}\right)\lambda_2\\
 &\leq\lambda_1\left(\frac{18}{25}-\frac{2\ell_+}{5}-\frac{6\ell_-}{5}\right)\\
 &=\lambda_1\left[\frac3{25}(1-e^{-L})-\frac{9L}{50}\right]
 \leq-\frac{3\lambda_1L}{50}<0.
\end{align*}
The first equality follows from \eqref{eq:mlrp_tail_probabilities}, and the second uses the cutoff equations \eqref{eq:mlrp_cut} in the form $\lambda_1\ell_+=1-\lambda_2\ell_-$ and $\lambda_2\ell_+(U)=1-\lambda_1\ell_-e^{L-U}$. The first inequality uses $e^{2(L-U)}\geq2e^{L-U}-1$; the second uses $e^{L-U}\leq1$ and the positivity of the coefficient of $e^{L-U}$, which exceeds $1/2-29/150>0$ because the cutoff identity gives $\lambda_1\ell_-<\ell_-/\ell_+\leq1/15$. The next equality replaces $-2/5$ by $-2(\lambda_1\ell_++\lambda_2\ell_-)/5$, the third inequality uses $\lambda_2\leq\lambda_1$, and the last step substitutes $\ell_+=3/2+9L/20$ and $\ell_-=e^{-L}/10$ and uses $1-e^{-L}\leq L$. Since $1+e^{L-U}>0$, the desired inequality follows.
\end{proof}

\begin{lemma}\label{lem:mlrp_symmetric_herding}
For any initial prior in $\mathcal D$, herding occurs.
\end{lemma}
\nopagebreak[4]

\begin{proof}
Let $\lambda_1=\lambda_2=r\in(0,5/8)$, and let $R>0$ solve $r[\ell_+(R)+\ell_-(R)]=1$. By \eqref{eq:mlrp_tail_probabilities}, $\psi_2^{a_1}=e^{-R}$ and $T_{a_1}(r,r)=(r',r')$, where
\[
 r'=\frac r2\left[\ell_+(R)+\frac9{20}+\frac{\ell_-(R)}2\right].
\]
Substitution gives
\[
 0<\frac{5/8-r'}{5/8-r}
 =\frac{9R-6(1-e^{-R})}{45R-10(1-e^{-R})}<\frac15,
\]
using $0<1-e^{-R}<R$.
Thus, along repeated $a_1$ observations, $r_n\uparrow5/8$ with $5/8-r_n\leq5^{-n}(5/8-r)$. For the corresponding cutoffs $R_n$,
\[
 \frac7{20}R_n\leq\frac1{r_n}-\frac85
 =\frac{8(5/8-r_n)}{5r_n}\leq\frac{8(5/8-r_n)}{5r},
\]
using $1-e^{-R_n}\leq R_n$. Hence $\sum_nR_n<\infty$, so $\mathrm{Pr}((a_1,a_1,\ldots)\mid\omega=2)=\exp(-\sum_nR_n)>0$.
Since $\mu_0(2)>0$, the public belief converges with positive probability to the full-support cascade belief with odds $(5/8,5/8)$. Herding follows from \Cref{lem:learning_limit}.
\end{proof}

\begin{lemma}\label{lem:mlrp_leave_Z0}
Fix $(\lambda_{1,0},\lambda_{2,0})\in Z_0$ with $\lambda_{1,0}>\lambda_{2,0}$, and write $(\lambda_{1,t},\lambda_{2,t}):=T_{a_1}^t(\lambda_{1,0},\lambda_{2,0})$. Then $(\lambda_{1,t},\lambda_{2,t})\in\mathcal W\setminus Z_0$ and $\lambda_{1,t}>\lambda_{2,t}$ for some $t\geq1$.
\end{lemma}

\begin{proof}
By \Cref{lem:mlrp_W_invariant}, the odds remain in $\mathcal W$. While they lie in $Z_0$ with $\lambda_1>\lambda_2$, the cutoffs satisfy $L<U$ and
\[
 \psi_3^{a_2}-\psi_1^{a_2}
 =\frac12\int_L^U[\ell_+(s)-\ell_-(s)]e^{-s}\,\mathrm ds>0.
\]
Thus $\lambda_1/\lambda_2$ strictly increases. Suppose that the public belief does not leave $Z_0$. The ratio stays at least $\lambda_{1,1}/\lambda_{2,1}>1$. For $t\geq1$, the overturning principle gives $\lambda_{1,t}+\lambda_{2,t}\geq1$. Together with $3\lambda_{1,t}/2+\lambda_{2,t}/10<1$, this implies $\lambda_{1,t}<9/14$, $\lambda_{2,t}>5/14$, and $\lambda_{1,t}/\lambda_{2,t}<9/5$. The odds therefore lie in the compact set defined by
\[
 \lambda_1+\lambda_2\geq1,\qquad 3\lambda_1/2+\lambda_2/10\leq1,\qquad \frac{\lambda_1}{\lambda_2}\geq\frac{\lambda_{1,1}}{\lambda_{2,1}}>1.
\]
The integral is also positive when $3\lambda_1/2+\lambda_2/10=1$, where $L=0$. Hence the continuous ratio $\psi_1^{a_1}/\psi_3^{a_1}$ has a minimum greater than $1$ on this set. This forces $\lambda_{1,t}/\lambda_{2,t}$ to grow geometrically, contradicting its bound $9/5$.
\end{proof}

\begin{lemma}\label{lem:mlrp_exceptional_null}
$\mathcal N$ is meager and has Lebesgue measure zero in $\mathbb R^2$.
\end{lemma}

\begin{proof}
By \Cref{SA:mlrp_prelim}, both update maps are real analytic on $Z_0,Z_1,Z_2,Z_1',Z_2'$, which cover $\mathcal W$ up to a null set. The sets $Z_0,Z_1$ are convex. Solving the cutoff equations gives the following continuous map from $\{(L,U):0<L<U\}$ onto $Z_2$:
\[
 \lambda_2=\left[\frac32+\frac9{20}\left(L+\frac{U-L}{1-e^{-(U-L)}}\right)\right]^{-1},
 \qquad
 \lambda_1=\frac{9(U-L)e^L}{2(1-e^{-(U-L)})}\lambda_2.
\]
Thus $Z_2$ is connected, and so are its image $Z_2'$ and the convex set $Z_1'$. By \Cref{lem:analytic_null_preimages}, it suffices to find one point in each region where the Jacobian determinant is nonzero for each action. Since $T_a\circ P=P\circ T_a$, we have $\det DT_a\circ P=\det DT_a$, so it suffices to check $Z_0,Z_1,Z_2$.

The Jacobian $DT_a$ is given by \eqref{eq:counter_update_jacobian}. Implicit differentiation of the cutoff equations yields
\begin{align*}
 \nabla L&=
 \begin{cases}
 -\dfrac{(\ell_+(L),\ell_-(L))}{9\lambda_1/20-\lambda_2\ell_-(L)},&\text{on }Z_0,\\[7pt]
 -\dfrac{(\ell_-(L),\ell_+(L))}{9\lambda_2/20-\lambda_1\ell_-(L)},&\text{on }Z_2,
 \end{cases}\\[4pt]
 \nabla U&=-\frac{(\ell_-(U),\ell_+(U))}{9\lambda_2/20-\lambda_1\ell_-(U)}
 \qquad\text{on }Z_0,Z_1,Z_2.
\end{align*}
Differentiating \eqref{eq:mlrp_tail_probabilities} then gives
\[
 \nabla\psi_\omega^{a_2}=
 \begin{cases}
 f_3(L\mid4-\omega)\nabla L+f_3(U\mid\omega)\nabla U,&\text{on }Z_0,\\
 f_3(U\mid\omega)\nabla U,&\text{on }Z_1,\\
 -f_3(L\mid\omega)\nabla L+f_3(U\mid\omega)\nabla U,&\text{on }Z_2,
 \end{cases}
 \qquad
 \nabla\psi_\omega^{a_1}=-\nabla\psi_\omega^{a_2}.
\]

At the following interior points, the cutoffs are $L=U=\log2$ on $Z_0$, $U=\log2$ on $Z_1$, and $L=\log2,U=2\log2$ on $Z_2$. Substitution into \eqref{eq:counter_update_jacobian} and $2/3<\log2<7/10$ give the indicated bounds:
\[
\renewcommand{\arraystretch}{1.3}
\begin{array}{c|c|c|c}
 \text{region}&(\lambda_1,\lambda_2)&\det DT_{a_1}\in&\det DT_{a_2}\in\\ \hline
 Z_0&\left(\dfrac{20}{31+9\log2},\dfrac{20}{31+9\log2}\right)
 &\left(\dfrac12,1\right)&(-2,-1)\\
 Z_1&\left(1,\dfrac{19}{30+9\log2}\right)
 &(3,4)&\left(\dfrac1{20},\dfrac1{10}\right)\\
 Z_2&\left(\dfrac{120\log2}{10+9\log2},\dfrac{20}{3(10+9\log2)}\right)
 &(4,8)&\left(-\dfrac1{40},-\dfrac1{200}\right)
\end{array}
\]
In particular, both determinants are nonzero in every row.

The rest of the argument is the same as in the proof of \Cref{lem:counter_exceptional_null}, with the five regions in place of $R_0,R_1,R_2$ and \Cref{lem:mlrp_W_invariant} in place of \Cref{lem:counter_W_invariant}.
\end{proof}

\begin{proof}[Proof of \Cref{prop:mlrp}]
Invariance follows from \Cref{lem:mlrp_W_invariant}. Fix a full-support prior in $\mathcal W\setminus\mathcal N$. By \Cref{lem: martingale}, public beliefs converge almost surely to a stationary belief. Nondegenerate limits on the edges $\{1,2\}$ and $\{2,3\}$ are excluded by pairwise distinguishability. A nondegenerate limit on $\{1,3\}$ would make both odds diverge, eventually giving $m\geq(3/2)\min\{\lambda_1,\lambda_2\}\geq1$, contrary to invariance.

The remaining possible limits are the three point masses and full-support beliefs in $C_3$. On $C_3$, $\lambda_1+\lambda_2\geq5/4$, whereas every $a_2$ observation leaves $\lambda_1+\lambda_2\leq1$. Thus convergence to a belief in $C_3$ requires all sufficiently late observations to be $a_1$. Since $\mathcal D$ is never reached from $\mathcal W\setminus\mathcal N$, \Cref{lem:mlrp_leave_Z0,lem:mlrp_positive} instead imply convergence along such a history to $\delta_1$ or $\delta_3$, a contradiction.

Hence the limit is almost surely a point mass. Ties have probability zero, so \Cref{lem:learning_limit} gives adequate learning in every equilibrium. By \Cref{lem:mlrp_exceptional_null}, $\mathcal N$ is meager and null. The coordinate map $(\lambda_1,\lambda_2)\mapsto(\lambda_1,1,\lambda_2)/(1+\lambda_1+\lambda_2)$ is a diffeomorphism from $(0,\infty)^2$ onto the full-support simplex, so the corresponding exceptional priors are also meager and null. This establishes learning for generic full-support priors outside $C_3$. Finally, \Cref{lem:mlrp_symmetric_herding} proves herding for priors in $\mathcal D$.
\end{proof}

\subsection{Gaussian Example}\label{SA:ex_counter_gaussian}

\begin{example}\label{ex:counter_gaussian}
  Let $\Omega = \{1, 2, 3\}$ and $A = \{a_1, a_2\}$. Preferences are the same as in \Cref{ex:counter}. Let the information structure be $F_2(\cdot \mid \omega) \sim N(\omega, 1)$.
\end{example}

In this example, the full-support cascade belief set is $C_2:=\{\mu:\mu(2)>0,\ \mu(2)\leq2e^{-1/2}\sqrt{\mu(1)\mu(3)}\}$. We depict optimal action regions and cascade beliefs in \Cref{fig:counterexample_gaussian_regions}.  Despite pairwise distinguishability, there is herding for every full-support prior in this example.

\begin{proposition}\label{prop:pairwise_no_learning}
  In \Cref{ex:counter_gaussian}, from every full-support prior outside $C_2$ there is a finite history of the form $h = (a_2, \ldots, a_2, a_1, \ldots, a_1)$ that occurs with positive probability and along which the public belief enters $C_2$. Consequently, there is herding for every full-support prior.
\end{proposition}

\subsubsection{Proof of \texorpdfstring{\Cref{prop:pairwise_no_learning}}{Proposition \ref*{prop:pairwise_no_learning}}}

For a full-support belief $\mu$, write $x=\log[\mu(1)/\mu(3)]$, $y=\log[\mu(2)/\sqrt{\mu(1)\mu(3)}]$, and $\bar y=\log(2/\sqrt e)$. Here $|x|$ measures asymmetry between states $1$ and $3$, while $y$ measures the relative weight on state $2$; the cascade region $C_2$ is given by $y\leq\bar y$. Let $(x_t,y_t)$ denote these coordinates after $t$ observations, with $t=0$ at the start of each block. The two block estimates below are proved in \Cref{SA:gaussian_a1,SA:gaussian_a2}.

\begin{proposition}[The $a_1$ block]\label[proposition]{prop:a1_sketch_ylogy}
There are constants $C>0$ and $Y_*>1$ such that, for every $Y\geq Y_*$,
if $\bar y<y_0\leq Y$ and $|x_0|\leq e^{-CY\log Y}$, then starting from $(x_0,y_0)$, repeated $a_1$ observations make $y\leq\bar y$ in at most $CY$ steps.
\end{proposition}

\begin{proposition}[The $a_2$ block]\label[proposition]{prop:a2_exp(y^2)}
Fix a full-support prior with $y_0>\bar y$. Along repeated $a_2$
observations, $y_t\to\infty$, and there is a constant $\kappa>0$, possibly depending on the prior, such that $|x_t|\leq e^{-\kappa y_t^2}$ for all sufficiently large $t$.
\end{proposition}

\begin{proof}[Proof of \Cref{prop:pairwise_no_learning}]
Fix any full-support prior $\mu_0$. If $y_0\leq\bar y$, the initial belief is already in $C_2$. Otherwise, let $C,Y_*$ be as in \Cref{prop:a1_sketch_ylogy} and let
$\kappa$ be as in \Cref{prop:a2_exp(y^2)}. Choose $T$ sufficiently large
that $Y:=y_T\geq Y_*$ and $|x_T|\leq e^{-\kappa Y^2}\leq e^{-CY\log Y}$. This is possible because $y_T\to\infty$ and $Y^2/(Y\log Y)\to\infty$. By \Cref{prop:a1_sketch_ylogy}, a subsequent $a_1$ block of length at most $CY$ makes $y\leq\bar y$. We thus have a finite history along which the public belief enters $C_2$. This history has positive probability. On $C_2$, $a_1$ is the unique optimal action for a.e.\ signal, so in any equilibrium, the public belief does not update and stays at a full-support belief without adequate knowledge. Since $\mu_0$ is arbitrary, there is herding for every full-support prior.
\end{proof}

\subsubsection{Preliminaries}
Let $\Phi$ and $\phi$ denote the standard normal distribution and density functions. For $y\geq\bar y$, define $\theta(y)=\mathrm{arcosh}\left(\frac{\sqrt e}{2}e^y\right)$; in particular, $\theta(\bar y)=0$.

\paragraph{Belief Updating.}
Writing $f=f_2$, at a full-support public belief $\mu$ the agent strictly prefers $a_1$ if and only if
\begin{align*}
    &\mu(1) f(s\mid1) + \mu(3) f(s\mid3) > \mu(2) f(s\mid2) \\
    \iff& \mu(1) \exp\left(-\frac{(s-1)^2}{2}\right) + \mu(3) \exp\left(-\frac{(s-3)^2}{2}\right) > \mu(2) \exp\left(-\frac{(s-2)^2}{2}\right)\\
    \iff& \sqrt{\mu(1)\mu(3)} \left( \exp\left(\frac{x}{2} + 2 - s\right) + \exp\left(-\frac{x}{2} + s - 2\right) \right) > \mu(2) \exp\left(\frac{1}{2}\right) \\
    \iff& \cosh\left(\frac{x}{2} + 2 - s\right) > \frac{\sqrt{e}}{2} \exp(y).
\end{align*}
Thus the cascade region is $\{y\leq\bar y\}$.

Outside this region, $a_2$ is chosen for signals in $[L(x,y),U(x,y)]$, up to endpoint ties, where
\[
 L(x,y)=2+\frac{x}{2}-\theta(y),\qquad
 U(x,y)=2+\frac{x}{2}+\theta(y).
\]
Ties have probability zero in every state, so tie-breaking does not affect action probabilities. Writing $\psi_\omega^a(x,y)$ for the probability
of action $a$ in state $\omega\in\{1,2,3\}$, we obtain
\begin{equation}\label{eq:gaussian-action-probabilities}
 \begin{aligned}
 \psi_\omega^{a_1}(x,y)&=\Phi(L(x,y)-\omega)+1-\Phi(U(x,y)-\omega),\\
 \psi_\omega^{a_2}(x,y)&=\Phi(U(x,y)-\omega)-\Phi(L(x,y)-\omega).
 \end{aligned}
\end{equation}
Both probabilities are strictly positive when $y>\bar y$. After $a$ is observed, the public belief is updated according to Bayes' rule:
\[
 \Delta_x^a(x,y)=\log\frac{\psi_1^a(x,y)}{\psi_3^a(x,y)},\qquad
 \Delta_y^a(x,y)=\log\frac{\psi_2^a(x,y)}
 {\sqrt{\psi_1^a(x,y)\psi_3^a(x,y)}}.
\]
After observing $a$, we denote the updated coordinates by $x^+=x+\Delta_x^a(x,y)$ and $y^+=y+\Delta_y^a(x,y)$. On $\{y\leq\bar y\}$, action $a_1$ is chosen almost surely and the public belief remains unchanged.

\paragraph{Elementary Estimates.} Throughout, $C$ and $k$ denote positive constants that may change from line to line; subscripts indicate dependence, and constants do not depend on $t$ or $Y$ unless stated.

\begin{lemma}\label{lem:prelim_estimates}
\begin{enumerate}[label=(\roman*)]
\item For $y>\bar y$, $y+\frac12-\log2\le \theta(y)\le y+\frac12$
and $\theta'(y)=\coth \theta(y)$. Thus $\theta(y)\sim y$ as $y\to\infty$,
and $\theta'(y)\le2$ for sufficiently large $\theta(y)$.

\item For $0\le w<1$, $\log(1-w)\le-w$; for $0\le w\le1/2$,
$w\le-\log(1-w)\le2w$.

\item For some universal $C>0$ and all $z\in\mathbb R$,
\[
 \frac{\phi(z)}{1-\Phi(z)}
 \le C(1+z_+)\le C(1+|z|), \qquad z_+=\max\{z,0\}.
\]
Also, $1-\Phi(z)\le\phi(z)/z$ for $z>0$.

\item For $z,r>0$,
$[1-\Phi(z+r)]/[1-\Phi(z)]\le e^{-zr-r^2/2}$.
\end{enumerate}
\end{lemma}

\begin{proof}
(i) Write $\theta=\theta(y)>0$, so that $\cosh\theta=e^{y+1/2}/2$. Since $e^\theta/2\le\cosh\theta\le e^\theta$, taking logarithms gives $\theta-\log2\le y+\frac12-\log2\le\theta$, which is the stated bound; in particular, $\theta(y)/y\to1$. Differentiating $\cosh\theta(y)=e^{y+1/2}/2$ gives $\sinh\theta(y)\,\theta'(y)=\cosh\theta(y)$, so $\theta'(y)=\coth\theta(y)$, which tends to $1$ as $\theta(y)\to\infty$.

(ii) For $0\le w<1$, $-\log(1-w)=\int_0^w(1-v)^{-1}\,\mathrm{d}v$. The integrand is at least $1$, which gives $-\log(1-w)\ge w$. If $w\le1/2$, the integrand is also at most $2$ on $[0,w]$, which gives $-\log(1-w)\le2w$.

(iii) Write $Q=1-\Phi$. For $z>0$, the Mills-ratio bounds give
\[
 \frac{z}{1+z^2}\phi(z)\le Q(z)\le\frac{\phi(z)}{z}.
\]
The upper bound is the second claim. The lower bound gives $\phi(z)/Q(z)\le z+z^{-1}\le z+1$ for $z\ge1$. For $z\le1$, $\phi(z)\le\phi(0)$ and $Q(z)\ge Q(1)$ give $\phi(z)/Q(z)\le\phi(0)/Q(1)$. Hence $C=\max\{1,\phi(0)/Q(1)\}$ works.

(iv) Fix $z,r>0$. Since $\phi(v+r)=\phi(v)e^{-vr-r^2/2}$,
\[
 1-\Phi(z+r)
 =\int_z^\infty\phi(v)e^{-vr-r^2/2}\,\mathrm{d}v
 \le e^{-zr-r^2/2}\bigl(1-\Phi(z)\bigr),
\]
because $v\ge z$ on the domain of integration. Dividing by $1-\Phi(z)>0$ gives the bound.
\end{proof}

\subsubsection{The \texorpdfstring{$a_1$}{a1} block}\label{SA:gaussian_a1}

We prove that a block of $a_1$ observations brings $y$ to or below $\bar y$, provided that $x$ is initially sufficiently small.  The estimates are stated with explicit dependence on $Y$, which is the upper bound on $y$ at the start of the $a_1$ block.  

\begin{lemma}\label{lem:a1-centered-journal}
For every $\varepsilon>0$ there exists $m_\varepsilon>0$, independent of $Y$, such that $\Delta_y^{a_1}(0,y)\leq -m_\varepsilon$ for all $y\geq \bar y+\varepsilon$.
\end{lemma}

\begin{proof}
Let $\theta=\theta(y)$. At $x=0$,
\[
 \psi_2^{a_1}(0,y)=2\Phi(-\theta), \quad \psi_1^{a_1}(0,y)=\psi_3^{a_1}(0,y)=\Phi(-\theta+1)+\Phi(-\theta-1).
\]
Define $H(\theta):=[\Phi(-\theta+1)+\Phi(-\theta-1)]/[2\Phi(-\theta)]$, so $-\Delta_y^{a_1}(0,y)=\log H(\theta(y))$. Since $\theta(y)$ is increasing, it suffices to bound $\log H$ away from zero on $[\theta_\varepsilon,\infty)$, where $\theta_\varepsilon:=\theta(\bar y+\varepsilon)>0$.

For every $\theta>0$, symmetry of $\phi$ gives
\[
 \Phi(-\theta+1)+\Phi(-\theta-1)-2\Phi(-\theta)
 =\int_0^1[\phi(\theta-v)-\phi(\theta+v)]\,\mathrm{d}v>0,
\]
because $|\theta-v|<\theta+v$ for $v>0$ and $\phi$ strictly decreases with the absolute value of its argument. Thus $\log H(\theta)>0$. For $\theta\geq2$, monotonicity of $\phi$ on $[0,\infty)$ and $\phi'(z)=-z\phi(z)$ give
\begin{align*}
 \Phi(-\theta+1)&\geq\int_{\theta-1}^{\theta}\phi(z)\,\mathrm{d}z\geq\phi(\theta),\\
 \Phi(-\theta)&=\int_\theta^\infty\phi(z)\,\mathrm{d}z
 \leq\int_\theta^\infty\frac{z}{\theta}\phi(z)\,\mathrm{d}z=\frac{\phi(\theta)}{\theta}.
\end{align*}
Hence $H(\theta)\geq\Phi(-\theta+1)/[2\Phi(-\theta)]\geq \theta/2$, so $\log H(\theta)\geq\log2$ for $\theta\geq4$.

If $\theta_\varepsilon\geq4$, take $m_\varepsilon=\log2$. Otherwise, continuity and strict positivity of $\log H$ on the compact interval $[\theta_\varepsilon,4]$ imply
\[
 m_\varepsilon:=\min\left\{\min_{\theta\in[\theta_\varepsilon,4]}\log H(\theta),\log2\right\}>0.
\]
In either case, $\log H(\theta)\geq m_\varepsilon$ for all $\theta\geq \theta_\varepsilon$, and $m_\varepsilon$ does not depend on $Y$.
\end{proof}

\begin{lemma}\label{lem:a1-derivatives-journal}
There is a constant $C<\infty$ such that, for every $Y\geq\bar y$,
\[
 \sup_{\substack{|x|\leq1\\ \bar y\leq y\leq Y}} \left( \left|\partial_x\Delta_x^{a_1}(x,y)\right| + \left|\partial_x\Delta_y^{a_1}(x,y)\right| \right) \leq C(1+Y).
\]
\end{lemma}

\begin{proof}
We give the estimate for $\partial_x\Delta_y^{a_1}$; the estimate for $\partial_x\Delta_x^{a_1}$ is similar.  Let $L=L(x,y)$ and $U=U(x,y)$ to simplify notation. Since $ \Delta_y^{a_1}=\log\psi_2^{a_1}-\frac12\log\psi_1^{a_1}-\frac12\log\psi_3^{a_1}$,
we have
\begin{align*}
|\partial_x\Delta_y^{a_1}(x,y)| &\leq \frac12\frac{|\phi(L-2)-\phi(U-2)|}{\Phi(L-2)+1-\Phi(U-2)}  \\
&\quad +\frac14\frac{|\phi(L-1)-\phi(U-1)|}{\Phi(L-1)+1-\Phi(U-1)}  \\
&\quad +\frac14\frac{|\phi(L-3)-\phi(U-3)|}{\Phi(L-3)+1-\Phi(U-3)} .
\end{align*}
Each ratio is bounded by the corresponding sum of two hazard ratios.  For example,
\begin{align*}
\frac{|\phi(L-2)-\phi(U-2)|}{\Phi(L-2)+1-\Phi(U-2)} &\leq \frac{\phi(L-2)}{\Phi(L-2)}+\frac{\phi(U-2)}{1-\Phi(U-2)} \\
&= \frac{\phi(2-L)}{1-\Phi(2-L)}+\frac{\phi(U-2)}{1-\Phi(U-2)} \\
&\leq C\bigl(1+|2-L|+|U-2|\bigr),
\end{align*}
where the last inequality comes from \Cref{lem:prelim_estimates}(iii). If $|x|\leq1$ and $y\leq Y$, then
\begin{align*}
 |\omega-L(x,y)|+|U(x,y)-\omega| &= |\omega-2-x/2+\theta(y)|+|\omega-2-x/2-\theta(y)| \\
 &\leq 2|\omega-2| + |x| + 2\theta(y)\\
 & \leq C(1+\theta(Y))\leq C(1+Y)
\end{align*}
for $\omega=1,2,3$, with a suitable choice of $C$. This proves the bound for $\partial_x\Delta_y^{a_1}$.  The formula
\[
 \partial_x\Delta_x^{a_1} =\frac{\partial_x\psi_1^{a_1}}{\psi_1^{a_1}} -\frac{\partial_x\psi_3^{a_1}}{\psi_3^{a_1}}
\]
and the same hazard-ratio estimate give the bound for $\partial_x\Delta_x^{a_1}$.
\end{proof}

\begin{lemma}\label{lem:terminal-journal}
There exist $\varepsilon_*>0$ and $\delta_*>0$ such that if $|x|\leq\delta_*$ and $\bar y<y\leq\bar y+\varepsilon_*$, then $y+\Delta_y^{a_1}(x,y)<\bar y$.
\end{lemma}

\begin{proof}
Write $y(\theta)=\bar y+\log\cosh \theta$ and $\Gamma(\theta,x):=\log\cosh \theta+\Delta_y^{a_1}(x,y(\theta))=y^+-\bar y$. At $\theta=0$ the belief is stationary, so $\Gamma(0,x)=0$. Writing $\psi_\omega=\psi_\omega^{a_1}(x,y(\theta))$, differentiation gives
\[
 \partial_\theta \Gamma(\theta,x)=\tanh \theta+\frac{\partial_\theta\psi_2}{\psi_2}
 -\frac12\left(\frac{\partial_\theta\psi_1}{\psi_1}+\frac{\partial_\theta\psi_3}{\psi_3}\right).
\]
At $\theta=0$, each $\psi_\omega=1$, and the right derivatives are
\begin{align*}
 \left.\partial_\theta\psi_1\right|_{\theta=0}=-2\phi(1+x/2), \quad
 \left.\partial_\theta\psi_2\right|_{\theta=0}=-2\phi(x/2), \quad
 \left.\partial_\theta\psi_3\right|_{\theta=0}=-2\phi(1-x/2).
\end{align*}
Since $\tanh0=0$, it follows that
\[
 \partial_\theta \Gamma(0,x)=-2\phi(x/2)+\phi(1+x/2)+\phi(1-x/2).
\]
In particular, $\partial_\theta \Gamma(0,0)=-2(\phi(0)-\phi(1))<0$. By continuity of $\partial_\theta \Gamma$, there exist $\theta_*,\delta_*,\gamma>0$ such that $\partial_\theta \Gamma(\theta,x)\leq-\gamma$ whenever $0\leq \theta\leq \theta_*$ and $|x|\leq\delta_*$. Thus, for $0<\theta\leq \theta_*$ and $|x|\leq\delta_*$,
\[
 \Gamma(\theta,x)=\int_0^\theta\partial_v\Gamma(v,x)\,\mathrm{d}v\leq-\gamma \theta<0.
\]
Set $\varepsilon_*:=\log\cosh \theta_*>0$. Then the condition $\bar y<y\leq\bar y+\varepsilon_*$ is equivalent to $0<\theta(y)\leq \theta_*$.
\end{proof}

\begin{samepage}
\begin{proposition}\label{prop:a1_exp(ylogy)}
Fix $0<\varepsilon\leq\varepsilon_*$ and $Y\geq\bar y+\varepsilon$. There exist $\delta_Y,\delta_Y^{\mathrm{init}}>0$ and $N_Y<\infty$ satisfying
\[
 \delta_Y\geq\frac{k_\varepsilon}{1+Y}, \quad N_Y=O_\varepsilon(Y), \quad \log\frac1{\delta_Y^{\mathrm{init}}}=O_\varepsilon(Y\log Y),
\]
such that, whenever $\bar y+\varepsilon\leq y_0\leq Y$ and $|x_0|\leq\delta_Y^{\mathrm{init}}$, repeated $a_1$ observations yield $y_\tau\leq\bar y$ for some $\tau\leq N_Y$, with $|x_t|\leq\delta_Y$ for all $0\leq t\leq\tau$.
\end{proposition}
\end{samepage}

Here $\delta_Y$ bounds the asymmetry throughout the block, while $\delta_Y^{\mathrm{init}}$ bounds its initial value.

\begin{proof}
By the mean-value theorem and \Cref{lem:a1-centered-journal,lem:a1-derivatives-journal}, for $|x|\leq1$ and $\bar y+\varepsilon\leq y\leq Y$,
\[
 \Delta_y^{a_1}(x,y)\leq\Delta_y^{a_1}(0,y)+C(1+Y)|x|\leq-m_\varepsilon+C(1+Y)|x|.
\]
Set $\delta_Y:=\min\{1,\delta_*,m_\varepsilon/[2C(1+Y)]\}$. Then $\Delta_y^{a_1}(x,y)\leq-m_\varepsilon/2$ whenever $|x|\leq\delta_Y$ and $\bar y+\varepsilon\leq y\leq Y$. The same derivative bound and $\Delta_x^{a_1}(0,y)=0$ imply
\begin{equation}\label{eq:a1-x-growth}
 |x+\Delta_x^{a_1}(x,y)|\leq\Lambda_Y|x|, \quad \Lambda_Y:=1+C(1+Y),
\end{equation}
for $|x|\leq\delta_Y$ and $\bar y\leq y\leq Y$. Define
\[
 n_Y:=\left\lceil\frac{2(Y-\bar y-\varepsilon)}{m_\varepsilon}\right\rceil, \quad
 N_Y:=n_Y+1, \quad \delta_Y^{\mathrm{init}}:=\delta_Y\Lambda_Y^{-N_Y}.
\]
Let $\tau_0:=\inf\{t\geq0:y_t\leq\bar y+\varepsilon\}$. If $|x_0|\leq\delta_Y^{\mathrm{init}}$, induction using \eqref{eq:a1-x-growth} and the decrease in $y$ gives
\[
 |x_t|\leq\Lambda_Y^t|x_0|\leq\Lambda_Y^{N_Y}\delta_Y^{\mathrm{init}}=\delta_Y
\]
for all $t\leq\min\{\tau_0,N_Y\}$, while $y_{t+1}\leq y_t-m_\varepsilon/2$ for $t<\min\{\tau_0,N_Y\}$. If $\tau_0>n_Y$, then $y_{n_Y}\leq Y-n_Ym_\varepsilon/2\leq\bar y+\varepsilon$, a contradiction. Hence $\tau_0\leq n_Y$. If $y_{\tau_0}\leq\bar y$, take $\tau=\tau_0$. Otherwise, $\bar y<y_{\tau_0}\leq\bar y+\varepsilon\leq\bar y+\varepsilon_*$ and $|x_{\tau_0}|\leq\delta_Y\leq\delta_*$. By \Cref{lem:terminal-journal}, one further $a_1$ observation yields $y_{\tau_0+1}<\bar y$; the bound on $x_{\tau_0+1}$ follows from \eqref{eq:a1-x-growth} and the same induction. Taking $\tau=\tau_0+1\leq N_Y$ proves the claim in this case.

Finally, since $\delta_*$ is fixed, $\delta_Y\geq k_\varepsilon/(1+Y)$. Also, $N_Y=O_\varepsilon(Y)$ and $\log\Lambda_Y=O(\log Y)$. Hence $\log(1/\delta_Y^{\mathrm{init}})=\log(1/\delta_Y)+N_Y\log\Lambda_Y\leq C_\varepsilon Y\log Y$ for all sufficiently large $Y$.
\end{proof}

\begin{proof}[Proof of \Cref{prop:a1_sketch_ylogy}]
Fix $\varepsilon=\varepsilon_*$ in \Cref{prop:a1_exp(ylogy)}. Its bounds on $N_Y$ and $\log(1/\delta_Y^{\mathrm{init}})$ allow us to choose $C\geq1$ such that $N_Y\leq CY$ and $\delta_Y^{\mathrm{init}}\geq e^{-CY\log Y}$ for all sufficiently large $Y$. Since $e^{-CY\log Y}\to0$, we may choose $Y_*>\max\{1,\bar y+\varepsilon_*\}$ large enough that both bounds hold and $e^{-CY\log Y}\leq\delta_*$ for every $Y\geq Y_*$. These choices are independent of $(x_0,y_0)$.

Now let $Y\geq Y_*$, $\bar y<y_0\leq Y$, and $|x_0|\leq e^{-CY\log Y}$. If $y_0\leq\bar y+\varepsilon_*$, then $|x_0|\leq\delta_*$, so \Cref{lem:terminal-journal} gives $y_1<\bar y$ after one $a_1$ observation, and $1\leq CY$. Otherwise, $\bar y+\varepsilon_*<y_0\leq Y$ and $|x_0|\leq\delta_Y^{\mathrm{init}}$. By \Cref{prop:a1_exp(ylogy)}, repeated $a_1$ observations give $y_\tau\leq\bar y$ for some $\tau\leq N_Y\leq CY$.
\end{proof}

\subsubsection{The \texorpdfstring{$a_2$}{a2} block}\label{SA:gaussian_a2}

Starting from $y_0>\bar y$, repeated $a_2$ observations send $y_t$ to infinity and eventually reduce asymmetry to $|x_t|\leq e^{-\kappa y_t^2}$, for some $\kappa>0$ that may depend on the initial prior. With $Y=y_t$, this bound is eventually smaller than $e^{-CY\log Y}$, as required by the $a_1$ block. We first establish monotonicity, then show that the path enters a region where $\theta(y)|x|$ is bounded, and finally derive the contraction rate within that region.

\begin{lemma}[Monotonicity]\label{lem:a2-global-journal}
If $y_0>\bar y$, repeated $a_2$ observations satisfy $|x_{t+1}|\leq|x_t|\leq|x_0|$ and $y_{t+1}>y_t$ for all $t$, with $y_t\to\infty$ as $t\to\infty$.
\end{lemma}

\begin{proof}
For $\theta=\theta(y)>0$, define $G_\theta(r):=\Phi(\theta+r)+\Phi(\theta-r)-1$. Then
\[
 \psi_1^{a_2}=G_\theta(1+x/2), \quad \psi_2^{a_2}=G_\theta(x/2), \quad \psi_3^{a_2}=G_\theta(x/2-1).
\]
Since $G_\theta$ is even and decreasing in $|r|$, $x>0$ implies $\psi_1^{a_2}\leq\psi_3^{a_2}$ and $\Delta_x^{a_2}\leq0$. To show that the update cannot cross zero, we next establish $0\leq x+\Delta_x^{a_2}(x,y)\leq x$ for $x>0$.
\[
 G_\theta(r)=\int_{-\theta}^\theta\phi(v-r)\,\mathrm{d}v
 =e^{-r^2/2}\int_{-\theta}^\theta\phi(v)e^{rv}\,\mathrm{d}v,
\]
so $e^{r^2/2}G_\theta(r)=2\int_0^\theta\phi(v)\cosh(rv)\,\mathrm{d}v$ is increasing in $|r|$. Hence, for $x>0$,
\[
 e^{-x}=\exp\left\{-\frac{(1+x/2)^2-(x/2-1)^2}{2}\right\}
 \leq\frac{\psi_1^{a_2}}{\psi_3^{a_2}}\leq1,
\]
which gives $0\leq x+\Delta_x^{a_2}(x,y)\leq x$. The case $x<0$ is symmetric, and $x=0$ remains fixed by symmetry. Thus $|x_t|\leq|x_0|$ for all $t$.

To verify strict log-concavity of $G_\theta$, let $p_r(v)=\phi(v-r)/G_\theta(r)$ on $[-\theta,\theta]$, with mean $m_r\in(-\theta,\theta)$. Differentiation under the integral and integration by parts give
\[
 (\log G_\theta)''(r)=\operatorname{Var}_{p_r}(v)-1
 =-\frac{(\theta-m_r)\phi(\theta-r)+(\theta+m_r)\phi(\theta+r)}{G_\theta(r)}<0.
\]
Since $x/2=\frac12(1+x/2)+\frac12(x/2-1)$,
\[
 \log\psi_2^{a_2}\geq\frac12\log\psi_1^{a_2}+\frac12\log\psi_3^{a_2},
\]
strictly for finite $\theta$. Thus $\Delta_y^{a_2}(x,y)>0$ for $y>\bar y$. If $y_t$ were bounded above, its increase and $|x_t|\leq|x_0|$ would keep the path in a compact subset of $\{y>\bar y\}$. Continuity would then give a strictly positive minimum for $\Delta_y^{a_2}$ on that set, contradicting boundedness. Hence $y_t\to\infty$.
\end{proof}

\begin{lemma}\label{lem:a2-entrance-journal}
Fix an initial prior $\mu_0$ with $y_0>\bar y$. There exist $0<M<\infty$ and $\theta_0=\theta_0(\mu_0,M)$ such that, for every $\theta^*\geq \theta_0$, repeated $a_2$ observations eventually yield $\theta(y)|x|\leq M$ and $\theta(y)\geq \theta^*$.
\end{lemma}

\begin{proof}
Fix $\eta\in(0,1/2)$. Choose $M$ large and then $\theta_0>\max\{2,1+|x_0|/2\}$ large so that, whenever $\theta\geq \theta_0$, $0<x\leq|x_0|$, and $\theta x\geq M$,
\begin{equation}\label{eq:a2-tail-domination}
 \psi_2^{a_1}(x,y)\leq\eta\psi_1^{a_1}(x,y), \quad \psi_3^{a_1}(x,y)\leq\eta\psi_1^{a_1}(x,y).
\end{equation}
Indeed, \Cref{lem:prelim_estimates}(iv) gives
\begin{align*}
 &\frac{1-\Phi(\theta+1+x/2)}{1-\Phi(\theta-1-x/2)}
 \leq\frac{1-\Phi(\theta+x/2)}{1-\Phi(\theta-1-x/2)}
 \leq\frac{1-\Phi(\theta-1+x/2)}{1-\Phi(\theta-1-x/2)}
 \leq e^{-(\theta-1)x},\\
 &\frac{1-\Phi(\theta+1-x/2)}{1-\Phi(\theta-1-x/2)}
 \leq\frac{1-\Phi(\theta-x/2)}{1-\Phi(\theta-1-x/2)}
 \leq e^{(1+x-2\theta)/2}.
\end{align*}
For $\theta\geq2$ and $\theta x\geq M$, $e^{-(\theta-1)x}\leq e^{-M/2}$, while $e^{(1+x-2\theta)/2}\leq e^{(1+|x_0|-2\theta)/2}$. Choosing $M$ and then $\theta_0$ so that these bounds are each at most $\eta/2$, and summing the two tails in each action probability, proves \eqref{eq:a2-tail-domination}.

Let $\varepsilon_\omega:=\psi_\omega^{a_1}(x,y)=1-\psi_\omega^{a_2}(x,y)$. Increasing $\theta_0$ if necessary makes $\varepsilon_\omega\leq1/2$ uniformly for $|x|\leq|x_0|$. In the stated region, \eqref{eq:a2-tail-domination} gives $\varepsilon_2,\varepsilon_3\leq\eta\varepsilon_1$, so the logarithmic bounds in \Cref{lem:prelim_estimates}(ii) imply
\begin{align*}
 -\Delta_x^{a_2}(x,y)
 &=\log(1-\varepsilon_3)-\log(1-\varepsilon_1)
 \geq\varepsilon_1-2\varepsilon_3\geq(1-2\eta)\varepsilon_1,\\
 \Delta_y^{a_2}(x,y)
 &=\log(1-\varepsilon_2)-\frac12\log(1-\varepsilon_1)-\frac12\log(1-\varepsilon_3)\\
 &\leq-\frac12\log(1-\varepsilon_1)-\frac12\log(1-\varepsilon_3)
 \leq\varepsilon_1+\varepsilon_3\leq(1+\eta)\varepsilon_1.
\end{align*}
Thus
\begin{equation}\label{eq:a2-descent-comparison}
 -\Delta_x^{a_2}(x,y)\geq\rho\Delta_y^{a_2}(x,y), \quad \rho:=\frac{1-2\eta}{1+\eta}>0,
\end{equation}
where $\rho$ is independent of $t,\theta,x,y$. The case $x<0$ is symmetric. By \Cref{lem:a2-global-journal}, the update cannot cross zero. Consequently, whenever $\theta(y_t)\geq \theta_0$ and $\theta(y_t)|x_t|\geq M$,
\begin{equation}\label{eq:a2-linear-descent}
 |x_{t+1}|\leq|x_t|-\rho(y_{t+1}-y_t).
\end{equation}

Fix $\theta^*\geq \theta_0$. By \Cref{lem:a2-global-journal}, $y_t\to\infty$, so there is $t_0$ such that $\theta(y_t)\geq \theta^*$ for all $t\geq t_0$. If $\theta(y_t)|x_t|\geq M$ for every $t\geq t_0$, summing \eqref{eq:a2-linear-descent} gives
\[
 |x_T|\leq|x_{t_0}|-\rho(y_T-y_{t_0})
\]
for every $T>t_0$, a contradiction as $y_T\to\infty$. Hence some $t\geq t_0$ satisfies $\theta(y_t)|x_t|<M$ and $\theta(y_t)\geq \theta^*$.
\end{proof}

We next consider beliefs satisfying $\theta(y)|x|\leq M$.

\begin{lemma}\label{lem:a2-tube-journal}
Fix $0\leq M<\infty$. There exist $\theta_1(M)<\infty$ and $C,C_M>0$ such that, whenever $\theta:=\theta(y)\geq \theta_1(M)$ and $\theta|x|\leq M$, the $a_2$ update satisfies
\begin{align}
 |x^+|&\leq(1-C\phi(\theta-1))|x|,\label{eq:tube-x-journal}\\
 0<y^+-y&\leq C_M\frac{\phi(\theta-1)}{\theta},\label{eq:tube-y-journal}\\
 \theta(y^+)|x^+|&\leq \theta(y)|x|.\label{eq:tube-inv-journal}
\end{align}
\end{lemma}

\begin{proof}
Write $Q=1-\Phi$ and suppress $(x,y)$ in the action probabilities. For $x>0$, \Cref{lem:a2-global-journal} gives $x+\Delta_x^{a_2}(x,y)\geq0$. Since $\Delta_x^{a_2}(0,y)=0$, the mean-value theorem therefore gives \eqref{eq:tube-x-journal} once we show
\begin{equation}\label{eq:a2-deltax-derivative}
 \partial_x\Delta_x^{a_2}(x,y)\leq-C\phi(\theta-1)
\end{equation}
for $x>0$ satisfying $\theta x\leq M$; the case $x<0$ follows by symmetry. Direct differentiation gives
\[
 \partial_x\Delta_x^{a_2}(x,y)
 =\frac12\left(\frac{\phi(\theta+1+x/2)-\phi(\theta-1-x/2)}{\psi_1^{a_2}}
 -\frac{\phi(\theta-1+x/2)-\phi(\theta+1-x/2)}{\psi_3^{a_2}}\right).
\]
For large $\theta$ with $\theta|x|\leq M$, we have $|x|<1$ and
\[
 \frac{\phi(\theta+1+x/2)}{\phi(\theta-1-x/2)}=e^{-(2+x)\theta}\leq\frac12, \quad
 \frac{\phi(\theta+1-x/2)}{\phi(\theta-1+x/2)}=e^{-(2-x)\theta}\leq\frac12.
\]
Since $0<\psi_\omega^{a_2}\leq1$, it follows that
\begin{align*}
 \partial_x\Delta_x^{a_2}(x,y)
 &\leq-\frac12\bigl[\phi(\theta-1-x/2)-\phi(\theta+1+x/2)\\
 &\hspace{5em}+\phi(\theta-1+x/2)-\phi(\theta+1-x/2)\bigr]\\
 &\leq-\frac14\bigl[\phi(\theta-1-x/2)+\phi(\theta-1+x/2)\bigr]\\
 &=-\frac12\phi(\theta-1)e^{-x^2/8}\cosh((\theta-1)x/2)\leq-C\phi(\theta-1),
\end{align*}
where the last bound uses $|x|<1$ and $\cosh\geq1$. This proves \eqref{eq:a2-deltax-derivative} and hence \eqref{eq:tube-x-journal}.

For \eqref{eq:tube-y-journal}, increase $\theta_1(M)$ so that $\psi_\omega^{a_1}\leq1/2$ whenever $\theta\geq\theta_1(M)$ and $\theta|x|\leq M$; this is possible because $|x|\leq M/\theta$. Using $\psi_\omega^{a_2}=1-\psi_\omega^{a_1}$ and the logarithmic bounds then gives
\begin{align*}
 \Delta_y^{a_2}(x,y)
 &=\log(1-\psi_2^{a_1})-\frac12\log(1-\psi_1^{a_1})-\frac12\log(1-\psi_3^{a_1})\\
 &\leq C(\psi_1^{a_1}+\psi_3^{a_1})\\
 &=C\bigl[Q(\theta-1-x/2)+Q(\theta+1+x/2)+Q(\theta-1+x/2)+Q(\theta+1-x/2)\bigr]\\
 &\leq C\bigl[Q(\theta-1-|x|/2)+Q(\theta-1+|x|/2)\bigr]\\
 &\leq C\left[\frac{\phi(\theta-1-|x|/2)}{\theta-1-|x|/2}+\frac{\phi(\theta-1+|x|/2)}{\theta-1+|x|/2}\right]\\
 &\leq C\frac{\phi(\theta-1-|x|/2)+\phi(\theta-1+|x|/2)}{\theta-1-|x|/2}\\
 &=C\frac{\phi(\theta-1)}{\theta-1-|x|/2}e^{-x^2/8}\,2\cosh((\theta-1)|x|/2)
 \leq C_M\frac{\phi(\theta-1)}{\theta}.
\end{align*}
The first inequality uses \Cref{lem:prelim_estimates}(ii); the second uses monotonicity of $Q$, enlarging $C$; and the third uses the Mills-ratio bound in \Cref{lem:prelim_estimates}(iii). The last bound follows from $(\theta-1)|x|/2\leq \theta|x|/2\leq M/2$ and $\theta-1-|x|/2\geq \theta/2$ for large $\theta$. Positivity of $y^+-y$ follows from \Cref{lem:a2-global-journal}.

Finally, the mean-value theorem and $\theta'(y)\leq2$ for large $\theta$ (\Cref{lem:prelim_estimates}(i)) give, for some $\tilde y\in(y,y^+)$,
\[
 \theta(y^+)-\theta(y)=\theta'(\tilde y)\Delta_y^{a_2}(x,y)\leq C_M\frac{\phi(\theta-1)}{\theta}.
\]
Increase $\theta_1(M)$, if necessary, so that $C_M/\theta^2\leq C$ and $C\phi(\theta-1)\leq1/2$ for every $\theta\geq\theta_1(M)$. Then
\begin{align*}
 \theta(y^+)|x^+|
 &\leq\theta|x|\left(1+C_M\frac{\phi(\theta-1)}{\theta^2}\right)(1-C\phi(\theta-1))\\
 &\leq\theta|x|(1+C\phi(\theta-1))(1-C\phi(\theta-1))\\
 &=\theta|x|\bigl(1-C^2\phi(\theta-1)^2\bigr)\leq\theta|x|.
\end{align*}
The first inequality combines the preceding bound on $\theta(y^+)-\theta(y)$ with \eqref{eq:tube-x-journal}; the second uses $C_M/\theta^2\leq C$ and $1-C\phi(\theta-1)\geq0$. Thus the update preserves $\theta(y)|x|\leq M$, proving \eqref{eq:tube-inv-journal}.
\end{proof}

\begin{proof}[Proof of \Cref{prop:a2_exp(y^2)}]
By \Cref{lem:a2-entrance-journal}, choose its constant $M<\infty$ and a finite time $t_0$ along repeated $a_2$ observations such that $\theta(y_{t_0})|x_{t_0}|\leq M$ and $\theta(y_{t_0})\geq \theta_1(M)$. By \Cref{lem:a2-tube-journal}, $\theta(y_t)|x_t|\leq M$ and $\theta(y_t)\geq\theta_1(M)$ for all $t\geq t_0$. Set $\theta_t:=\theta(y_t)$. By \Cref{lem:a2-global-journal}, $y_t\to\infty$ and hence $\theta_t\to\infty$. If $x_t=0$ for some $t\geq t_0$, then $x$ stays at zero by symmetry. Otherwise, \eqref{eq:tube-x-journal} gives
\begin{equation}\label{eq:a2-log-x-step}
 \log|x_{t+1}|-\log|x_t|\leq\log(1-C\phi(\theta_t-1))\leq-C\phi(\theta_t-1).
\end{equation}
Moreover, \eqref{eq:tube-y-journal} and $\theta'(y)\leq2$ give $\theta_{t+1}-\theta_t\leq C_M\phi(\theta_t-1)/\theta_t$, hence $\phi(\theta_t-1)\geq \theta_t(\theta_{t+1}-\theta_t)/C_M$. Combining this with \eqref{eq:a2-log-x-step} yields
\begin{equation}\label{eq:a2-log-x-vs-A}
 \log|x_{t+1}|-\log|x_t|\leq-k_M \theta_t(\theta_{t+1}-\theta_t), \quad k_M:=C/C_M>0.
\end{equation}

The increment bound and $\theta_t\to\infty$ allow us to increase $t_0$ so that $\theta_{t+1}\leq2\theta_t$ for all $t\geq t_0$. Then $\theta_{t+1}^2-\theta_t^2=(\theta_{t+1}+\theta_t)(\theta_{t+1}-\theta_t)\leq3\theta_t(\theta_{t+1}-\theta_t)$, and summing gives $\sum_{t=t_0}^{T-1}\theta_t(\theta_{t+1}-\theta_t)\geq(\theta_T^2-\theta_{t_0}^2)/3$. Summing \eqref{eq:a2-log-x-vs-A} over the same range thus yields
\begin{align*}
 \log|x_T|-\log|x_{t_0}|
 &\leq-k_M\sum_{t=t_0}^{T-1}\theta_t(\theta_{t+1}-\theta_t)
 \leq-\frac{k_M}{3}(\theta_T^2-\theta_{t_0}^2),\\
 |x_T|&\leq|x_{t_0}|\exp\left\{-\frac{k_M}{3}(\theta_T^2-\theta_{t_0}^2)\right\}
 =C_{t_0}e^{-k_M'\theta_T^2},
\end{align*}
where $k_M'=k_M/3$ and $C_{t_0}=|x_{t_0}|e^{k_M'\theta_{t_0}^2}$. Since $\theta(y)\sim y$ and $y_T\to\infty$, any fixed $0<\kappa<k_M'$ satisfies $C_{t_0}e^{-k_M'\theta_T^2}\leq e^{-\kappa y_T^2}$ for all sufficiently large $T$.
\end{proof}

\clearpage
\subsection{Additional Gaussian Calculations}\label{SA:gaussian_additional}

\begin{table}[htbp]
\centering
\begin{tabular}{c|c|ccc}
\hline
$t$ & Observed action & $\mu_t(1)$ & $\mu_t(2)$ & $\mu_t(3)$ \\
\hline
0 & Initial prior & 0.5000 & 0.4000 & 0.1000 \\
1 & $a_2$ & 0.2610 & 0.5626 & 0.1764 \\
2 & $a_2$ & 0.2027 & 0.6286 & 0.1688 \\
3 & $a_2$ & 0.1726 & 0.6726 & 0.1548 \\
4 & $a_2$ & 0.1532 & 0.7045 & 0.1423 \\
5 & $a_2$ & 0.1392 & 0.7289 & 0.1319 \\
6 & $a_1$ & 0.3221 & 0.3986 & 0.2793 \\
7 & $a_1$ & 0.3496 & 0.3579 & 0.2926 \\
\hline
\end{tabular}
\caption{Beliefs along the history $a_2^5a_1^2$ in \Cref{fig:counterexample_gaussian_entry}.}
\label{tab:gaussian_path}
\end{table}

\subsubsection{Repeated \texorpdfstring{$a_1$}{a1} observations without a cascade}\label{SA:gaussian_a1_learning}

The following lemma establishes the path shown in \Cref{fig:counterexample_gaussian_a1}.

\begin{lemma}[Repeated $a_1$ observations]\label{lem:gaussian_a1_learning}
In \Cref{ex:counter_gaussian}, let $\mu_t$ be the public belief after the history $(a_1,\ldots,a_1)$ of length $t$, starting from $\mu_0=(1/2,2/5,1/10)$. Then $\mu_t\notin C_2$ for every $t\geq0$, and $\mu_t\to\delta_1$ as $t\to\infty$.
\end{lemma}

\begin{proof}[Proof of \Cref{lem:gaussian_a1_learning}]
Fix $\theta>0$ and write $G=G_\theta$, where $G_\theta(r)=\Phi(\theta+r)+\Phi(\theta-r)-1$ is the function used in the proof of \Cref{lem:a2-global-journal}. For $r>0$, differentiation gives
\begin{align*}
 -G'(r)&=\phi(r-\theta)-\phi(r+\theta)
 =2e^{-\theta^2/2}\phi(r)\sinh(r\theta)>0,\\
 \frac{G''(r)}{G'(r)}&=-r+\theta\coth(r\theta),\\
 \left[-G'(r)-(1-G(r))\frac{G''(r)}{G'(r)}\right]'
 &=(1-G(r))\left(1+\frac{\theta^2}{\sinh^2(r\theta)}\right)>0.
\end{align*}
We also have
\[
 \frac{\phi(\theta)\sinh^2\theta}{\theta\cosh\theta-\sinh\theta}
 >\int_\theta^\infty\phi(v)\cosh v\,\mathrm{d}v.
\]
To verify this inequality, let $H(\theta)$ denote the difference between the two sides. Direct differentiation gives
\[
 H'(\theta)
 =-\frac{\phi(\theta)\cosh\theta\,(\sinh^2\theta-\theta^2)}
 {(\theta\cosh\theta-\sinh\theta)^2}<0,
\]
where the strict inequality follows from $\sinh\theta>\theta$ for $\theta>0$. Both terms defining $H(\theta)$ tend to zero as $\theta\to\infty$, so $H(\theta)\to0$. Since $H$ is strictly decreasing, $H(\theta)>0$ for every $\theta>0$.

For $r\geq1$, we therefore have
\begin{align*}
 (\log(1-G))''(r)
 &=\frac{G'(r)}{(1-G(r))^2}
 \left[-G'(r)-(1-G(r))\frac{G''(r)}{G'(r)}\right]\\
 &\leq\frac{G'(r)}{(1-G(r))^2}
 \left[-G'(1)-(1-G(1))\frac{G''(1)}{G'(1)}\right]\\
 &=\frac{2e^{-1/2}G'(r)}{(1-G(r))^2}
 \left[\phi(\theta)\sinh\theta
 -(\theta\coth\theta-1)\int_\theta^\infty\phi(v)\cosh v\,\mathrm{d}v\right]\\
 &<0.
\end{align*}
The first inequality holds because the bracketed expression is increasing in $r$ and its prefactor $G'(r)/(1-G(r))^2$ is negative. The last equality substitutes the derivative formulas above and $1-G(1)=2e^{-1/2}\int_\theta^\infty\phi(v)\cosh v\,\mathrm{d}v$; the final bracket is positive by the preceding integral inequality, since $\theta\coth\theta-1>0$.

For $x\geq4$, $y>\bar y$, and $\theta=\theta(y)$, \eqref{eq:gaussian-action-probabilities} gives
\begin{gather*}
 \psi_1^{a_1}=1-G_\theta(1+x/2)
 >\psi_2^{a_1}=1-G_\theta(x/2)
 >\psi_3^{a_1}=1-G_\theta(x/2-1),\\
 x^+-x=\Delta_x^{a_1}(x,y)
 =\log\frac{\psi_1^{a_1}}{\psi_3^{a_1}}>0,\\
 y^+-y=\Delta_y^{a_1}(x,y)
 =\log\frac{1-G_\theta(x/2)}
 {\sqrt{[1-G_\theta(1+x/2)][1-G_\theta(x/2-1)]}}>0.
\end{gather*}
The last inequality uses strict concavity of $\log(1-G_\theta)$, since $x/2-1\geq1$. Hence $\{x\geq4,\ y>\bar y\}$ is invariant under repeated $a_1$ observations and lies outside $C_2$.

Along any path of repeated $a_1$ observations starting in this region, the belief odds $\lambda_{1,t},\lambda_{2,t}$ of \Cref{app:counter_prelim} satisfy, by Bayes' rule,
\[
 \frac{\lambda_{1,t+1}}{\lambda_{1,t}}
 =\frac{\psi_1^{a_1}(x_t,y_t)}{\psi_2^{a_1}(x_t,y_t)}>1,
 \qquad
 0<\frac{\lambda_{2,t+1}}{\lambda_{2,t}}
 =\frac{\psi_3^{a_1}(x_t,y_t)}{\psi_2^{a_1}(x_t,y_t)}<1.
\]
Suppose $\lambda_{1,t}\to\lambda_{1,\infty}<\infty$. If $\lambda_{2,t}\to\lambda_{2,\infty}>0$, the coordinates converge to a finite limit $(x_\infty,y_\infty)$ with $x_\infty\geq4$ and $y_\infty\geq y_0>\bar y$, so
\[
 1=\lim_{t\to\infty}\frac{\lambda_{1,t+1}}{\lambda_{1,t}}
 =\frac{\psi_1^{a_1}(x_\infty,y_\infty)}{\psi_2^{a_1}(x_\infty,y_\infty)}>1.
\]
The equalities use $0<\lambda_{1,\infty}<\infty$ and continuity, respectively, giving a contradiction.

If $\lambda_{2,t}\to0$, the coordinate definitions and $2\cosh\theta(y)=e^{y+1/2}$ give
\begin{align*}
 y_t&=-\tfrac12\log(\lambda_{1,t}\lambda_{2,t})\longrightarrow\infty,\\
 L(x_t,y_t)&=2+x_t/2-\theta(y_t)\\
 &=\tfrac32+\log\lambda_{1,t}+\log(1+e^{-2\theta(y_t)})
 \longrightarrow\tfrac32+\log\lambda_{1,\infty},\\
 U(x_t,y_t)&=L(x_t,y_t)+2\theta(y_t)\longrightarrow\infty.
\end{align*}
Substituting these limits into \eqref{eq:gaussian-action-probabilities} yields the contradiction
\[
 1=\lim_{t\to\infty}\frac{\lambda_{1,t+1}}{\lambda_{1,t}}
 =\frac{\Phi(\tfrac12+\log\lambda_{1,\infty})}
 {\Phi(-\tfrac12+\log\lambda_{1,\infty})}>1,
\]
where the strict inequality follows because $\Phi$ is strictly increasing. Thus $\lambda_{1,t}\to\infty$, and
\[
 0\leq1-\mu_t(1)
 =\frac{1+\lambda_{2,t}}{1+\lambda_{1,t}+\lambda_{2,t}}
 \leq\frac{1+\lambda_{2,0}}{\lambda_{1,t}}\longrightarrow0.
\]
The inequality uses $\lambda_{2,t}\leq\lambda_{2,0}$; hence $\mu_t\to\delta_1$.

Finally, from $\mu_0=(1/2,2/5,1/10)$, we compute the first four updates:
\[
\begin{array}{c|cc}
 t & x_t & y_t\\ \hline
 0 &(1.6094,1.6095)&(0.5815,0.5816)\\
 1 &(2.4292,2.4293)&(0.4831,0.4832)\\
 2 &(3.1815,3.1816)&(0.5586,0.5587)\\
 3 &(3.9497,3.9498)&(0.7139,0.7140)\\
 4 &(4.6851,4.6852)&(0.8990,0.8991)
\end{array}
\]
The displayed bounds give
\[
 \min_{0\leq t\leq4}y_t>0.4831>\frac15>\bar y,
 \qquad x_4>4.6851>4.
\]
Thus the path stays outside $C_2$ and reaches the invariant region, proving the lemma.
\end{proof}

\end{document}